\documentclass[ba]{imsart}
\pubyear{2022}
\volume{TBA}
\issue{TBA}
\arxiv{2110.10216}
\firstpage{1}
\lastpage{1}

\usepackage{amsthm}
\usepackage{amsmath}
\usepackage{natbib}
\usepackage[colorlinks,citecolor=blue,urlcolor=blue,filecolor=blue,backref=page]{hyperref}
\usepackage{graphicx}

\usepackage{amsfonts} 
\usepackage{graphicx,psfrag,epsf}
\usepackage{enumerate}
\usepackage{natbib}
\usepackage{url} 
\usepackage{hyperref}
\usepackage{graphicx}
\usepackage[english]{babel}
\usepackage{bbm}
\usepackage{soul}
\usepackage{dutchcal}
\usepackage{adjustbox}
\usepackage{booktabs}
\usepackage{enumitem}

\startlocaldefs
\numberwithin{equation}{section}
\theoremstyle{plain}

\newtheorem{assumption}{Assumption}

\newcommand{\E}{\mathbb{E}}

\endlocaldefs

\begin{document}
	
	
	\begin{frontmatter}
		\title{A Bayesian Analysis of Two-Stage Randomized Experiments in the Presence of Interference, Treatment Nonadherence, and Missing Outcomes}
		\runtitle{}
		
		\begin{aug}
			\author{\fnms{Yuki} \snm{Ohnishi}\thanksref{addr1,t1,t2,m1}\ead[label=e1]{yohnishi@purdue.edu}} 
			\and
            \author{\fnms{Arman} \snm{Sabbaghi}\thanksref{addr1,t3,m1,m2}\ead[label=e2]{sabbaghi@purdue.edu}}
            
            \runauthor{Y. Ohnishi and A. Sabbaghi.}
            
            \address[addr1]{Purdue University, West Lafayette, IN, USA,
                \printead{e1} 
                \printead*{e2}
            }
			
		\end{aug}
		
		\begin{abstract}
			
			Three critical issues for causal inference that often occur in modern, complicated experiments are interference, treatment nonadherence, and missing outcomes. A great deal of research efforts has been dedicated to developing causal inferential methodologies that address these issues separately. However, methodologies that can address these issues simultaneously are lacking. We propose a Bayesian causal inference methodology to address this gap. Our methodology extends existing causal frameworks and methods, specifically, two-staged randomized experiments and the principal stratification framework. In contrast to existing methods that invoke strong structural assumptions to identify principal causal effects, our Bayesian approach uses flexible distributional models that can accommodate the complexities of interference and missing outcomes, and that ensure that principal causal effects are weakly identifiable. We illustrate our methodology via simulation studies and a re-analysis of real-life data from an evaluation of India’s National Health Insurance Program. Our methodology enables us to identify new active causal effects that were not identified in past analyses. Ultimately, our simulation studies and case study demonstrate how our methodology can yield more informative analyses in modern experiments with interference, treatment nonadherence, missing outcomes, and complicated outcome generation mechanisms.
			
		\end{abstract}
		
		
	\end{frontmatter}
	
	\section{Introduction}
	\label{sec:intro}
	
	Causal inference is a fundamental consideration across a wide range of domains in science, technology, engineering, and medicine \citep{Pearl2009, imbens_rubin_2015}. A traditional gold standard for performing causal inference is the classical randomized experiment \citep{Rubin2008, imbens_rubin_2015}. In this type of experiment, a great deal of control and precautions can be taken so as to eliminate events that would introduce instabilities and biases in causal inferences. However, modern experiments, e.g., social experiments, can become so complicated that it may be difficult to institute such control and precautions. Three significant sources of complications that are increasingly of interest are interference among experimental units, nonadherence/noncompliance of experimental units to their assigned treatments, and unintended missing outcomes of experimental units. Interference exists if the outcome of an experimental unit depends not only on its assigned treatment, but also on the assigned treatments for other units. It arises when limited controls are placed on the interactions of experimental units with one another, or when competition for a limited set of resources exists,
	during the course of an experiment. Treatment nonadherence  frequently occurs in human subject experiments, as it can be unethical to force an individual to take their assigned treatment.
	Clinical trials in particular typically have subjects that do not adhere to their assigned treatments due to adverse side effects or intercurrent events \citep{Little2012}. Missing outcomes commonly occur in human studies. For example, respondents may refuse to report sensitive outcomes (e.g., income) after receiving a treatment in a study \citep[p.~3]{little2002statistical}. Failing to account for interference, nonadherence, and missing outcomes in modern experiments will generally yield unstable and biased inferences on treatment effects.
	
	A great deal of research efforts has been dedicated over the past three decades to developing causal inferential methodologies that address the first two issues separately. \citet{Hudgens2008} first introduced the concept of the two-stage randomized design for performing causal inference in the presence of interference when all experimental units adhere to their assigned treatments. In this design, experimental units belong to clusters, and randomizations are performed at both the cluster-level and experimental unit-level. Specifically, the clusters are first randomly assigned different probabilities for treatment assignment of their constituent experimental units, and then treatments are randomly assigned to the units within the clusters (with treatment assignment performed independently across clusters) based on the cluster-specific assignment probabilities. For example, in our case study, villages correspond to the clusters, and each household within a village corresponds to an experimental unit. In this manner, we can analyze the effectiveness of an insurance plan implemented in India under the two-stage randomized design. \citet{Hudgens2008} demonstrated how both direct treatment effects and indirect treatment effects (i.e., those effects that can be attributed to the treatments received by other units) can be inferred under this design in the presence of interference. This design and the corresponding causal inference methods that can be performed under it have been further studied by \citet{VanderWeele2011}, \citet{Tchetgen2012}, \citet{Liu2014}, and \citet{Basse_avi2018}. For experiments that have treatment nonadherence but not interference, one standard methodology that has been considered for their analyses is the intention-to-treat (ITT) method. Under this approach, the treatment received by an experimental unit is ignored, and instead only the treatment assigned is considered. This method follows the traditional principle of analyzing an experiment according to its physical randomization \citep[p.~14]{CoxReid2000} mechanism, and can yield valid causal inferences on the effects of treatment \emph{assigned} in certain situations. However, the ITT method will generally yield biased inferences on the effects of treatment \emph{received} because it does not account for the latent stratification of experimental units defined according to their adherence behaviors to the different treatment assignments, and consequently does not provide inferences for the target stratum of compliers. \citet{Angrist1996} and \citet{ImbensRubin1997} developed a framework to provide a more principled approach for the analyses of experiments with nonadherence, and \citet{Frangakis2002_1} extended those frameworks to develop the more general principal stratification framework. This framework has been applied to a wide variety of real-life problems involving complications such as censoring or truncation due to death \citep{ZhangRubin2003} and the occurrence of intermediate variables that are thought to mediate the effects of treatments on the outcome \citep{Gallop2009}. \citet{VanderWeele2011_2} provides a detailed review of principal stratification. 
	
	Recently, methodologies have been developed to address combinations of interference, nonadherence, and missing outcomes, but not all of these issues simultaneously. Under the assumption of fully observed outcomes, \citet{Imai2021} presented a nonparametric identification of the complier average direct and indirect effects, and proposed consistent estimators for them under the two-stage randomized design in the presence of interference and nonadherence. They derived large-sample nonparametric bounds for the causal effects, but their results are not necessarily valid for the finite-sample regime. In addition, the estimators of \citet{Imai2021} are sensitive to outliers. \citet{Gonzalo2021} built on the work of \citet{Imai2021} and analyzed spillover effects using instrumental variables in the two-stage randomized experiment with fully observed outcomes. They considered the identification of causal direct and spillover effects under one-sided noncompliance and demonstrated that these effects can be estimated using two-stage least squares (2SLS). However, their methods do not work in general under two-sided noncompliance or when units have multiple peers without a strong structural assumption about peers' compliance types. \citet{Kang2016} considered the peer encouragement design to study network treatment effects when treatment randomization cannot feasibly be forced on experimental units, and presented identification results only for the case of one-sided noncompliance. \citet{Forastiere2016} developed a Bayesian principal stratification method for causal inference in clustered encouragement designs (CEDs), where the assignment of treatment encouragement is performed at the cluster level. The CED is effectively a special case of the two-stage randomized experiment with no randomization within clusters, i.e., with treatment assignment probability for units within clusters being either zero or one. As there is no randomization performed at the unit level within clusters, their methodology cannot capture the complexity of adherence behaviors in the two-stage experiments, where units could possibly change their adherence behaviors depending on what assignment probabilities their clusters are assigned to. It is important to note that none of the above methods can easily accommodate missing outcomes in the presence of interference and nonadherence. Experiments with treatment nonadherence and missing outcomes have been analyzed by \citet{FrangakisRubin1999}, \citet{Mattei2007}, \citet{Frumento2012}, and \citet{Mattei2014}, but none of these studies involved interference.
	
	We develop a new Bayesian causal inferential methodology for two-stage randomized experiments with interference, noncompliance, and missing outcomes. Our methodology utilizes the principal stratification framework to address the identification issues arising from treatment nonadherence and missing outcomes in the presence of interference. To the best of our knowledge, none of the existing causal inference methods have applied Bayesian principal stratification to two-stage randomized designs and addressed these issues simultaneously as our method does. Not only does our Bayesian approach provide a principled framework to analyze two-stage randomized experiments in the presence of all of these complications, but it clarifies what can be learned when causal estimands are not identifiable but are instead weakly identifiable (i.e., when the likelihood functions of parameters and causal estimands have substantial regions of flatness). It is important to recognize that issues of identifiability under the Bayesian paradigm are distinct from those under the frequentist paradigm because the specification of proper prior distributions always yields proper posterior distributions \citep{ImbensRubin1997}. 
	In particular, our Bayesian method enables us to infer principal causal effects under two-sided nonadherence without the need for strong structural assumptions, and define new types of interpretable and informative causal estimands (such as the complier spillover and overall treatment effects) that would be of great interest to policy makers. Furthermore, our methodology highlights new assumptions on compliance types and missingness mechanisms for making causal inferences more efficient and stable in such complicated experiments. The use of Bayesian models in our methodology enables us to naturally accommodate complicated outcome generation mechanisms that frequently occur in modern experiments, such as heavy-tailed, skewed, zero-inflated, and/or multi-modal outcome distributions. 
	
	We proceed in Section \ref{sec:preliminaries} to review the Rubin Causal Model \citep{Rubin1974, Holland1986}, the principal stratification framework, and relevant assumptions and causal estimands for the two-stage randomized experiment. Section \ref{sec:methodology} introduces the models and computational algorithms involved in our Bayesian methodology. In Section \ref{sec:simulationstudies} we perform extensive simulation studies to investigate the frequentist performance of our Bayesian method for a heavy-tailed distribution with an excess of zero outcomes. These simulation studies effectively validate our methodology in a situation where the existing frequentist approach \citep{Imai2021} performs poorly in terms of bias and mean squared error (MSE). Finally, in Section \ref{sec:application} we apply our methodology to the real-life data from the evaluation of India’s National Health Insurance Program (RSBY) \citep{Nandi2015}. In our analysis of this case study we are able to uncover more definitive evidence of causal effects that were previously found to be insignificant in past analyses. Our concluding remarks are in Section \ref{sec:conclusion}.

	\section{Background}
	\label{sec:preliminaries}
	
	\subsection{Two-Stage Randomized Experiments with Interference and Noncompliance}
	\label{sec:two_stage_randomization}
	
	Throughout this manuscript we consider two-stage randomized experiments involving two treatments and $J$ clusters, with $N_j$ experimental units in cluster $j = 1, \ldots, J$ (each experimental unit belongs to only one cluster). We let $N = \sum_{j=1}^J N_j$ denote the total number of experimental units. The assignment mechanism in the two-stage randomized experiment is performed sequentially, with each stage involving a type of completely randomized design. In the first stage, $J_1$ clusters are randomly chosen to have a treatment assignment probability of $a_1 \in (0,1)$ for their constituent experimental units, and the remaining $J - J_1$ clusters have a treatment assignment probability of $a_0 \in (0,1)$ for their units. For each $j = 1, \ldots, J$ we let $A_j$ be the indicator for whether cluster $j$ was assigned $a_1$ ($A_j = 1)$ or $a_0$ ($A_j = 0$). We let $\mathbf{A} = (A_1, \ldots, A_J)^{\mathsf{T}}$, and without loss of generality we let $a_1 > a_0$. In the second stage, the experimental units within the clusters are randomly assigned treatment and control according to the treatment assignment probabilities assigned to their clusters, with the treatment assignment performed independently across clusters. For each cluster $j$ with $A_j = 1$, $N_ja_1$ of their experimental units are randomly assigned treatment and the remaining $N_j(1-a_1)$ are assigned control. Similarly, for each cluster $j'$ with $A_{j'} = 0$, $N_{j'}a_0$ of their experimental units are randomly assigned treatment and the remaining $N_{j'}(1-a_0)$ are assigned control. We assume that $N_ja_1$ and $N_ja_0$ are integers for all $j = 1, \ldots, J$. We let $Z_{i,j}$ denote the treatment assignment indicator for unit $i$ in cluster $j$, with $Z_{i,j} = 1$ if it is assigned treatment and $Z_{i,j} = 0$ otherwise. We let $\mathbf{Z}_j = (Z_{1,j}, \ldots, Z_{N_j,j})^{\mathsf{T}}$ denote the vector of treatment assignment indicators for all units in cluster $j$, and $\mathbf{Z}_{-i,j}$  denote the subvector of $\mathbf{Z}_j$ with the $i$th entry removed. Other assignment mechanisms for two-stage randomized experiments are provided by \citet{VanderWeele2011}, but we do not consider them here.
	
	As we consider two-stage randomized experiments with interference, nonadherence and missingness under the Rubin Causal Model, we must introduce two types of potential outcomes for the treatment received by an experimental unit and the final outcome of interest that are functions of the experimental units' treatment assignments. We let $D_{i,j}(\mathbf{z})$ denote the treatment received for unit $i$ in cluster $j$ under treatment assignment $\mathbf{z} \in \{0,1\}^N$, $\mathbf{D}_j(\mathbf{z}) = (D_{1,j}(\mathbf{z}),...,D_{N_{j},j}(\mathbf{z}))^{\mathsf{T}}$ be the vector of treatments received by the units in cluster $j$, and $\mathbf{D}(\mathbf{z}) = (\mathbf{D}_1(\mathbf{z}), \ldots, \mathbf{D}_J(\mathbf{z}))^{\mathsf{T}}$. Furthermore, we let $Y_{i, j}(\mathbf{z}, \mathbf{D}(\mathbf{z}))$ denote the potential outcome for unit $i$ in cluster $j$ under treatment assignment vector $\mathbf{z}$ and treatment received vector $\mathbf{D}(\mathbf{z})$. Although the potential outcomes can be written solely as a function of $\mathbf{z}$ (because $\mathbf{D}(\mathbf{z})$ is a function of $\mathbf{z}$) we include $\mathbf{D}(\mathbf{z})$ in the notation for $Y_{i,j}(\mathbf{z}, \mathbf{D}(\mathbf{z}))$ to emphasize the existence of nonadherence. We let $\mathbf{D}$ and $\mathbf{Y}$ denote the matrices containing all the potential values of treatment receipts and outcomes, respectively, for all experimental units across all treatment assignments.
	
	Finally, we let $M_{i,j}(\mathbf{z})$ denote the missingness indicator for the realized outcome of unit $i$ in cluster $j$ under treatment assignment $\mathbf{z} \in \{0,1\}^N$, $\mathbf{M}_j(\mathbf{z}) = (M_{1,j}(\mathbf{z})$, $\ldots ,M_{N_{j},j}(\mathbf{z}))^{\mathsf{T}}$ be the vector of missingness indicators of the units in cluster $j$, and $\mathbf{M}(\mathbf{z}) = (\mathbf{M}_1(\mathbf{z}), \ldots, \mathbf{M}_J(\mathbf{z}))^{\mathsf{T}}$. We note that $Y_{i, j}(\mathbf{z}, \mathbf{D}(\mathbf{z}))$ is observed when $M_{i,j}(\mathbf{z}) = 0$, and is missing otherwise. It is important to distinguish between missing potential outcomes and unrealized potential outcomes. For a given treatment assignment vector $\mathbf{z}$ that was realized in an experiment, the potential outcomes $Y_{i,j}(\mathbf{z}', \mathbf{D}(\mathbf{z}'))$ for any other distinct treatment assignment vector $\mathbf{z}'$ are referred to as unrealized potential outcomes. Also, for a given treatment assignment vector $\mathbf{z}$ that was realized in an experiment, if $M_{i,j}(\mathbf{z}) = 1$ then $Y_{i,j}(\mathbf{z}, \mathbf{D}(\mathbf{z}))$ is not observed, although the potential outcome was realized in the experiment. 
	
	Besides the treatments, potential outcomes, and missingness indicators, we assume covariates are measured for the experimental units. These covariates are either measured prior to treatment assignment, or are measured afterwards but are not affected by treatment assignment. We denote the vector of covariates for unit $i$ in cluster $j$ by $\mathbf{X}_{i,j}$.
	
	\subsection{Assumptions on the Structure of Interference}
	\label{sec:assumptions_interference}
	
	We extend the assumptions proposed by \citep{Imai2021} for the two-stage randomized experiment with nonadherence. 
	We first invoke the partial interference assumption in which units in different clusters do not interact or affect one another. This assumption was first formulated by \citet{Hudgens2008}, and was extended to the noncompliance setting by \citet{Imai2021}. Partial interference facilitates causal inference in our setting of interest because an experimental unit $i$'s treatment received, missingness indicator will only be functions of treatment assignments for other units within the same cluster $j$ as unit $i$. 
	
	\begin{assumption}
		\label{asmp:partial}
		For all $\mathbf{z}, \mathbf{z}' \in \{0, 1\}^N$ such that $\mathbf{z}_j = \mathbf{z}_j'$ for a cluster $j$, then $D_{i,j}(\mathbf{z}) = D_{i,j}(\mathbf{z}')$ and $M_{i,j}(\mathbf{z}) = M_{i,j}(\mathbf{z}')$ for all experimental units $i$ in cluster $j$. 
	\end{assumption}
	
	The next assumption that we consider is the stratified interference assumption of \citet{Hudgens2008}. This assumption imposes further structure on interference by having the treatment received, the missingness indicator, and the potential outcome for an experimental unit being a function of just the number of experimental units assigned treatment within the same cluster. This assumption was also considered by \citet{Forastiere2016} and \citet{Imai2021}. It is important to recognize that a great deal of work has been conducted to move beyond this condition and consider more flexible structures of interference \citep{Aronow2012, Manski2013, Basse2015, Aronow2017, Baird2017, Basse_Airoldi2018, Athey2018, Basse2019, Leung2020, Forastiere2021, Fredrik2021}. However, none of these works considered noncompliance and missing outcomes.
	
	\begin{assumption}
		\label{asmp:stratified}
		For a cluster $j$ and experimental unit $i$ in cluster $j$, if $\mathbf{z}, \mathbf{z}' \in \{0,1\}^N$ such that $z_{i,j} = z_{i,j}'$ and $\mathbf{z}_j^{\mathsf{T}}\mathbf{1} = (\mathbf{z}_j')^{\mathsf{T}}\mathbf{1}$, then $D_{i,j}(\mathbf{z}) = D_{i,j}(\mathbf{z}')$ and $M_{i,j}(\mathbf{z}) = M_{i,j}(\mathbf{z}')$.
	\end{assumption}
	
	\noindent We invoke the same assumptions on outcomes.
	
	\begin{assumption}
		\label{asmp:partial_Y}
		For all $\mathbf{z}, \mathbf{z}' \in \{0, 1\}^N$ such that $\mathbf{z}_j = \mathbf{z}_j'$ for a cluster $j$, then  $Y_{i,j}(\mathbf{z}, \mathbf{D}(\mathbf{z})) = Y_{i,j}(\mathbf{z}', \mathbf{D}(\mathbf{z}'))$ for all experimental units $i$ in cluster $j$. 
	\end{assumption}
	
	\begin{assumption}
		\label{asmp:stratified_Y}
		For a cluster $j$ and experimental unit $i$ in cluster $j$, if $\mathbf{z}, \mathbf{z}' \in \{0,1\}^N$ such that $z_{i,j} = z_{i,j}'$ and $\mathbf{z}_j^{\mathsf{T}}\mathbf{1} = (\mathbf{z}_j')^{\mathsf{T}}\mathbf{1}$,  $Y_{i,j}(\mathbf{z}, \mathbf{D}(\mathbf{z})) = Y_{i,j}(\mathbf{z}', \mathbf{D}(\mathbf{z}'))$.
	\end{assumption}
	
	Finally, we assume the exclusion restriction with interference between units in the two-stage randomized experiment. In this case, the potential outcome of a unit $i$ in cluster $j$ only depends on the treatments received by units within cluster $j$. 
	
	\begin{assumption}
		\label{asmp:exclusion_intf}
		For any treatment assignment vectors $\mathbf{z}, \mathbf{z}' \in \{0, 1\}^N$ with $\mathbf{D}_j(\mathbf{z}) = \mathbf{D}_j(\mathbf{z}')$, $Y_{i,j}(\mathbf{z}, \mathbf{D}(\mathbf{z})) = Y_{i,j}(\mathbf{z}',\mathbf{D}(\mathbf{z}'))$ for all experimental units $i$ in cluster $j$.
	\end{assumption}
	
	These five assumptions imply that for two-stage randomized experiments in which the number of treated units within any cluster is fixed by design, the potential outcomes and missingness indicator for an experimental unit are a function of its own treatment assignment and the treatment assignment probability for its cluster. We accordingly slightly abuse the notation to write $D_{i,j}(\mathbf{z})$, $M_{i,j}(\mathbf{z})$  and $Y_{i,j}(\mathbf{z}, \mathbf{D}(\mathbf{z}))$ as $D_{i,j}(z,a)$, $M_{i,j}(z,a)$ and $Y_{i,j}(z,a)$, respectively, where $z$ denotes the treatment assignment for the experimental unit and $a$ denotes the treatment assignment probability for the unit's cluster.
	\subsection{Principal Strata, Monotonicity, and the Exclusion Restriction for Two-Stage Randomized Experiments}
	\label{sec:principal_strata}
	
	Under the principal stratification framework, we stratify the experimental units according to their values of $D_{i,j}(z,a)$ under the different possible treatment assignments $z \in \{0,1\}$ and treatment assignment probabilities $a \in \{a_0, a_1\}$ for the clusters. There exist four such potential values: $D_{i,j}(0,a_0), D_{i,j}(1,a_0), D_{i,j}(0,a_1)$, and $D_{i,j}(1,a_1)$. A unique feature of our consideration of nonadherence for the two-stage randomized design is that, according to Assumption \ref{asmp:stratified} and \ref{asmp:stratified_Y}, units can have different compliance behaviors under different assignment probabilities for their clusters. We formally define the compliance behavior of unit $i$ in cluster $j$ under each treatment assignment probability $a \in \{a_0, a_1\}$ for the cluster as
	\begin{equation*}
	G_{i,j}(a) = 
	\begin{cases}
	\mathcal{n} & \text{if $D_{i,j}(0,a) = D_{i,j}(1,a) = 0$,} \\
	\mathcal{c} & \text{if $D_{i,j}(0,a) = 0$, $D_{i,j}(1,a) = 1$,} \\
	\mathcal{d} & \text{if $D_{i,j}(0,a) = 1$, $D_{i,j}(1,a) = 0$,} \\
	\mathcal{a} & \text{if $D_{i,j}(0,a) = D_{i,j}(1,a) = 1$,}\\
	\end{cases}
	\end{equation*}
	where $\mathcal{n}, \mathcal{c}, \mathcal{d}$, and $\mathcal{a}$ denote never-takers, compliers, defiers, and always-takers, respectively. Finally, the compliance behavior for unit $i$ in cluster $j$ is defined according to the pair of compliance indicators $G_{i,j} = ( G_{i,j}(a_0), G_{i,j}(a_1))$. 
	
	A standard assumption for the compliance behavior is monotonicity. 
	\begin{assumption}
		\label{asmp:monotonocity_fixed}
		For all units $i = 1, \ldots, N_j$ in cluster $j =  1, \ldots, J$ and any $a \in \{a_0, a_1\}$, $D_{i,j}(1,a) \geq D_{i,j}(0,a)$, and strict inequality exists for at least one experimental unit.
	\end{assumption}
	\noindent This assumption was also considered by \citet{Imai2021} and \citet{Forastiere2016}. Monotonicity eliminates the possibility of defiers under either treatment assignment probabilities $a_0$ and $a_1$. It also reduces the number of principal strata from sixteen to nine.
	
	In addition to this existing monotonicity assumption, we also consider the following new assumption on compliance behaviors across the different treatment assignment probabilities that can be assigned to the clusters.
	\begin{assumption}
		\label{asmp:monotonocity_across_mech}
		The set $\{\mathcal{n}, \mathcal{c}, \mathcal{a}\}$ is a partial order set. For $a_0<a_1$, units whose clusters have been assigned $a_1$ would have a non-strictly lower compliance type of $\{\mathcal{n}, \mathcal{c}, \mathcal{a}\}$ than the current compliance type if their clusters were assigned $a_0$.  Units whose clusters have been assigned $a_0$ would have a non-strictly greater compliance type of $\{\mathcal{n}, \mathcal{c}, \mathcal{a}\}$ than the current compliance type if their clusters were assigned $a_1$.  
	\end{assumption}
	
	\noindent This assumption further reduces the number of principal strata. To see this, we first recognize that there are only six possible definitions of partial orders for $\{\mathcal{n}, \mathcal{c}, \mathcal{a}\}$: $\mathcal{n} \leq \mathcal{c} \leq \mathcal{a}$, $\mathcal{n} \leq \mathcal{a} \leq \mathcal{c}$, $\mathcal{c} \leq \mathcal{n} \leq \mathcal{a}$, $\mathcal{c} \leq \mathcal{a} \leq \mathcal{n}$, $\mathcal{a} \leq \mathcal{n} \leq \mathcal{c}$, and $\mathcal{a} \leq \mathcal{c} \leq \mathcal{n}$. We adopt the partial orders $\mathcal{n} \leq \mathcal{c} \leq \mathcal{a}$ as well as $a_0 \leq a_1$ throughout. These correspond to units being more likely to take treatment if a larger proportion of their neighbors take treatment. Alternatively, if a cluster is assigned $a_1$ then its units are more likely to receive treatment, no matter what the units are assigned, compared to the case if the cluster is assigned $a_0$. The combination of Assumptions \ref{asmp:monotonocity_fixed} and \ref{asmp:monotonocity_across_mech} thus reduces the number of principal strata to six: $\{(\mathcal{n},\mathcal{n}),(\mathcal{c},\mathcal{c}),(\mathcal{a},\mathcal{a}),(\mathcal{n},\mathcal{c}),(\mathcal{n},\mathcal{a}),(\mathcal{c},\mathcal{a})\}$. 
	
	
	Our definition of principal strata based on Assumptions \ref{asmp:monotonocity_fixed} and \ref{asmp:monotonocity_across_mech} is more general than existing definitions. For example, \citet{Imai2021} considered the monotonicity assumption with respect to the treatment assignment probability. Their assumption corresponds to the partial order $\mathcal{n} \leq \mathcal{c} \leq \mathcal{a}$ but can not express other orderings of monotonicity, e.g., $\mathcal{a} \leq \mathcal{c} \leq \mathcal{n}$. \citet{Forastiere2016} only defined three principal strata based on the treatment uptake status, excluding defiers. Also, \citet{Gonzalo2021} defined five principal strata by posing monotonicity directly on treatment received, which led to the removal of strata $(\mathcal{n}, \mathcal{a})$ from consideration. These distinctions exist because we consider the two-stage randomized design, and we define the monotonicity assumptions on the compliance behaviors with respect to the treatment probabilities for clusters. \citet{Forastiere2016} considered clustered encouragement designs (CEDs) in which encouragement is randomized at the level of clusters but with no randomization carried out within clusters. In contrast, the two-stage randomized design has encouragement randomized at the level of units within clusters, with clusters assigned a treatment probability that governs the proportion of treated units within the cluster. The latter design generates more complicated structures for the units' compliance behaviors because some units might behave differently based on the treatment probabilities assigned to their clusters. This distinction is critical because of our need to capture behavioral shifts of units between such treatment probabilities, and because of the direct and spillover causal estimands of interest that are defined in Section \ref{sec:estimand1}. 
	Our definitions of the monotonicity assumptions provide more flexible orderings of compliance behaviors than that of \citet{Gonzalo2021} because we permit more relevant orderings.

	In addition to monotonicity, we also consider the exclusion restriction for certain principal strata in the case of nonadherence.
	\begin{assumption}
		\label{asmp:exclusion}
		For any unit $i = 1, \ldots, N_j$ in cluster $j = 1, \ldots, J$:
		\begin{itemize}
			\item if $G_{i,j} \in \{ (\mathcal{a},\mathcal{a}), (\mathcal{n}, \mathcal{n}), (\mathcal{n}, \mathcal{a})\}$, then $Y_{i,j}(0,a) = Y_{i,j}(1,a)$ for $a \in \{a_0, a_1\}$, 
			\item if $G_{i,j} = (\mathcal{c}, \mathcal{a})$, $Y_{i,j}(0,a_1) = Y_{i,j}(1,a_1)$, and
			\item if $G_{i,j} = (\mathcal{n}, \mathcal{c})$, $Y_{i,j}(0,a_0) = Y_{i,j}(1,a_0)$.
		\end{itemize}
	\end{assumption}
	\noindent This assumption captures the idea that treatment assignment has no effect on the outcome if the unit is either an always-taker or a never-taker under each treatment assignment probability. It also corresponds to Assumption \ref{asmp:exclusion_intf}, from which the outcome of an experimental unit is determined only through the treatment received by units within the same cluster. This assumption is equivalent to a modified form of the restricted interference assumption of \citet{Imai2021} that is stated in terms of principal strata.
	
	\subsection{Direct, Spillover, and Overall Causal Estimands}
	\label{sec:estimand1}
	
	We consider the finite-population framework for causal estimands under the Rubin Causal Model. In this case, estimands are defined in terms of comparisons of potential outcomes for the $N$ experimental units. Our approach for defining direct, spillover, and overall causal estimands in two-stage randomized experiments with interference and nonadherence follows that of \citet{Hudgens2008} and \citet{Imai2021}. 
	
	To simplify our expressions of the causal estimands, we first write the treatment received and potential outcome for unit $i$ in cluster $j$ under treatment assignment $\mathbf{z}$ as $D_{i,j}(\mathbf{z})$ and $Y_{i,j}(\mathbf{z}, \mathbf{D}(\mathbf{z}))$, respectively. We let $K_j(a)$ denote the number of treated units in cluster $j$ under treatment assignment probability $a \in \{a_0, a_1\}$, and $\mathcal{Z}_{-i,j}$ denote the set of all subvectors of $\mathbf{z}_{j} \in \{0, 1\}^{N_j}$ with the $i$th element removed such that $\sum_{i=1}^{N_j} z_{i,j} = K_{j}(a)$. Under Assumptions \ref{asmp:partial} - \ref{asmp:stratified_Y}, we recognize that $\Bar{D}_{i,j}(z,a)  = \sum_{\mathbf{z}_{-i,j} \in \mathcal{Z}_{-i,j}} D_{i,j}(\mathbf{z}) p(\mathbf{Z}_{-i,j}=\mathbf{z}_{-i,j} \mid Z_{i,j} = z, A_j=a) = D_{i,j}(z, a)$ and $\Bar{Y}_{i,j}(z,a)  = \sum_{\mathbf{z}_{-i,j} \in \mathcal{Z}_{-i,j}} Y_{i,j}(\mathbf{z}, \mathbf{D}(\mathbf{z})) p(\mathbf{Z}_{-i,j}=\mathbf{z}_{-i,j} \mid Z_{i,j}=z, A_j=a) = Y_{i,j}(z, a)$.
	
	One set of causal estimands that we consider are the unit-level direct intention-to-treat (ITT) effects of treatment assignment on treatment received and the final outcome under treatment assignment probability $a \in \{a_0, a_1\}$, i.e., $\mathrm{ITT}_{D,i,j}(a) = \Bar{D}_{i,j}(1,a) - \Bar{D}_{i,j}(0,a)$ and $\mathrm{ITT}_{Y,i,j}(a) = \Bar{Y}_{i,j}(1,a) - \Bar{Y}_{i,j}(0,a)$, respectively. These estimands capture the adherence behavior and changes in the final outcome under the same treatment assignment probability $a$ when unit $i$ in cluster $j$ is assigned treatment as opposed to control. We average the unit-level effects to define the cluster-level and finite population-level ITT effects as $\mathrm{ITT}_{D, \cdot, j}(a) = \sum_{i=1}^{N_j} \mathrm{ITT}_{D,i,j}(a)/N_j$, $\mathrm{ITT}_{D, \cdot, \cdot}(a) = \sum_{j=1}^{J} N_{j}\mathrm{ITT}_{D, \cdot, j}(a)/N$, $\mathrm{ITT}_{Y, \cdot, j}(a) \sum_{i=1}^{N_j} \mathrm{ITT}_{Y, i, j}(a)/N_j$, $\mathrm{ITT}_{Y, \cdot, \cdot}(a) = \sum_{j=1}^{J} N_{j}\mathrm{ITT}_{Y, \cdot, j}(a)/N$.
	
	Another causal estimand of interest in the presence of interference is the spillover (indirect) effect of treatments assigned to other units on the potential outcomes for a particular experimental unit. Following \citet{Imai2021} we define unit-level spillover effects on treatment receipt and outcome as $S_{D, i, j}(z) = \Bar{D}_{i,j}(z,a_1) - \Bar{D}_{i,j}(z,a_0)$ and $S_{Y,i,j}(z) = \Bar{Y}_{i,j}(z,a_1) - \Bar{Y}_{i,j}(z,a_0)$. In these two estimands, we consider the average potential outcomes corresponding to the two different treatment assignment probabilities $a_0$ and $a_1$ for cluster $j$ but the same treatment assignment $z$ for unit $i$ in cluster $j$. These estimands quantify the intuition that, if differences exist in potential outcomes under the same treatment assignment $z$, then they can be attributed to the first stage of the experiment that governs the proportion of treated units in each cluster. This is because under Assumption \ref{asmp:stratified} and \ref{asmp:stratified_Y} the treated units in a particular unit $i$'s cluster are the only factor besides the assigned treatment that can affect unit $i$'s outcomes, and so the $S_{D, i, j}(z)$ and $S_{Y, i, j}(z)$ can be reasonably regarded as spillover effects. We define cluster-level and population-level spillover effects by averaging the $S_{D,i,j}(z)$ and $S_{Y,i,j}(z)$ across $i$ and $j$, respectively. 
	
	The final estimand that we consider is the overall effect of the first stage of the experiment, i.e., the effect of having treatment assignment probability $a_1$ versus $a_0$ for a cluster. This effect is usually of greatest interest for policy makers. For example, infectious disease experts may be interested in comparing infection rates under two different vaccine allocation plans (e.g., $40\%$ and $80\%$) within each cluster. We define unit-level overall effects of the first stage on treatment receipt and outcome as $O_{D,i,j} = \Bar{D}_{i,j}(a_1)-\Bar{D}_{i,j}(a_0)$ and $O_{Y,i,j} = \Bar{Y}_{i,j}(a_1)-\Bar{Y}_{i,j}(a_0)$, respectively, where $\Bar{Y}_{i,j}(a) = \sum_{\mathbf{z}_{j} \in \mathcal{Z}_{j}} Y_{i,j}(\mathbf{z})
	p(\mathbf{Z}_{j} = \mathbf{z}_{j} \mid A_j=a)$ and $\mathcal{Z}_{j}$ is a set of all possible assignment vectors $\mathbf{z}_{j}$. We have that $\Bar{Y}_{i,j}(a)$ is effectively the average value of the individual's outcome under the treatment assignment probability $a \in \{a_0, a_1\}$. It is decomposed into
	\begin{equation}
	\label{eq:Y_a}
	\begin{split}
	\Bar{Y}_{i,j}(a)  = \left ( \frac{K_j(a)}{N_j} \right ) \Bar{Y}_{i,j}(1,a) + \left ( \frac{N_j-K_j(a)}{N_j} \right ) \Bar{Y}_{i,j}(0,a). \\
	\end{split}
	\end{equation}
	The proof of equation \eqref{eq:Y_a} is provided in the appendix. Plugging this into $O_{Y,i,j}$, we have $O_{Y,i,j} = \left \{ K_j(a_1)/N_j \right \} \mathrm{ITT}_{Y,i,j}(a_1) - \left \{ K_j(a_0)/N_j \right \} \mathrm{ITT}_{Y,i,j}(a_0) + S_{Y,i,j}(0)$. We average the $O_{Y,i,j}$ to define the cluster-level and population-level overall effects as $O_{Y,\cdot,j} = \sum_{i=1}^{N_j}O_{Y,i,j}/N_j$ and  $O_{Y,\cdot,\cdot} = \sum_{j=1}^{J} N_{j}O_{Y,\cdot,j}/N$, respectively. \citet{VanderWeele2011} provided the same decomposition but with a slightly different proof. The overall effect is expressed as the sum of the spillover effect and the contrast of two ITT effects under the treatment assignment probability $a_1$ and under the treatment assignment probability $a_0$, each of which are multiplied by the proportion of treated units under that assignment probability.
	
	\subsection{Principal Causal Estimands}
	\label{sec:estimands2}
	
	In addition to defining direct, spillover, and overall causal estimands, we define new principal causal estimands. These estimands extend those in Section \ref{sec:estimand1} to account for compliers under the different treatment probabilities $a_0$ and $a_1$ that can be assigned to clusters. 
	
	\citet{Imai2021} used Assumption \ref{asmp:stratified} and \ref{asmp:stratified_Y} to define the complier average direct effect under treatment probability $a \in \{a_0, a_1\}$ as
	\begin{equation}
	\label{eq:CADE_1}
	\left [ \sum_{j=1}^J \sum_{i=1}^{N_j} \{Y_{i,j}(1,a) - Y_{i,j}(0,a)\} \mathbbm{1}(G_{i,j}(a)=\mathcal{c}) \right ] \bigg{/} \sum_{j=1}^J \sum_{i=1}^{N_j} \mathbbm{1}(G_{i,j}(a)=\mathcal{c}).
	\end{equation} 
	\noindent where $\mathbbm{1}(\cdot)$ denotes the indicator for an event. We define a new complier average direct effect by generalizing the estimand in equation (\ref{eq:CADE_1}) to account for how experimental units comply under the different treatment probabilities assigned to clusters. This new definition is motivated by the practical interest in modifying the effect of $a_1$ to account for the units who would comply under $a_0$, and vice versa. To define this new complier average direct effect, we introduce the notions of the ``base cluster assignment'' as the treatment probability for a cluster under which compliers are defined, and the ``target cluster assignment'' as the treatment probability for a cluster under which causal effects are considered. It is important to separate the base from the target cluster assignments because the treatments received by units are influenced by the treatments assigned to other units, and because compliance behaviors vary with the treatment probabilities $a_0$ and $a_1$. For a target cluster assignment $a$ and base cluster assignment $a'$, our new population-level complier average direct effect is defined as
	\begin{equation}
	\label{eq:CADE_new}
	\mathrm{CADE}(a,a') = \left [ \sum_{j=1}^J \sum_{i=1}^{N_j} \left \{ Y_{i,j}(1,a) - Y_{i,j}(0,a) \right \} \mathbbm{1}(G_{i,j}(a') = \mathcal{c}) \right ] \bigg{/} \sum_{j=1}^J \sum_{i=1}^{N_j} \mathbbm{1}(G_{i,j}(a')=\mathcal{c}).
	\end{equation}
	Equation (\ref{eq:CADE_1}) is a special case of equation (\ref{eq:CADE_new}) with $a = a'$.
	
	The complier average spillover effect for treatment $z$ defined by \citet{Imai2021} is the ratio of $\left [ \sum_{j=1}^{J}\sum_{i=1}^{N_j} \left \{ Y_{i,j}(z,a_1) - Y_{i,j}(z,a_0) \right \} \mathbbm{1}(D_{i,j}(z,a_1)=1, D_{i,j}(z,a_0)=0 ) \right ]$ and $\sum_{j=1}^{J}\sum_{i=1}^{N_j} \mathbbm{1}(D_{i,j}(z,a_1)=1, D_{i,j}(z,a_0)=0)$, which is equivalent to
	\begin{align}
	\label{eq:LATE_estimand}
	& \left [ \sum_{j=1}^J \sum_{i=1}^{N_j} \left \{ Y_{i,j}(z,a_1) - Y_{i,j}(z,a_0) \right \} \mathbbm{1}(G_{i,j}\in \{(\mathcal{c},\mathcal{a}), (\mathcal{n},\mathcal{a})\}) \right ] \bigg{/} \sum_{j=1}^J \sum_{i=1}^{N_j} \mathbbm{1}(G_{i,j}\in \{(\mathcal{c},\mathcal{a}), (\mathcal{n},\mathcal{a})\}).
	\end{align}
	The estimand in equation (\ref{eq:LATE_estimand}) is interpreted as a local average treatment effect for units who comply with the treatment probability assigned to the cluster. These estimands are not clearly interpretable as complier spillover effects because the compliance is defined for the compliance behaviors with respect to treatment probability, not the actual treatment assignment. We define a new complier average spillover effect using base and target cluster assignments. For treatment $z$ and base cluster assignment $a'$, we define $\mathrm{CASE}(z, a')$ as the ratio of $\left [ \sum_{j=1}^J \sum_{i=1}^{N_j} \left \{ Y_{i,j}(z,a_1) - Y_{i,j}(z,a_0) \right \} \mathbbm{1}(G_{i,j}(a')=\mathcal{c}) \right ]$ and $\sum_{j=1}^J \sum_{i=1}^{N_j} \mathbbm{1}(G_{i,j}(a')=\mathcal{c})$. Similar to the population-level spillover effect on outcome defined in Section \ref{sec:estimand1}, $\mathrm{CASE}(z, a')$ captures the compliers' local average spillover effect by comparing the two sets of potential outcomes under the two cluster assignments $a_0$ and $a_1$ and treatment $z$. In this estimand, the compliers are defined under the base cluster assignment $a'$, and it is interpretable as an average effect for the compliers of the treatment received by others in the same cluster.
	
	We finally define the finite-population complier average overall effect by combining equation \eqref{eq:Y_a} with the unit-level complier average overall effect under the base cluster assignment $a'$, $\{\bar{Y}_{i,j}(a_1) - \bar{Y}_{ij}(a_0)\}\mathbbm{1}(G_{i,j}(a')=\mathcal{c})$, and decomposing the effect according to the sum of the unit-level complier average direct and spillover effects. Specifically, we have
	\begin{align}
	\label{eq:CAOE_estimand}
	\mathrm{CAOE}_{i,j}(a') &= \left \{ \frac{K_j(a_1)}{N_j} \right \} \left \{ Y_{i,j}(1,a_1) - Y_{i,j}(0,a_1) \right \} \mathbbm{1}(G_{i,j}(a') = \mathcal{c}) \nonumber \\
	& \ \ \ - \left \{ \frac{K_j(a_0)}{N_j} \right \} \left \{ Y_{i,j}(1,a_0) - Y_{i,j}(0,a_0) \right \} \mathbbm{1}(G_{i,j}(a') = \mathcal{c}) \nonumber \\
	& \ \ \ + \left \{ Y_{i,j}(0,a_1) - Y_{i,j}(0,a_0) \right \} \mathbbm{1}(G_{i,j}(a')=\mathcal{c}).
	\end{align}
	The population-level effect is then the average of the unit-level effects in equation (\ref{eq:CAOE_estimand}) over compliers under the base cluster assignment $a'$, i.e., 
	\begin{equation}
	\label{eq:CAOE_population_estimand}
	\mathrm{CAOE}(a') = \sum_{j=1}^J \sum_{i=1}^{N_j} \mathrm{CAOE}_{i,j}(a') \bigg{/} \sum_{j=1}^J \sum_{i=1}^{N_j} \mathbbm{1}(G_{i,j}(a') = \mathcal{c})
	\end{equation}
	\noindent To facilitate references to causal estimands in this manuscript, Table \ref{tab:abbreviation} summarizes the estimands that we defined in Sections \ref{sec:estimand1} and \ref{sec:estimands2}.
	
	\begin{table*}[t]
		\centering
		\caption{Causal estimands defined in Sections \ref{sec:estimand1} and \ref{sec:estimands2}.}
		\label{tab:abbreviation}
		\begin{tabular}{ll}
			\toprule
			$\mathrm{ITT_Y}$ & Direct effect on the outcome\\  
			$\mathrm{ITT_D}$ & Direct effect on the treatment receipt\\  
			$\mathrm{S_Y}$ & Spillover effect on the outcome\\  
			$\mathrm{S_D}$ & Spillover effect on the treatment receipt\\  
			$\mathrm{O_Y}$ & Overall effect on the outcome\\  
			$\mathrm{O_D}$ & Overall effect on the treatment receipt\\  
			CADE & Complier average direct effect on the outcome\\ 
			CASE & Complier average spillover effect on the outcome\\
			CAOE & Complier average overall effect on the outcome\\
			\bottomrule
		\end{tabular}
	\end{table*}
	
	\section{Bayesian Model and Inferential Approach}
	\label{sec:methodology}
	
	\subsection{Overview of Methodology}
	\label{sec:methodology_overview}
	
	Our Bayesian methodology for performing causal inferences on two-stage randomized experiments with interference, nonadherence, and missing outcomes is based on model-based imputation of missing and unrealized potential outcomes to derive the posterior distributions of the finite-population causal estimands. To describe this more formally, we let $\tau$ denote one of the causal estimands from Sections \ref{sec:estimand1} and \ref{sec:estimands2}, $\mathbf{X}$ the $N \times P$ matrix of covariates for all experimental units, $\mathbf{A}^{\mathrm{o}}$ the $N \times 1$ vector of treatment probabilities that were assigned to the clusters in the experiment, $\mathbf{Z}^{\mathrm{o}}$ the $N \times 1$ vector of treatments that were assigned to the units in the experiment, $\mathbf{D}^{\mathrm{o}}$ the $N \times 1$ vector of treatments that were received by the units in the realized experiment, $\mathbf{M}^{\mathrm{o}}$ the $N \times 1$ vector of missingness indicators for the outcomes in the realized experiment, and $\mathbf{Y}^{\mathrm{obs}}$ the vector of observed outcomes in the realized experiment. Then the Bayesian methodology calculates the distribution $p \left ( \tau \mid \mathbf{X}, \mathbf{A}^{\mathrm{o}}, \mathbf{Z}^{\mathrm{o}}, \mathbf{D}^{\mathrm{o}}, \mathbf{M}^{\mathrm{o}}, \mathbf{Y}^{\mathrm{obs}} \right )$. As causal estimands are functions of observed, missing, and unrealized potential outcomes, we calculate the posterior distribution above by integrating over the unrealized treatments received, which we denote by $\mathbf{D}^{\mathrm{u}}$, and both the missing and unrealized potential outcomes, which we denote by $\mathbf{Y}^{\mathrm{mis}}$ and $\mathbf{Y}^{\mathrm{u}}$ respectively, according to
	\begin{align*}
	 &p \left ( \tau \mid \mathbf{X}, \mathbf{A}^{\mathrm{o}}, \mathbf{Z}^{\mathrm{o}}, \mathbf{M}^{\mathrm{o}},\mathbf{D}^{\mathrm{o}}, \mathbf{Y}^{\mathrm{obs}} \right ) = p \left ( \tau \mid  \mathbf{O} \right )\\
	 &=  \int  p \left ( \tau \mid \mathbf{O}, \mathbf{D}^{\mathrm{u}}, \mathbf{Y}^{\mathrm{mis}}, \mathbf{Y}^{\mathrm{u}} \right ) 
	  p \left ( \mathbf{D}^{\mathrm{u}}, \mathbf{Y}^{\mathrm{mis}}, \mathbf{Y}^{\mathrm{u}}  \mid \mathbf{O} \right ) d\mathbf{D}^{\mathrm{u}} d\mathbf{Y}^{\mathrm{mis}} d\mathbf{Y}^{\mathrm{u}} \\
	&= \int   p \left ( \tau \mid \mathbf{O}, \mathbf{G}, \mathbf{Y}^{\mathrm{mis}}, \mathbf{Y}^{\mathrm{u}} \right )  p \left ( \mathbf{G}, \mathbf{Y}^{\mathrm{mis}}, \mathbf{Y}^{\mathrm{u}} \mid \mathbf{O} \right ) d\mathbf{G} d\mathbf{Y}^{\mathrm{mis}} d\mathbf{Y}^{\mathrm{u}}, 
	\end{align*}
	where $\mathbf{O}=(\mathbf{X}, \mathbf{A}^{\mathrm{o}}, \mathbf{Z}^{\mathrm{o}}, \mathbf{M}^{\mathrm{o}},\mathbf{D}^{\mathrm{o}}, \mathbf{Y}^{\mathrm{obs}})$ and  $\mathbf{G}$ is an $N \times 1$ vector of (latent) principal strata memberships for the experimental units. This motivates our Bayesian methodology of first deriving a Monte Carlo approximation for the posterior predictive distribution of the missing principal strata and outcomes conditional on the observed data, and then using that posterior predictive distribution to derive a Monte Carlo approximation for the posterior distribution of $\tau$. Our methodology is distinct from existing methods for two-stage randomized experiments in that we impute missing principal strata memberships and at most four missing/unrealized potential outcomes for each experimental unit according to the assumptions in Section \ref{sec:preliminaries}. 
	
	Our imputation approach, and the latent ignorability assumption that underlies it, is in Section \ref{sec:imputation_principal_strata}. In Section \ref{sec:bayesianinference} we provide additional details on our Bayesian modeling approach and the unconfoundedness condition underlying it, which is justified by the design of the two-stage randomized experiment. The Gibbs sampling algorithm that we use to derive a Monte Carlo approximation to the posterior distribution of the causal estimand is in Section \ref{sec:Gibbs_sampling}. We provide a justification of exchangeability in our Bayesian methodology in the appendix.

	\subsection{Imputation of Missing Values}
	\label{sec:imputation_principal_strata}
	
	By virtue of Assumptions \ref{asmp:monotonocity_fixed} and \ref{asmp:monotonocity_across_mech}, any experimental unit in the two-stage randomized experiment belongs to one of six principal strata. Table \ref{tab:comp_obs_missingrate} illustrates the correspondence between the possible values of $A_{j}^{\mathrm{o}}, Z_{i,j}^{\mathrm{o}}$, and $D_{i,j}^{\mathrm{o}}$ and the principal strata memberships, along with the missingness rates for the outcomes in the RSBY study that we analyze in Section \ref{sec:application}. We let $\mathcal{G}(a,z,d)$ denote a set of possible principal strata for units with $(A_{j}^{\mathrm{o}}, Z_{i,j}^{\mathrm{o}}, D_{i,j}^{\mathrm{o}}) = (a,z,d)$. Assumption \ref{asmp:monotonocity_across_mech} in particular helps to narrow down the possible strata that a unit could belong to based on the observed data. For example, if an experimental unit $i$ in cluster $j$ has $A_j^{\mathrm{o}} = a_0, Z_{i,j}^{\mathrm{o}} = 0$, and $D_{i,j}^{\mathrm{o}} = 1$, then $\mathcal{G}(a_0,0,1) = \{(\mathcal{a}, \mathcal{a})\}$, and so $G_{i,j} = (\mathcal{a}, \mathcal{a})$. Once $G_{i,j}$ is fixed, we can immediately impute all unrealized $D_{i,j}^{\mathrm{u}}$.
	
	We also observe in Table \ref{tab:comp_obs_missingrate} that different strata, e.g., $(\mathcal{a},\mathcal{a})$ and $(\mathcal{n},\mathcal{n})$, exhibit different missingness rates. This corresponds to more general phenomenon in modern experiments in which different strata have different missing data mechanisms for the final outcomes. This phenomenon is intuitive for certain strata. For example, those units in the $(\mathcal{n},\mathcal{n})$ strata may be more reluctant to take any actions compared to units in other strata, and thus their outcomes are more likely to be missing. As the missingness mechanism depends on the latent strata, the traditional missing completely at random (MCAR) and missing at random (MAR) assumptions may not be appropriate. In contrast to other methods, our Bayesian methodology accommodates a class of missing not at random (MNAR) mechanisms \citep[p.~351]{little2002statistical} that correspond to the following latent ignorability \citep[LI]{FrangakisRubin1999} assumption.
	
	\begin{assumption}[Latent Ignorability of Missing Data]
		\label{asmp:latentignorable}
		For any unit $i = 1, \ldots, N_j$ in cluster $j = 1, \ldots, J$,
		\begin{equation*}
		\begin{split}
		p(\mathbf{M}_{i,j} \mid \mathbf{D}_{i,j}, \mathbf{Y}_{i,j}, X_{i,j}) =p(\mathbf{M}_{i,j} \mid \mathbf{D}_{i,j}, X_{i,j}), \ p(\mathbf{M}_{i,j} \mid G_{i,j}, \mathbf{Y}_{i,j}, X_{i,j}) =p(\mathbf{M}_{i,j} \mid G_{i,j}, X_{i,j}),
		\end{split}
		\end{equation*}
		where $\mathbf{D}_{i,j} = (D_{i,j}(0,a_0)$, $D_{i,j}(0,a_1)$, $D_{i,j}(1,a_0)$, $D_{i,j}(1,a_1))$, $\mathbf{M}_{i,j} = (M_{i,j}(0,a_0)$, $M_{i,j}(0,a_1)$, $M_{i,j}(1,a_0)$, $M_{i,j}(1,a_1))$, and $\mathbf{Y}_{i,j} = (Y_{i,j}(0,a_0)$, $Y_{i,j}(0,a_1)$, $Y_{i,j}(1,a_0)$, $Y_{i,j}(1,a_1))$.
	\end{assumption}
	
	\noindent We formulate LI to mean that if we knew the latent strata $G_{i,j}$ of each unit, the missingness mechanism would be ignorable. Assumptions \ref{asmp:exclusion} and \ref{asmp:latentignorable} facilitate and unify our imputations of missing and unrealized potential outcomes. For those units with $M_{i,j}^{\mathrm{o}} = 0$, as one of their potential outcomes is observed we need to impute at most three unrealized potential outcomes for each of them. In the case of a unit whose principal stratum is $(\mathcal{c}, \mathcal{c})$, we need to impute three unrealized potential outcomes based on our Bayesian model. For another unit whose principal stratum is  $(\mathcal{a}, \mathcal{a})$, we only need to impute one unrealized potential outcome, as the potential outcomes are restricted according to Assumption \ref{asmp:exclusion} for this stratum. For a unit with $M_{i,j}^{\mathrm{o}} = 1$, we impute at most four potential outcomes based on their imputed principal strata $G_{i,j}$.

	\begin{table*}[t]
		\centering
		\caption{Illustration of the possible principal strata that an experimental unit can belong to under different combinations of $A_j^{\mathrm{o}}, Z_{i,j}^{\mathrm{o}}$, and $D_{i,j}^{\mathrm{o}}$. We provide examples of the number of units and the missingness rates of the outcomes for these different sets of principal strata based on data from the RSBY study in Section \ref{sec:application}.}
		\label{tab:comp_obs_missingrate}
		\begin{tabular}{lllrrr}
			\toprule
			$A_j^{\mathrm{o}}$& $Z_{i,j}^{\mathrm{o}}$ & $D_{i,j}^{\mathrm{o}}$ &  \text{$\mathcal{G}(A_j^{\mathrm{o}},Z_{i,j}^{\mathrm{o}},D_{i,j}^{\mathrm{o}})$} & Number of units & Missing rates  \\
			\hline
			$a_0$ & $0$ & $0$ & $(\mathcal{c},\mathcal{c}),(\mathcal{n},\mathcal{n}), (\mathcal{c},\mathcal{a}), (\mathcal{n},\mathcal{c}), (\mathcal{n},\mathcal{a})$ & $2258$ & $0.095$  \\  
			$a_0$ & $1$ & $0$ & $(\mathcal{n},\mathcal{n}), (\mathcal{n},\mathcal{c}), (\mathcal{n},\mathcal{a})$ & $545$ & $0.119$   \\  
			$a_0$ & $0$ & $1$ & $(\mathcal{a},\mathcal{a})$ & $911$ & $0.043$    \\  
			$a_0$ & $1$ & $1$ & $(\mathcal{c},\mathcal{c}), (\mathcal{a},\mathcal{a}), (\mathcal{c},\mathcal{a})$ & $1554$ & $0.062$  \\  
			$a_1$ & $0$ & $0$ & $(\mathcal{c},\mathcal{c}), (\mathcal{n},\mathcal{n}), (\mathcal{n},\mathcal{c})$ & $779$ & $0.105$   \\  
			$a_1$ & $1$ & $0$ & $(\mathcal{n},\mathcal{n})$ & $1028$ & $0.135$  \\  
			$a_1$ & $0$ & $1$ & $(\mathcal{a},\mathcal{a}), (\mathcal{c},\mathcal{a}), (\mathcal{n},\mathcal{a})$ & $350$ & $0.031$ \\  
			$a_1$ & $1$ & $1$ & $(\mathcal{c},\mathcal{c}), (\mathcal{a},\mathcal{a}), (\mathcal{c},\mathcal{a}), (\mathcal{n},\mathcal{c}), (\mathcal{n},\mathcal{a})$ & $3454$ & $0.059$ \\  
			\bottomrule
		\end{tabular}
	\end{table*}

	\subsection{Bayesian Model Building}
	\label{sec:bayesianinference}
	
	Following the Bayesian paradigm of \citet{imbens_rubin_2015} and \citet{gelman_et_al_2013}, five inputs are necessary for deriving the posterior distributions of the causal estimands. The first input is knowledge of the assignment mechanisms as encoded in the probability mass functions $p \left ( \mathbf{A} \mid \mathbf{X} \right )$ and $p \left ( \mathbf{Z} \mid \mathbf{X}, \mathbf{A} \right )$. These two probability mass functions are obtained immediately by virtue of the known design of the two-stage randomized experiment. In addition, these two probability mass functions are independent of the experimental units' covariates $\mathbf{X}$, so that $p \left ( \mathbf{A} \mid \mathbf{X} \right ) = p \left ( \mathbf{A} \right )$ and $p \left ( \mathbf{Z} \mid \mathbf{X}, \mathbf{A} \right ) = p \left ( \mathbf{Z} \mid \mathbf{A} \right )$. The second input is the model for treatment received conditional on $\mathbf{X}, \mathbf{A}, \mathbf{Z}$. We denote this input by the probability mass function $p \left ( \mathbf{D} \mid \mathbf{X}, \mathbf{A}, \mathbf{Z}, \boldsymbol{\psi} \right )$ with parameter vector $\boldsymbol{\psi}$. An equivalent input is the model for principal strata memberships of all experimental units, $\mathbf{G}$, conditional on $\mathbf{X}, \mathbf{A}, \mathbf{Z}$, because there is a one-to-one mapping between $\mathbf{G}$ and $\mathbf{D}$. To simplify our notation we denote this input as $p \left ( \mathbf{G} \mid \mathbf{X}, \mathbf{A}, \mathbf{Z}, \boldsymbol{\psi} \right )$ with the understanding that $\boldsymbol{\psi}$ is generic notation for the parameter vector associated with either of these two models. The third input is the model for the potential outcomes conditional on $\mathbf{X}, \mathbf{A}, \mathbf{Z}, \mathbf{D}$. We denote this model via a probability density/mass function $p \left ( \mathbf{Y} \mid \mathbf{X}, \mathbf{A}, \mathbf{Z}, \mathbf{D}, \boldsymbol{\theta} \right )$ with parameter vector $\boldsymbol{\theta}$. This model can also be equivalently formulated using $\mathbf{G}$ as $p \left ( \mathbf{Y} \mid \mathbf{X}, \mathbf{A}, \mathbf{Z}, \mathbf{G}, \boldsymbol{\theta} \right )$. The fourth input is the model for the missingness indicators $\mathbf{M}$ conditional on $\mathbf{X}, \mathbf{A}, \mathbf{Z}, \mathbf{D}$, $\mathbf{Y}$, denoted by $p \left ( \mathbf{M} \mid \mathbf{X}, \mathbf{A}, \mathbf{Z}, \mathbf{D}, \mathbf{Y}, \boldsymbol{\phi} \right )$ with parameter vector $\boldsymbol{\phi}$. Under the LI assumption, this model will not involve $\mathbf{Y}$ conditional on the other variables. The final input is the joint prior distribution for $\boldsymbol{\psi}$, $\boldsymbol{\phi}$, and $\boldsymbol{\theta}$, denoted by $p \left ( \boldsymbol{\psi}, \boldsymbol{\phi}, \boldsymbol{\theta} \right )$. We assume throughout that $\boldsymbol{\psi}$, $\boldsymbol{\phi}$, and $\boldsymbol{\theta}$ are distinct and do not share any common parameters.
	
	Specifying the second and third inputs is facilitated by the fact that the two-stage randomized experiment has an unconfounded assignment mechanism at both cluster and individual experimental unit levels. Unconfoundedness is formally expressed via the following assumption.
	
	\begin{assumption}
		\label{asmp:unconfoundedness}
		For experimental unit $i = 1, \ldots, N_j$ in cluster $j = 1, \ldots, J$, $p(A_j, Z_{i,j} \mid X_{i,j}, \mathbf{D}_{i,j}) = p(A_j, Z_{i,j} \mid X_{i,j})$, $p(A_j, Z_{i,j} \mid X_{i,j}, \mathbf{Y}_{i,j}, \mathbf{D}_{i,j}) = p( A_j, Z_{i,j}\mid X_{i,j}, \mathbf{D}_{i,j})$, $p( A_j, Z_{i,j} \mid X_{i,j}, \mathbf{Y}_{i,j}, G_{i,j}) = p(A_j, Z_{i,j} \mid X_{i,j}, G_{i,j})$  and $p(A_j, Z_{i,j}, \mathbf{M}_{i,j} \mid X_{i,j}, \mathbf{D}_{i,j}, \mathbf{Y}_{i,j}) = p(A_j, Z_{i,j} \mid X_{i,j}, \mathbf{D}_{i,j}, \mathbf{Y}_{i,j})p(\mathbf{M}_{i,j} \mid X_{i,j}, \mathbf{D}_{i,j}, \mathbf{Y}_{i,j})$.
	\end{assumption}

	\noindent Assumption \ref{asmp:unconfoundedness} implies that the treatment assignment mechanism is ignorable. In addition, unit exchangeability implies by de Finetti's Theorem that we can specify parametric models on the level of the individual experimental units for the principal strata memberships $G_{i,j}$, the potential outcomes, and the missingness indicators to derive the joint model for $\mathbf{G}$, $\mathbf{M}$, and $\mathbf{Y}$. A justification of unit exchangeability in the two-stage randomized experiment is in the appendix. Hence we have that
	\begin{equation}
	\label{eq:outcome_receipt_missingness_model}
	\begin{split}
	&p \left (\mathbf{Y},\mathbf{D}, \mathbf{M}  \mid \mathbf{X}, \boldsymbol{\psi}, \boldsymbol{\theta}, \boldsymbol{\phi} \right) \\
	&= \prod_{i,j}p(\mathbf{D}_{i,j} \mid X_{i,j},\boldsymbol{\psi})p(\mathbf{Y}_{i,j} \mid X_{i,j}, \mathbf{D}_{i,j},\boldsymbol{\theta})p(\mathbf{M}_{i,j} \mid X_{i,j}, \mathbf{D}_{i,j},\mathbf{Y}_{i,j},\boldsymbol{\phi})\\
	&= \prod_{i,j}p(\mathbf{D}_{i,j} \mid X_{i,j},\boldsymbol{\psi})p(\mathbf{Y}_{i,j} \mid X_{i,j}, \mathbf{D}_{i,j},\boldsymbol{\theta})p(\mathbf{M}_{i,j} \mid X_{i,j}, \mathbf{D}_{i,j},\boldsymbol{\phi}).
	\end{split}
	\end{equation}
	\noindent The last line in equation (\ref{eq:outcome_receipt_missingness_model}) follows from Assumption \ref{asmp:latentignorable}. 
	The posterior distribution $p \left ( \boldsymbol{\psi}, \boldsymbol{\phi} ,\boldsymbol{\theta} \mid \mathbf{X}, \mathbf{A}^{\mathrm{o}}, \mathbf{Z}^{\mathrm{o}}, \mathbf{D}^{\mathrm{o}}, \mathbf{M}^{\mathrm{o}}, \mathbf{Y}^{\mathrm{obs}} \right )$ is proportional to
	\begin{equation*}
	\begin{split}
	&p \left ( \boldsymbol{\psi}, \boldsymbol{\phi} ,\boldsymbol{\theta} \mid \mathbf{X}, \mathbf{A}^{\mathrm{o}}, \mathbf{Z}^{\mathrm{o}}, \mathbf{D}^{\mathrm{o}}, \mathbf{M}^{\mathrm{o}}, \mathbf{Y}^{\mathrm{obs}} \right ) \\ &\propto p(\boldsymbol{\psi}, \boldsymbol{\phi},\boldsymbol{\theta})  \int\prod_{i,j}p(\mathbf{D}_{i,j} \mid X_{i,j},\boldsymbol{\psi})p(\mathbf{Y}_{i,j} \mid X_{i,j}, \mathbf{D}_{i,j},\boldsymbol{\theta}) p(M_{i,j}^{\mathrm{o}} \mid X_{i,j}, \mathbf{D}_{i,j}, \boldsymbol{\theta})d\mathbf{D}^{\mathrm{u}} d\mathbf{Y}^{\mathrm{mis}}d\mathbf{Y}^{\mathrm{u}}\\
	&=p(\boldsymbol{\psi}, \boldsymbol{\phi},\boldsymbol{\theta})  \int\prod_{i,j}p(\mathbf{G}_{i,j} \mid X_{i,j},\boldsymbol{\psi})p(\mathbf{Y}_{i,j} \mid X_{i,j}, \mathbf{G}_{i,j},\boldsymbol{\theta}) p(M_{i,j}^{\mathrm{o}} \mid X_{i,j}, \mathbf{G}_{i,j}, \boldsymbol{\theta})d\mathbf{G}d\mathbf{Y}^{\mathrm{mis}}d\mathbf{Y}^{\mathrm{u}} \\
	\end{split}
	\end{equation*}
	\noindent To express the posterior distribution of the model parameters, let $O(a, z, d, m)$ denote the set of experimental units whose realized values of $A_j^{\mathrm{o}}, Z_{i,j}^{\mathrm{o}}, D_{i,j}^{\mathrm{o}}$, and $M_{i,j}^{\mathrm{o}}$ are $a, z, d$, and $m$, respectively. For each unit $i$ in cluster $j$ and each principal strata $g$, define $w_{i,j}^{(g)}= \mathrm{Pr} \left ( G_{i,j} = g \mid X_{i,j}, \boldsymbol{\psi} \right)$, $\rho_{i,j}^{(g,a,z)} = \mathrm{Pr}(M_{i,j} = 1 \mid G_{i,j} = g, A_{j} = a, Z_{i,j} = z, X_{i,j}, \boldsymbol{\phi})$, and $f_{i,j}^{(g,a,z)}$ be the probability mass/density function of $Y_{i,j}(z,a)$ for $g \in \{ (\mathcal{n},\mathcal{n}), (\mathcal{c},\mathcal{c}), (\mathcal{a},\mathcal{a}), (\mathcal{n},\mathcal{c}), (\mathcal{n},\mathcal{a}), (\mathcal{c},\mathcal{a})\}$. 
 Under Assumptions \ref{asmp:monotonocity_fixed} - \ref{asmp:exclusion}, the posterior distribution of $\boldsymbol{\psi}$, $\boldsymbol{\phi}$, and $\boldsymbol{\theta}$ can be written as
	\begin{equation}
	\label{eq:posterior_params}
	\begin{split}
	p & \left ( \boldsymbol{\psi}, \boldsymbol{\phi},\boldsymbol{\theta} \mid \mathbf{X}, \mathbf{A}^{\mathrm{o}}, \mathbf{Z}^{\mathrm{o}}, \mathbf{D}^{\mathrm{o}},         \mathbf{M}^{\mathrm{o}}, \mathbf{Y}^{\mathrm{obs}} \right )
	\propto p(\boldsymbol{\psi}, \boldsymbol{\phi} ,\boldsymbol{\theta})\\
	&\times \prod_{a \in \{a_0,a_1\}}\prod_{z \in \{0,1\}}\prod_{d \in \{0,1\}}\prod_{(i,j) \in O(a,z,d,0)}\sum_{g \in \mathcal{G}(a,z,d)}(1-\rho_{i,j}^{g,z,d})w_{i,j}^{g} f_{i,j}^{g,z,d} \\
	&\times \prod_{a \in \{a_0,a_1\}}\prod_{z \in \{0,1\}}\prod_{d \in \{0,1\}} \prod_{(i,j) \in O(a,z,d,1)}\sum_{g \in \mathcal{G}(a,z,d)}\rho_{i,j}^{g,z,d}w_{i,j}^{g} f_{i,j}^{g,z,d}.
	\end{split}
	\end{equation}
	\noindent Our model-based Bayesian method ultimately utilizes the above mixture model that requires the specification of $w_{i,j}^{(g)}$, $\rho_{i,j}^{(g,a,z)}$ and $f_{i,j}^{(g,a,z)}$. Specific models and prior distributions are provided in our simulation studies in Section \ref{sec:simulationstudies} and our case study in Section \ref{sec:application}. 
	Our Bayesian methodology also enables us to infer principal causal effects under two-sided noncompliance, whereas existing methods require additional strict assumptions to perform inferences for principal causal effects under two-sided noncompliance including always-takers and social-interaction compliers (i.e., the strata $(\mathcal{a}, \mathcal{a})$ and $(\mathcal{c}, \mathcal{a})$) \citep{Gonzalo2021}.

	\subsection{Gibbs Sampling Algorithm}
	\label{sec:Gibbs_sampling}
	
	We utilize the Gibbs sampling algorithm \citep{Geman_Geman_1984, Gelfand1990,ImbensRubin1997} to obtain the posterior predictive distributions of $\mathbf{Y}^{\mathrm{mis}}$ and $\mathbf{Y}^{\mathrm{u}}$ given the observed data, so as to impute all missing and unrealized outcomes and derive the posterior distributions for the causal estimands. Specifically, we iterate between drawing from the conditional distributions of $(\boldsymbol{\psi}, \boldsymbol{\phi}, \boldsymbol{\theta})$ and $\mathbf{G}$ given the other variables, respectively. Then for each iteration we impute $\mathbf{Y}^{\mathrm{mis}}$ and $\mathbf{Y}^{\mathrm{u}}$, and calculate the causal estimands of interest to effectively get a draw from their respective posterior distributions. The essential algorithm is outlined below.
	
	\begin{enumerate}
		\item Initialize parameters $\boldsymbol{\psi}^{(0)}$, $\boldsymbol{\phi}^{(0)}$, $\boldsymbol{\theta}^{(0)}$.
		
		\item For $t = 0, 1, \ldots$:
		
		\begin{enumerate}
			
			\item Draw $\mathbf{G}^{ (t+1)} \sim p \left ( \mathbf{G} \mid \mathbf{X}, \mathbf{A}^{\mathrm{o}},\mathbf{Z}^{\mathrm{o}},\mathbf{D}^{\mathrm{o}},\mathbf{M}^{\mathrm{o}},\mathbf{Y}^{\mathrm{obs}}, \boldsymbol{\psi}^{(t)} , \boldsymbol{\phi}^{(t)} , \boldsymbol{\theta}^{(t)} \right )$.
			
			\item Draw $(\mathbf{Y}^{\mathrm{mis}, (t+1)}, \mathbf{Y}^{\mathrm{u}, (t+1)}) \sim p \left ( \mathbf{Y}^{\mathrm{mis}}, \mathbf{Y}^{\mathrm{u}} \mid \mathbf{X}, \mathbf{A}^{\mathrm{o}},\mathbf{Z}^{\mathrm{o}}, \mathbf{M}^{\mathrm{o}},\mathbf{Y}^{\mathrm{obs}}, \mathbf{G}^{(t+1)}, \boldsymbol{\psi}^{(t)} , \boldsymbol{\phi}^{(t)} , \boldsymbol{\theta}^{(t)} \right )$.
			
			\item Calculate the causal estimand based on the observed and imputed data.
			
			\item Draw $\left ( \boldsymbol{\psi}^{(t+1)}, \boldsymbol{\phi}^{(t+1)}, \boldsymbol{\theta}^{(t+1)} \right ) \sim p \left ( \boldsymbol{\psi}, \boldsymbol{\phi}, \boldsymbol{\theta} \mid  \mathbf{X}, \mathbf{A}^{\mathrm{o}}, \mathbf{Z}^{\mathrm{o}}, \mathbf{M}^{\mathrm{o}}, \mathbf{Y}^{\mathrm{obs}},
			\mathbf{G}^{ (t+1)} \right )$.
			
			\item Repeat steps 2(a) - 2(e).
			
		\end{enumerate}
		
	\end{enumerate}
	
	This Gibbs sampler enables us to obtain posterior draws of all the unobserved potential outcomes and principal strata for all units. We then perform causal inferences by means of the posterior distributions. For example, points estimates of the estimands can be obtained via their posterior means or medians, and intervals for the estimands can be obtained via the central credible intervals. Detailed derivations for each step of the Gibbs sampler are provided in the supplementary material. 

	\section{Simulation Studies}
	\label{sec:simulationstudies}

	\subsection{Evaluation Metrics and Data Generating Mechanisms}
	\label{sec:simulation_studies_overview}
	
	We evaluate the frequentist properties of our Bayesian methodology with respect to those of the method of \citet{Imai2021}, which we implement using its R package \texttt{experiment} \citep{experiment}. \citet{Imai2021} implicitly assumed MCAR data in their analyses, and so removed rows with missing outcomes from their analyses because their methodology does not accommodate missing outcomes. Therefore, to ensure fair comparisons, in our first simulation studies we have either no missing data or MCAR data. Our Bayesian methodology easily accommodates the MCAR mechanism without major modifications, as this would just entail removing units that have missing outcomes and ignoring the missing data mechanism model in the derivations of the likelihood functions and posteriors.
	
	The evaluation metrics that we consider are bias and mean square error (MSE) in estimating a causal estimand, coverage of an interval estimator for a causal estimand, and the interval length. Bias and MSE are generally defined as $\sum_{m=1}^M \left ( \tau - \hat{\tau}_m \right )/M$ and $\sum_{m=1}^M \left ( \tau - \hat{\tau}_m \right )^2/M$ respectively, where $M$ denotes the number of simulated datasets, $\tau$ denotes the true causal estimand, and $\hat{\tau}_m$ denotes the estimate of the causal estimand in dataset $m = 1, \ldots, M$. For our Bayesian method, the point estimator is the median of the posterior distribution of a causal estimand, and the interval estimator is the $95\%$ central credible interval. Our summary of the interval length is the median of the lengths of the credible intervals computed from $M$ simulated datasets. The median is desirable as it is more robust for the chosen data generating mechanism. The data generating mechanisms that we consider in our study are specified to simulate data resembling those in our case study in Section \ref{sec:application}.
	
	For the two-stage randomized experiments in our simulation study, the clusters have equal numbers of units (i.e., $N_j=N/J$), and we specify $p(A_j = a_0) = p(A_j = a_1) = 0.5, p(Z_{i,j}=1 \mid A_j = a_0) = 0.4$, and $p(Z_{i,j}=1 \mid A_j = a_1) = 0.8$ for any experimental unit $i = 1, \ldots, N_j$ in cluster $j = 1, \ldots, J$. Each experimental unit belongs to just one of the latent principal strata $\{ (\mathcal{n}, \mathcal{n}), (\mathcal{c}, \mathcal{c}), (\mathcal{a}, \mathcal{a}), (\mathcal{n}, \mathcal{c}), (\mathcal{n}, \mathcal{a}), (\mathcal{c}, \mathcal{a}) \}$. The principal strata memberships are generated according to $G_{i,j} \sim \mathrm{Multinomial} ( \pi_{(\mathcal{n}, \mathcal{n})}$, $\pi_{(\mathcal{c}, \mathcal{c})}$, $\pi_{(\mathcal{a}, \mathcal{a})}$, $\pi_{(\mathcal{n}, \mathcal{c})}$, $\pi_{(\mathcal{n}, \mathcal{a})}$, $\pi_{(\mathcal{c},\mathcal{a})} )$. Evaluation for different values of $J$ is presented in the appendix.
	
	The potential outcomes in our simulation study are continuous and are generated to have a right-skewed distribution with an excess of zeros. The specific generation mechanism that we implement is a zero-inflated Log-Normal distribution, with the parameters of the underlying Bernoulli and Log-Normal random variables (representing the excess zeros and heavy tail of the outcomes, respectively) specified to be distinct for each strata and treatment. More formally, for $a \in \{a_0, a_1\}$ and $z \in \{0, 1\}$, the potential outcomes for unit $i$ in cluster $j$ are generated by first sampling $W_{i,j}(z, a) \sim \mathrm{Bernoulli}(p_{z, a, G_{i,j}})$, then sampling $\tilde{Y}_{i,j}(z,a) \sim$ Log-Normal$(\mu_{z, a, G_{i,j}}, \sigma_{z, a, G_{i,j}}^2)$, and finally generating $Y_{i,j}(z,a) = \left \{ 1 - W_{i,j}(z,a) \right \}\tilde{Y}_{i,j}(z,a)$. For this simulation study, we assume that potential outcomes are generated independently of one another. In addition, Assumption \ref{asmp:exclusion} applies throughout the data generating mechanism, including for $W_{i,j}(z,a)$ and $\tilde{Y}_{i,j}(z,a)$. The generated $W_{i,j}(z,a)$ and $\tilde{Y}_{i,j}(z,a)$ are not recorded as data.  Finally, we use conjugate prior distributions for all parameters, that is, $ \left ( \pi_{(\mathcal{n},\mathcal{n})}, \pi_{(\mathcal{c},\mathcal{c})}, \pi_{(\mathcal{a},\mathcal{a})}, \pi_{(\mathcal{n},\mathcal{c})}, \pi_{(\mathcal{n},\mathcal{a})}, \pi_{(\mathcal{c},\mathcal{a})} \right ) \sim \mathrm{Dirichlet}(2,2,2,2,2,2)$, $p_{z, a, G_{i,j}} \sim \mathrm{Beta}(1,1)$, $\mu_{z, a, G_{i,j}} \sim \mathrm{Normal}(0, 10^2)$ and $\sigma_{z, a, G_{i,j}}^2 \sim \mathrm{InverseGamma}(1, 1)$. Detailed algorithms for the Gibbs sampler are provided in the appendix. We simulate $500$ datasets for each $N = 5000, 10000, 50000$ with a fixed cluster size of $J = 100$.  In the appendix we provide all parameter values that are utilized in our simulation study  the frequentist evaluations for  the super-population versions of the causal estimands, and an additional simulation study using the Gamma distribution for the outcome model.
	
	\subsection{Results}
	\label{sec:simulation_wellspec_results}
	
	Table \ref{tab:simulation_fn_lognormal} summarizes the results of our simulation study in the case of the Log-Normal distribution. The method of \citet{Imai2021} is abbreviated as ``IJM'' in this table. We observe that both methods perform well with respect to coverage and bias, with the IJM method exhibiting less bias than our Bayesian method for small sample sizes. This difference in bias can be attributed to the effect of the prior distributions, which is not negligible for small $N$. In addition, the posterior median is not an unbiased estimator of the mean of the Log-Normal distribution, although it is arguably a desirable estimator as it should be more robust for the chosen data generating mechanism. This difference in bias should also be expected as the emphasis of the Bayesian approach is on the posterior distribution and whether it is well-calibrated. Finally, we recognize that our Bayesian methodology outperforms the IJM method with respect to MSE under all conditions. The differences in MSE imply that the Bayesian point estimator is less likely to deviate from the true causal effect over multiple experiments as compared to the IJM method, that the Bayesian estimator has less variability, and that the Bayesian intervals have smaller widths. Ultimately, the Bayesian methodology yields more precise causal inferences. The IJM method is sensitive to the shape of the distribution and exhibits greater variability and MSEs as it attempts to accommodate outliers.  
	
	\begin{table*}
		\centering
		\caption{Evaluation metrics for our Bayesian methodology versus the method of \citet{Imai2021} (abbreviated as ``IJM'') under the Log-Normal data-generating process and the finite-population perspective.}
		\begin{adjustbox}{width=13.3cm}
			\begin{tabular}{lrrrrrrrrr}
				\toprule
				& & \multicolumn{2}{c}{Coverage}   & \multicolumn{2}{c}{Bias}   & \multicolumn{2}{c}{MSE}   & \multicolumn{2}{c}{Interval Width} \\
				\cmidrule(lr){3-4} \cmidrule(lr){5-6} \cmidrule(lr){7-8} \cmidrule(lr){9-10} 
				&N  & IJM & Bayes & IJM & Bayes & IJM & Bayes & IJM & Bayes \\ \hline
				&5000  & 96\% & 99\% & 4.43E+02 & 5.02E+02 & 3.60E+07 & 4.57E+06 & 1.91E+04 & 9.11E+03 \\ 
				$\mathrm{CADE}(0,a_0)$&10000 & 96\% & 98\% & -2.52E+02 & 2.99E+02 & 1.39E+07 & 1.94E+06 & 1.36E+04 & 5.41E+03 \\ 
				&50000 & 94\% & 95\% & -1.22E+02 & 1.59E+02 & 3.58E+06 & 2.71E+05 & 6.61E+04 & 2.09E+03 \\
				\hline
				&5000  & 92\% & 97\% & -1.05E+03 & 1.20E+03 & 2.28E+08 & 5.97E+06 & 4.52E+04 & 1.67E+04 \\ 
				$\mathrm{CADE}(1,a_1)$&10000 & 93\% & 97\% & 2.12E+02 & 8.92E+02 & 1.29E+08 & 3.61E+06 & 3.38E+04& 1.13E+04 \\ 
				&50000 & 94\% & 97\% & 4.81E+02 & 2.98E+02 & 2.18E+07 & 7.75E+05 & 1.68E+05 &4.81E+03 \\ 
				\hline
				&5000  & 96\% & 98\% & 1.68E+02 & 2.46E+02 & 9.11E+06 & 1.20E+06 & 9.63E+03 &4.68E+03 \\ 
				$\mathrm{ITT}_{\mathrm{Y},\cdot,\cdot}(a_0)$&10000 & 96\% & 98\% & -1.21E+02 & 1.46E+02 & 3.49E+06 & 4.86E+05 & 6.86E+03 & 2.80E+03 \\ 
				& 50000 & 93\% & 96\% & -5.61E+01 & 8.06E+01 & 9.08E+05 & 7.39E+04 & 3.32E+04 &1.08E+03 \\ 
				\hline
				& 5000  & 92\% & 97\% & -4.54E+02 & 6.00E+02 & 4.62E+07 & 1.42E+06 & 2.03E+04 & 7.76E+03 \\ 
				$\mathrm{ITT}_{\mathrm{Y},\cdot,\cdot}(a_1)$& 10000 & 94\% & 97\% & 1.18E+02 & 4.41E+02 & 2.64E+07 & 8.34E+05 &  1.52E+04 &5.30E+03 \\ 
				& 50000 & 94\% & 97\% & 2.31E+02 & 1.42E+02 & 4.45E+06 & 1.69E+05 & 7.54E+04&2.23E+03 \\ 
				\hline
				&5000  & 87\% & 95\% & -1.56E-02 & 2.60E-03 & 5.63E-04 & 2.22E-04 &6.96E-02& 5.70E-02 \\ 
				$\mathrm{ITT}_{\mathrm{D},\cdot,\cdot}(a_0)$&10000 & 93\% & 94\% & -2.26E-04 & -1.47E-03 & 1.74E-04 & 1.21E-04 & 4.92E-02 &4.06E-02 \\ 
				&50000 & 94\% & 95\% & -6.52E-04 & -1.52E-04 & 3.31E-05 & 2.20E-05 & 2.20E-02 &1.83E-02 \\ 
				\hline
				&5000  & 94\% & 94\% & -7.57E-03 & 4.87E-03 & 5.34E-04 & 4.10E-04 & 9.06E-02 & 7.64E-02 \\ 
				$\mathrm{ITT}_{\mathrm{D},\cdot,\cdot}(a_1)$&10000 & 93\% & 93\% & -2.34E-03 & 3.76E-03 & 2.85E-04 & 2.24E-04 & 6.40E-02 &5.51E-02\\ 
				&50000 & 94\% & 96\% & -2.36E-03 & 4.44E-04 & 5.29E-05 & 3.70E-05 & 2.86E-02 &2.53E-02 \\ 
				\bottomrule
			\end{tabular}
		\end{adjustbox}
		\label{tab:simulation_fn_lognormal}
	\end{table*}
	
	\subsection{Evaluation under Misspecification}
	\label{sec:simulation_misspecification}
	We evaluate our Bayesian methodology under model misspecification. To generate data that resemble those in our case study, we consider heavy-tailed distributions for the data-generating process. We use the Half-Student-t distribution with different values of the degrees of freedom, $\nu=1.5,2.0,3.0,5.0$ to evaluate how the outliers and skewness of the distribution influence our results. We did not choose $\nu=1$ because that corresponds to the Cauchy distribution, which does not have a finite first moment. The outcomes are generated by $\tilde{Y}_{i,j}(z,a) \sim C_{z, a, G_{i,j}} \times$ Half-t$(\nu)$, where $C_{z, a, G_{i,j}}$ is a scaling constant. We fit the Log-Normal model described in the previous section to the generated data. We fix the number of units and the number of clusters at $N=10000$ and $J=100$ respectively. The additional parameters $C_{z, a, G_{i,j}}$ are provided in Table \ref{tab:params_misspecification} in the appendix.
	
	Table \ref{tab:simulation_misspecification} summarizes the results of the simulation study in the case of model misspecification. Our Bayesian model outperforms the IJM method for the skewed distribution, $\nu = 1.5$. It also exhibits smaller MSE and slightly larger bias for $\nu = 2.0,3.0$ for the case of the CADE and $\mathrm{ITT_Y}$ estimands. The larger bias is attributable to the same reasons as those outlined in the explanation of the results for the first simulation study (Section \ref{sec:simulation_wellspec_results}). When the data distribution is less skewed and less heavy-tailed, i.e., when $\nu=5.0$, the IJM method performs well. This results suggest that one should adopt more appropriate outcome models for less skewed data. On the other hand, our model is assumed to be more appropriate for the data in the case study that is considered in Section \ref{sec:application}, as those data are heavily skewed and exhibit substantial outliers on the tail.

		\begin{table*}
		\centering
		\caption{Evaluation metrics for our Bayesian methodology versus the IJM method under model misspecification.}
		\begin{adjustbox}{width=13.3cm}
			\begin{tabular}{lrrrrrrrrr}
				\toprule
				& & \multicolumn{2}{c}{Coverage}   & \multicolumn{2}{c}{Bias}   & \multicolumn{2}{c}{MSE}   & \multicolumn{2}{c}{Interval Width} \\
				\cmidrule(lr){3-4} \cmidrule(lr){5-6} \cmidrule(lr){7-8} \cmidrule(lr){9-10} 
				& $\nu$  & IJM & Bayes & IJM & Bayes & IJM & Bayes & IJM & Bayes \\ \hline
				&1.5  & 94\% & 92\% & 8.38E-01 & 4.43E-01 & 3.52E+02 & 2.78E+01 & 2.08E+01 & 2.19E+01 \\ 
				$\mathrm{CADE}(0,a_0)$&2.0 & 92\% & 93\% & -7.55E-02 & -4.02E-01 & 7.21E+00 & 3.70E+00 & 7.77E+00 & 8.39E+00 \\ 
				&3.0 & 95\% & 96\% & -5.92E-02 & 2.51E-02 & 1.11E+00 & 8.78E-01 & 4.01E+00 & 4.10E+00 \\ 
				&5.0 & 93\% & 84\% & -4.01E-02 & 5.29E-01 & 5.65E-01 & 8.85E-01 & 2.84E+00 & 2.88E+00 \\
				\hline
				&1.5  & 97\% & 90\% & 2.01E+00 & -1.64E+00 & 7.36E+02 & 1.43E+02 & 2.58E+01 & 2.61E+01 \\ 
				$\mathrm{CADE}(1,a_1)$&2.0 & 98\% & 96\% & -6.03E-02 & -4.60E-01 & 1.24E+01 & 7.57E+00 & 1.07E+01& 1.08E+01 \\ 
				&3.0 & 96\% & 98\% & 3.72E-02 & -1.43E-01 & 2.10E+00 & 1.50E+00 & 5.42E+00 & 5.96E+00 \\ 
				&5.0 & 96\% & 93\% & -1.41E-01 & -6.70E-01 & 9.33E-01 & 1.62E+00 & 3.83E+00 & 4.39E+00 \\ 
				\hline
				&1.5  & 94\% & 92\% & 3.60E-01 & 1.61E-01 & 6.37E+01 & 5.02E+00 & 8.99E+00 & 9.50E+00 \\ 
				$\mathrm{ITT}_{\mathrm{Y},\cdot,\cdot}(a_0)$&2.0 & 93\% & 94\% & -2.74E-02 & -1.94E-01 & 1.34E+00 & 6.90E-01 & 3.37E+00 & 3.54E+00 \\ 
				&3.0 & 96\% & 97\% & -1.92E-02 & -1.10E-02 & 2.11E-01 & 1.61E-01 & 1.74E+00 & 1.75E+00 \\ 
				&5.0 & 93\% & 86\% & -1.21E-02 & 2.02E-01 & 1.07E-01 & 1.50E-01 & 1.23E+00 &1.21E+00 \\ 
				\hline
				&1.5  & 96\% & 89\% & 9.11E-01 & -7.06E-01 & 1.14E+02 & 2.15E+01 & 1.03E+01 & 9.90E+00 \\ 
				$\mathrm{ITT}_{\mathrm{Y},\cdot,\cdot}(a_1)$&2.0 & 98\% & 96\% & -3.04E-02 & -2.30E-01 & 2.01E+00 & 1.16E+00 &  4.24E+00 &4.28E+00 \\ 
				&3.0 & 95\% & 98\% & 1.23E-02 & -1.10E-01 & 3.37E-01 & 2.36E-01 & 2.20E+00 & 2.30E+00 \\ 
				&5.0 & 96\% & 87\% & -5.61E-02 & -3.10E-01 & 1.58E-01 & 2.79E-01 & 1.55E+00 & 1.64E+00 \\ 
				\hline
				&1.5  & 95\% & 93\% & 1.57E-03 & -4.14E-03 & 1.51E-04 & 1.97E-04 & 5.09E-02& 4.70E-02 \\ 
				$\mathrm{ITT}_{\mathrm{D},\cdot,\cdot}(a_0)$&2.0 & 95\% & 85\% & 1.57E-03 & -5.87E-03 & 1.51E-04 & 2.45E-04 & 5.09E-02 &4.66E-02 \\ 
				&3.0 & 95\% & 77\% & 1.57E-03 & -6.43E-03 & 1.51E-04 & 3.46E-04 & 5.09E-02 & 4.60E+00 \\ 
				&5.0 & 95\% & 67\% & 1.57E-03 & -7.54E-03 & 1.51E-04 & 4.85E-04 & 5.09E-02 & 4.46E-02 \\ 
				\hline
				&1.5  & 95\% & 76\% & -1.24E-03 & -1.27E-03 & 2.78E-04 & 6.22E-04 & 6.52E-02 & 5.90E-02 \\ 
				$\mathrm{ITT}_{\mathrm{D},\cdot,\cdot}(a_1)$&2.0 & 95\% & 69\% & -1.24E-03 & -1.53E-02 & 2.78E-04 & 7.32E-04 & 6.52E-02 &5.75E-02\\ 
				&3.0 & 95\% & 59\% & -1.24E-03 & -1.77E-02 & 2.78E-04 & 1.00E-03 & 6.52E-02 & 5.55E-02 \\ 
				&5.0 & 95\% & 47\% & -1.24E-03 & 2.43E-02 & 2.78E-04 & 1.53E-03 & 6.52E-02 & 5.29E-02 \\ 
				\bottomrule
			\end{tabular}
		\end{adjustbox}
		\label{tab:simulation_misspecification}
	\end{table*}
 
	\subsection{Evaluation under MNAR}
		\label{sec:MNAR_evaluation}
		Table \ref{tab:simulation_mnar_mcar_ijm} presents the simulation results under MNAR. The underlying parametrizations are the same as the Log-normal mechanism in Section \ref{sec:simulation_studies_overview}, except that the outcome is missing with probabilities given in Table \ref{tab:params_missingrate} in the appendix.  It is important to recognize that the IJM method is evaluated after applying listwise deletion of the rows with missing outcome. We see that for $N=5000$ the interval lengths of the Bayesian method are a bit larger than those in Table \ref{tab:simulation_fn_lognormal}. This is due to incorporating the missingness mechanism into our model. For sufficiently large data (i.e., $N=10000, 50000$), the results are similar to those in Table \ref{tab:simulation_fn_lognormal}.

	The performance of the IJM method in Table \ref{tab:simulation_mnar_mcar_ijm} is worse in all metrics compared to the IJM results in Table \ref{tab:simulation_fn_lognormal}. This is due to the bias resulting from applying listwise deletion of rows under the assumption of an MCAR mechanism when the actual missing data mechanism is MNAR. The bias becomes more severe when the missingness probabilities are significantly large. We present additional simulations with larger missing probabilities in the appendix.
		
		\begin{table*}
    \centering
    \caption{Evaluation for our Bayesian methodology under MNAR. The IJM method is evaluated after applying listwise deletion of rows with missingness.}
    \begin{adjustbox}{width=13.3cm}
    \begin{tabular}{lrrrrrrrrr}
        \toprule
        & & \multicolumn{2}{c}{Coverage}   & \multicolumn{2}{c}{Bias}   & \multicolumn{2}{c}{MSE}   & \multicolumn{2}{c}{Interval Width} \\
        \cmidrule(lr){3-4} \cmidrule(lr){5-6} \cmidrule(lr){7-8} \cmidrule(lr){9-10} 
        & N  & IJM & Bayes & IJM & Bayes & IJM & Bayes & IJM & Bayes \\ 
        \hline
        &5000   &  96\% & 98\% &  6.12E+02 & 4.65E+02 & 3.88E+07 & 4.77E+06 & 2.06E+04&  3.24E+04\\ 
        $\mathrm{CADE}(0,a_0)$&10000 & 98\% & 98\% & -2.52E+02 & 2.02E+02 & 1.58E+07 & 1.12E+06 & 1.48E+04 & 1.41E+04\\ 
        &50000   & 94\% & 96\% & -3.25E+02 & 1.22E+02 & 3.97E+06 &  5.21E+05 & 7.07E+03 &  2.87E+03\\ 
        \hline
        &5000 & 92\% & 96\% & -1.27E+03 & 1.84E+03 & 2.82E+08 & 9.11E+06 & 4.91E+04 &  5.43E+04\\ 
        $\mathrm{CADE}(1,a_1)$&10000 & 93\% & 97\% & 6.10E+02 & 1.25E+03  & 1.47E+08 & 6.07E+06 & 3.71E+04 & 3.92E+04\\ 
        &50000  & 96\% & 98\% & 3.65E+02 & 5.31E+03 & 2.56E+07  & 5.93E+06 & 1.84E+04 & 1.26E+04\\
        \hline
        &5000  & 96\% & 98\% & 3.34E+02 &  2.22E+02 & 9.50E+06 & 1.26E+06 & 1.00E+04 & 1.60E+04\\ 
        $\mathrm{ITT}_{Y, \cdot , \cdot}(a_0)$& 10000 & 97\% & 98\% & 1.94E+02 & 9.60E+01 & 3.83E+06 & 3.02E+05 & 7.30E+03 &  7.06E+03\\ 
        &50000  &  93\% & 96\% & -2.14E+02 & 6.24E+01 & 9.86E+05 & 1.38E+05& 3.47E+03 & 1.45E+03\\ 
        \hline
        & 5000 & 92\% & 95\% & -4.79E+02 & 9.28E+02 & 5.51E+07 & 2.25E+06 & 2.20E+04 &  2.50E+04 \\ 
        $\mathrm{ITT}_{Y, \cdot , \cdot}(a_1)$& 10000 & 93\% & 97\% & 2.66E+02 & 6.20E+02 & 2.83E+07 & 1.40E+06 & 1.63E+04 & 1.80E+04\\ 
        & 50000 & 96\% & 98\% & 1.30E+02 & 2.49E+02 & 5.03E+06 & 1.22E+06 & 8.17E+03  & 5.77E+03 \\ 
        \hline
        &5000 & 94\% & 91\% & 1.89E-03 & 4.81E-03 & 3.43E-04& 3.33E-04 & 7.10E-02 &  5.94E-02\\ 
        $\mathrm{ITT}_{D, \cdot , \cdot}(a_0)$&10000 & 72\% & 92\% & -1.71E-02 &  -2.36E-03 & 4.72E-04 &  1.62E-04 & 5.04E-02 &  4.23E-02\\ 
        &50000 & 47\% & 93\% & -1.13E-02 & -2.22E-04 & 1.62E-04 & 2.50E-05 & 2.24E-02 &  1.86E-02\\ 
        \hline
        &5000 & 96\% & 91\% & 8.62E-04 & 9.62E-03 & 5.47E-04 & 5.09E-04 & 9.83E-02 &  7.29E-02\\ 
        $\mathrm{ITT}_{D, \cdot , \cdot}(a_1)$&10000 & 89\% & 91\% & -1.09E-02 & 5.62E-03 & 4.42E-04 & 2.50E-04 & 6.90E-02  &  5.43E-02\\ 
        &50000 & 89\% & 95\% & -6.75E-03 &  7.27E-04 & 1.00E-04 & 3.80E-05 & 3.08E-02 & 2.49E-02\\
        \bottomrule
    \end{tabular}
    \end{adjustbox}
    \label{tab:simulation_mnar_mcar_ijm}
\end{table*}

	\section{Case Study: The Rashtriya Swasthya Bima Yojana Health Insurance Dataset}
	\label{sec:application}
	
	\subsection{Description of Experiment and Data}
	\label{sec:application_introduction}
	
	Approximately $63$ million people in India are below the poverty line due to health care expenditures. In 2008 a large-scale national hospital insurance plan for the poor was launched. This plan is known as the Rashtriya Swasthya Bima Yojana (RSBY). It is a large-scale national hospital insurance plan that households below the poverty line can join with a nominal co-payment. Under RSBY, these households can be covered for up to five people and more than $700$ medical treatments and procedures, with the price set by the government. Medical services are provided nationwide by government-contracted public and private hospitals. Beneficiaries use their RSBY biometric ID cards, eliminating the need for cash transactions and insurance claims. Additional information and references for RSBY are provided by \citet{Nandi2015}.
	
	\citet{Imai2021} conducted a two-stage randomized experiment to determine whether access to the national insurance plan provided by RSBY increases access to hospitals and reduces impoverishment due to high medical expenses. This experiment consisted of $N = 10,072$ households after \citet{Imai2021} applied listwise deletion for missing outcomes, with the households residing in $J = 435$ villages. Of these villages, $J_1 = 219$ were assigned treatment probability $a_1 = 0.8$ and the remaining $216$ villages were assigned treatment probability $a_0 = 0.4$. One concern in this experiment was the spillover effects between households, because one household's enrollment in RSBY may depend on the treatments assigned to other households. Another concern is that some households assigned treatment may decide to not enroll in RSBY, and some households assigned control may ultimately manage to enroll in RSBY. 
	
	\subsection{Bayesian Analyses of the Experiment}
	\label{sec:application_analyses}
	
	We utilize our Bayesian methodology to infer the direct and spillover effects accounting for interference, treatment nonadherence, and missing outcomes on the annual household hospital expenditure outcome (which ranges from $0$ to INR $500,000$). We first consider the case of \citet{Imai2021} in which the outcomes are assumed to be MCAR. In this case, listwise deletion of rows with missingness is performed. 
	We use the same mixture model as in Section \ref{sec:simulation_studies_overview} to analyze the data. The model for principal strata memberships is the Multinomial distribution, $G_{i,j} \sim \mathrm{Multinomial} \left ( \pi_{(\mathcal{n},\mathcal{n})}, \pi_{(\mathcal{c},\mathcal{c})}, \pi_{(\mathcal{a},\mathcal{a})}, \pi_{(\mathcal{n},\mathcal{c})}, \pi_{(\mathcal{n},\mathcal{a})}, \pi_{(\mathcal{c},\mathcal{a})} \right )$. For the potential outcomes we specify a zero-inflated Log-Normal distribution for each principal stratum, that is, for $a \in \{a_0, a_1\}$ and $z \in \{0, 1\}$, the potential outcomes for unit $i$ in cluster $j$ are determined by $Y_{i,j}(z,a) = \left \{ 1 - W_{i,j}(z,a) \right \}\tilde{Y}_{i,j}(z,a)$ where $W_{i,j}(z, a) \sim \mathrm{Bernoulli}(p_{z, a, G_{i,j}})$ and $\tilde{Y}_{i,j}(z,a) \sim$ Log-Normal$(\mu_{z, a, G_{i,j}}, \sigma_{z, a, G_{i,j}}^2)$. Finally, we use conjugate prior distributions for all parameters. $ \left ( \pi_{(\mathcal{n},\mathcal{n})}, \pi_{(\mathcal{c},\mathcal{c})}, \pi_{(\mathcal{a},\mathcal{a})}, \pi_{(\mathcal{n},\mathcal{c})}, \pi_{(\mathcal{n},\mathcal{a})}, \pi_{(\mathcal{c},\mathcal{a})} \right ) \sim \mathrm{Dirichlet}(\alpha_{(\mathcal{n},\mathcal{n})}
	,\alpha_{(\mathcal{c},\mathcal{c})}
	,\alpha_{(\mathcal{a},\mathcal{a})}
	,\alpha_{(\mathcal{n},\mathcal{c})}
	,\alpha_{(\mathcal{n},\mathcal{a})}
	,\alpha_{(\mathcal{c},\mathcal{a})})$
	, $p_{z, a, G_{i,j}} \sim \mathrm{Beta}(a_0,b_0)$, $\mu_{z, a, G_{i,j}} \sim \mathrm{Normal}(\mu_0, \sigma_0^2)$ and $\sigma_{z, a, G_{i,j}}^2 \sim \mathrm{Inverse Gamma}(k_0, \theta_0)$ where $\theta_0$ is a scale parameter. For the hyperparameters we choose $\alpha_{(\mathcal{n},\mathcal{n})}
	=\alpha_{(\mathcal{c},\mathcal{c})}
	=\alpha_{(\mathcal{a},\mathcal{a})}
	=\alpha_{(\mathcal{n},\mathcal{c})}
	=\alpha_{(\mathcal{n},\mathcal{a})}
	=\alpha_{(\mathcal{c},\mathcal{a})}=1$, $a_0=b_0=1$, $\mu_0=0$, $\sigma_0^2=5$, $k_0=0.1$ and  $\theta_0=1$. Note that  $\sigma_0^2=5$ is sufficiently large on a log scale. Our MCMC algorithm was performed for $100,000$ iterations with a burn-in of $50,000$ draws. 
	
	We consider the finite-population inference as was done by \citet{Imai2021}. One advantage of the alternative, super-population inference is that the estimands are free of the parameters governing the associations between potential outcomes. This is an advantage because the data are not informative about the associations between potential outcomes as all potential outcomes can never be jointly observed simultaneously for an experimental unit. \citet{Ding2018} suggested isolating the parameters that govern the marginal distributions from the parameters that govern the associations between potential outcomes, and performing sensitivity analyses for the association parameters. We omit the sensitivity analysis for these parameters and instead perform a sensitivity analysis for the prior distributions.
	
	\begin{table*}[t]
		\centering
		
		\caption{Comparisons of the causal inferences obtained from our Bayesian methodology with those obtained from the method of IJM.}
		
		\label{tab:resultcomparison}
		
		\begin{tabular}{ lrrrrrr}
			\toprule
			& Mean  & SD & Median & $95\%$ interval & IJM Est. & IJM SD  \\
			\hline
			$\mathrm{CADE}(a_1,a_1)$ & $-2041 $ & $7247$  & $-1813$ & $(-3782,-256)$ & $ -1649$ & $1061$  \\
			$\mathrm{CADE}(a_0, a_0)$ & $298  $ & $3912$ & $158$  & $(-1356,1982)$ &  $1984$ & $ 1215$ \\
			$\mathrm{ITT_{Y, \cdot, \cdot}}(a_1)$& $-853  $ & $3040$ & $-759$ & $(-1586, -106)$ & $-795 $ & $514$    \\
			$\mathrm{ITT_{Y, \cdot, \cdot}}(a_0)$& $139  $ & $1811$ & $74$ & $(-632,913)$ &  $875$ & $530$   \\
			$\mathrm{ITT_{D, \cdot, \cdot}}(a_1)$& $0.418  $ & $0.010$ & $0.418$ & $(0.397,0.438)$ &  $0.482$ & $0.023$   \\
			$\mathrm{ITT_{D, \cdot, \cdot}}(a_0)$& $0.465  $ & $0.009$ & $0.465$ & $(0.446, 0.483)$ &  $0.441$ & $0.021$   \\
			$\mathrm{S_{Y, \cdot, \cdot}}(1)$& $-1129  $ & $1795$ & $-1071$ & $(-1741,-459)$ &  $-1374$ & $823$   \\
			$\mathrm{S_{Y, \cdot, \cdot}}(0)$& $-136  $ & $3044$ & $-222$ & $(-1003,666)$ & $297$ & $858$    \\
			$\mathrm{S_{D, \cdot, \cdot}}(1)$& $0.030  $ & $0.007$ & $0.029$ & $(0.018,0.047)$ &   $0.086$ & $0.053$  \\
			$\mathrm{S_{D, \cdot, \cdot}}(0)$& $0.077  $ & $0.009$ & $0.077$ & $(0.060,0.095)$ &  $0.045$ & $0.028$ \\
			\bottomrule
		\end{tabular}
	\end{table*}
	
	Table \ref{tab:resultcomparison} compares the results obtained from our Bayesian methodology with those obtained by the method of \citet{Imai2021} in the case of MCAR outcomes. We do not consider the complier average spillover effects because we defined these estimands differently from \citet{Imai2021}, and because \citet{Imai2021} mentioned that these estimands were imprecisely estimated in their analyses. We observe in Table \ref{tab:resultcomparison} that our results are generally consistent with those of \citet{Imai2021}. However, differences exist because our Bayesian method is able to detect significance effects in $\mathrm{CADE}(a_1, a_1)$, $\mathrm{ITT_{Y,\cdot,\cdot}}(a_1)$, and $\mathrm{S_{Y,\cdot,\cdot}}(1)$. The negative spillover effects that our method detects indicate that treated households are more likely to be negatively affected by the shift from $a_0$ to $a_1$. Alternatively, assigning a greater proportion of households to treatment will cause another treated household in the same village to spend less. We also observe large posterior standard deviations, and that the posterior intervals always have smaller widths than IJM's confidence intervals. The new effects that our methodology detects can be attributed to the greater precision (hence power) that follows from the use of the Bayesian model. Furthermore, our Bayesian methodology's ability to consider a point estimator based on the median, and a model that accommodates both an abundance of zeros and heavy tails, is advantageous for analyzing the data as the inferences would be robust to outlying observations. In particular, there are $36$ observations greater than INR $100,000$, with the two largest ones being INR $403,000$ and $500,000$. For comparison, the median in the dataset is INR $1,000$. 
	
	\begin{table*}[t]
		\centering
		\caption{Additional causal estimands and their estimates for the RSBY data.}
		\label{tab:resultrest}
		
		\begin{tabular}{ lrrr}
			\toprule
			& Post. Mean & Post. Median & $95\%$ Credible Interval   \\
			\hline
			$\mathrm{O_{Y, \cdot, \cdot}}$& $-731$  & $-739$ & $(-1280,-181)$   \\
			$\mathrm{CAOE(a_0)}$& $-1242$ & $-1154$  & $(-2334,-30)$ \\
			$\mathrm{CAOE(a_1)}$& $-1234$ & $-1328$ & $(-2600, -75)$    \\
			$\mathrm{CADE(a_1,a_0)}$ & $-1647$ & $-1626$ & $(-3329,-230)$  \\
			$\mathrm{CADE(a_0,a_1)}$& $259$ & $237$ & $(-1444,2162)$ \\
			$\mathrm{CASE(0,a_0)}$& $-33$ & $-64$ & $(-1691, 1797)$\\
			$\mathrm{CASE(1,a_0)}$& $-2004$ & $-1870$ & $(-3484,-548)$\\
			$\mathrm{CASE(0,a_1)}$& $128$ & $-76$ & $(-1903,1981)$   \\
			$\mathrm{CASE(1,a_1)}$& $-2172$ & $-2160$ & $(-3831,-694)$\\
			\bottomrule
		\end{tabular}
	\end{table*}
	
	Table \ref{tab:resultrest} summarizes the results for the other causal estimands in the case of MCAR outcomes. Interestingly, the credible intervals of the overall effects for all units, as well as for compliers, lie below zero. The overall effect is of greatest interest to policy makers as it captures a pure impact of the intervention on all units. Our inferences on compliers imply that the overall effects are negative for units who comply with the assignment regardless of which treatment probability $a_0$ or $a_1$ is assigned to their respective cluster. We can also conclude from both Tables \ref{tab:resultcomparison} and \ref{tab:resultrest} that the direct effect of the treatment assignment under treatment probability $a_1$ is negative for compliers, regardless of their base cluster assignment. Finally, we can conclude that the spillover effects for compliers are negative when all units are assigned to treatment, regardless of whether they are compliers under $a_0$ or $a_1$. Combined with our inferences on $S_{Y, \cdot, \cdot}(1)$, the spillover effect of treatment assignment on units is negative, regardless of their principal strata.
	
	Table \ref{tab:results_mnar} presents the results in the case of MNAR outcomes. As the overall missingness rate is small ($7.8\%$), we do not observe significant changes from the results under the MCAR mechanism in Table \ref{tab:resultcomparison} and \ref{tab:resultrest}. This observation implies that assuming the MCAR mechanism would not hurt the final estimates for our analyses. Our Bayesian methodology enables us to explicitly assess the plausibility of missingness assumptions. This is another advantage of our methodology over the previous works \citep{Imai2021, Forastiere2016,Gonzalo2021}.
	
	\begin{table*}[t]
		\centering
		\caption{Results in the case of MNAR outcomes.}
		\label{tab:results_mnar}
		\begin{tabular}{ lrrr}
			\toprule
			& Post. Mean & Post. Median & $95\%$ Credible Interval   \\
			\hline
			$\mathrm{CADE}(a_1,a_1)$ & $-1951$ & $-1871$  & $(-3808,-361)$  \\
			$\mathrm{CADE}(a_0, a_0)$ & $127 $ & $78$ & $(-1473,1854)$  \\
			$\mathrm{ITT_{Y, \cdot, \cdot}}(a_1)$& $-814$ & $-783$ & $(-1597,-149)$    \\
			$\mathrm{ITT_{Y, \cdot, \cdot}}(a_0)$& $58$ & $34$ & $(-674,838)$  \\
			$\mathrm{ITT_{D, \cdot, \cdot}}(a_1)$& $0.417$ & $0.417$ & $(0.395,0.436)$  \\
			$\mathrm{ITT_{D, \cdot, \cdot}}(a_0)$& $0.456$ & $0.456$ & $(0.437,0.474)$  \\
			$\mathrm{S_{Y, \cdot, \cdot}}(1)$& $-1069$ & $-1038$ & $(-1726,-465)$  \\
			$\mathrm{S_{Y, \cdot, \cdot}}(0)$& $-196$ & $-209$ & $(-1011,700)$  \\
			$\mathrm{S_{D, \cdot, \cdot}}(1)$& $0.036$ & $0.036$ & $(0.024,0.051)$ \\
			$\mathrm{S_{D, \cdot, \cdot}}(0)$& $0.075$ & $0.075$ & $(0.059,0.092)$  \\
			$\mathrm{O_{Y, \cdot, \cdot}}$& $-729$  & $-714$ & $(-1284,-217)$   \\
			$\mathrm{CAOE(a_0)}$& $-1106$ & $-1067$  & $(-2291,-18)$ \\
			$\mathrm{CAOE(a_1)}$& $-1222$ & $-1213$ & $(-2455, -37)$    \\
			$\mathrm{CADE(a_1,a_0)}$ & $-1773$ & $-1710$ & $(-3456,-328)$  \\
			$\mathrm{CADE(a_0,a_1)}$& $143$ & $151$ & $(-1565,1965)$ \\
			$\mathrm{CASE(0,a_0)}$& $54$ & $37$ & $(-1595, 1939)$\\
			$\mathrm{CASE(1,a_0)}$& $-1846$ & $-1750$ & $(-3390,-451)$\\
			$\mathrm{CASE(0,a_1)}$& $50$ & $26$ & $(-1778,2097)$   \\
			$\mathrm{CASE(1,a_1)}$& $-2044$ & $-1990$ & $(-3657,-596)$\\
			\bottomrule
		\end{tabular}
	\end{table*}
	
	\subsection{Sensitivity Analysis to Prior Specifications}
	\label{sec:case_study_sensitivity_analysis}
	
	When performing Bayesian analyses with weakly identifiable models it is important to investigate the robustness of the results with respect to the prior specifications, so as to make inferences more reliable. The results in this section are derived using proper prior distributions. In particular, we use $\mu_{z, a, G_{i,j}} \sim \mathrm{Normal}(0, \sigma_0^2)$, $\sigma_{z, a, G_{i,j}}^2 \sim \mathrm{InverseGamma}(a_0,b_0)$, $(\pi_{(\mathcal{n},\mathcal{n})}$, $\pi_{(\mathcal{c},\mathcal{c})}$, $\pi_{(\mathcal{a},\mathcal{a})}$, $\pi_{(\mathcal{n},\mathcal{c})}$, $\pi_{(\mathcal{n},\mathcal{a})}$, $\pi_{(\mathcal{c},\mathcal{a})} ) \sim$ $\mathrm{Dirichlet}(\phi_{(\mathcal{c},\mathcal{c})}$,$\phi_{(\mathcal{a},\mathcal{a})}$,$\phi_{(\mathcal{n},\mathcal{n})}$,$\phi_{(\mathcal{n},\mathcal{c})}$,$\phi_{(\mathcal{n},\mathcal{a})}$,$\phi_{(\mathcal{c},\mathcal{a})})$, and $p_{z, a, G_{i,j}} \sim \mathrm{Beta}(\alpha_0,\beta_0)$ for the prior specifications. Table \ref{tab:sensitivity_parameters} presents the specifications of hyperparameters. Note that $\sigma_0^2=3$ is substantially different from $\sigma_0^2=5$ and $\sigma_0^2=7$ on a log scale.
	Table \ref{tab:sensitivity} reports the results for CADE. we observe that the extreme cases for $p$, specifically, Cases $10$ and $11$, lead to slight changes in the intervals, and that the other cases only exhibit minor changes in the results. However, the extreme values of $p$ indicate strong prior beliefs about whether the outcome is zero or not for each principle strata. In our case study, as we lack knowledge about the unobservable strata, we believe that such strong priors about these parameters are unreasonable.
	\begin{table}[t]
		\centering
		
		\caption{Sensitivity analysis to prior specifications. We present the posterior medians and $95\%$ central credible intervals of the causal estimands.}
		
		\label{tab:sensitivity}
		
		\begin{adjustbox}{width=13.3cm}
			\begin{tabular}{ lrrrrrrrrrrrr}
				\toprule
				& \multicolumn{3}{c}{$\mathrm{CADE}(a_0, a_0)$}   & \multicolumn{3}{c}{$\mathrm{CADE}(a_1, a_0)$}   & \multicolumn{3}{c}{$\mathrm{CADE}(a_0, a_1)$}   & \multicolumn{3}{c}{$\mathrm{CADE}(a_1, a_1)$} \\
        \cmidrule(lr){2-4} \cmidrule(lr){5-7} \cmidrule(lr){8-10} \cmidrule(lr){11-13} 
        Percentile & $2.5\%$ & $50\%$ & $97.5\%$  & $2.5\%$ & $50\%$ & $97.5\%$  & $2.5\%$ & $50\%$ & $97.5\%$  & $2.5\%$ & $50\%$ & $97.5\%$  \\
				\hline
				Case 1 & $-1346$ & $ 125$ & $ 1773$ & $-3329$ & $ -1593$ & $ -161$ & $-1423$ & $ 210$ & $ 2034$ & $-3691$ & $ -1770$ & $ -181$ \\
				Case 2 & $-1356$ & $158$ & $1982$  & $-3328$ & $ -1626$ & $ -229$ & $-1444$ & $ 236$ & $ 2162$ & $-3782$ & $ -1813$ & $ -256$  \\
				Case 3 & $-1265 $ & $ 165 $ & $ 2134$ & $-3431 $ & $ -1679 $ & $ -170$  & $-1357 $ & $ 212 $ & $ 2084$ &  $-3845 $ & $ -1873$ & $ -241$ \\
				Case 4 & $-1313$ & $ 192$ & $ 1944$ & $-3322$ & $ -1579$ & $ -179$ & $-1399$ & $ 262$ & $ 2094$ & $-3742$ & $ -1764$ & $ -205$ \\
				Case 5 & $-1351$ & $ 143$ & $ 2013$ & $-3458$ & $ -1682$ & $ -308$ & $-1458$ & $ 190$ & $ 2120$ & $-3905$ & $ -1881$ & $ -353$ \\
				Case 6 & $-1388$ & $ 141$ & $ 1783$ & $-3326$ & $ -1610$ & $ -192$  & $-1463$ & $ 233$ & $ 2055$ &  $-3718$ & $ -1782$ & $ -218$ \\
				Case 7 & $-1194$ & $ 259$ & $ 1884$ & $-3797$ & $ -2034$ & $ -576$ & $-1238$ & $ 261$ & $ 1938$ & $-3888$ & $ -2081$ & $ -596$ \\
				Case 8 & $-1323$ & $ 168$ & $ 2004$ & $-3447$ & $ -1652$ & $ -288$ & $-1442$ & $ 230$ & $ 2115$ & $-3874$ & $ -1853$ & $ -326$ \\
				Case 9 & $-1335$ & $ 176$ & $ 2052$ & $-3341$ & $ -1639$ & $ -195$ & $-1387$ & $ 255$ & $ 2147$ & $-3759$ & $ -1813$ & $ -239$ \\
				Case 10 & $-902$ & $ 484$ & $ 2128$ & $-3354$ & $ -1734$ & $ -326$ & $-990$ & $ 525$ & $ 2326$ & $-3627$ & $ -1881$ & $ -355$ \\
				Case 11 & $-1411$ & $ -10$ & $ 1629$ & $-3447$ & $ -1762$ & $ -447$ & $-1521$ & $ 21$ & $ 1770$ & $-3796$ & $ -1928$ & $ -490$ \\
				Case 12 & $-1384$ & $ 136$ & $ 1857$ & $-3298$ & $ -1660$ & $ -219$ & $-1445$ & $ 223$ & $ 2131$ & $-3632$ & $ -1813$ & $ -239$ \\
				\bottomrule
			\end{tabular}
		\end{adjustbox}
	\end{table}
	
	\begin{table}[t]
		\centering
		
		\caption{Parameter specifications for sensitivity analysis.}
		
		\label{tab:sensitivity_parameters}
		
		\begin{adjustbox}{width=13.3cm}
			\begin{tabular}{ lrrrrrrrrrrr}
				\toprule
				& \multicolumn{3}{c}{Log-Normal} & \multicolumn{6}{c}{Principal Strata}& \multicolumn{2}{c}{Excess Zeros}\\
				\cmidrule(lr){2-4} \cmidrule(lr){5-10} \cmidrule(lr){11-12} 
                 & $\sigma_0^2$ & $a_0$ & $b_0$  & $\phi_{(\mathcal{c},\mathcal{c})}$ & $\phi_{(\mathcal{a},\mathcal{a})}$ & $\phi_{(\mathcal{n},\mathcal{n})}$ & $\phi_{(\mathcal{n},\mathcal{c})}$ & $\phi_{(\mathcal{n},\mathcal{a})}$  & $\phi_{(\mathcal{c},\mathcal{a})}$ & $\alpha_0$ & $\beta_0$  \\
				\hline
				Case 1 & $3.0$ & $1.0$ & $1.0$  & $1.0$ & $1.0$ & $1.0$ & $1.0$ & $1.0$ & $1.0$ & $1.0$ & $1.0$   \\
				Case 2 & $5.0$ & $0.1$ & $1.0$  & $1.0$ & $1.0$ & $1.0$ & $1.0$ & $1.0$ & $1.0$ & $1.0$ & $1.0$   \\
				Case 3 & $7.0$ & $0.01$ & $1.0$  & $1.0$ & $1.0$ & $1.0$ & $1.0$ & $1.0$ & $1.0$ & $1.0$ & $1.0$   \\
				Case 4 & $5.0$ & $0.1$ & $1.0$  & $2.0$ & $2.0$ & $2.0$ & $2.0$ & $2.0$ & $2.0$ & $1.0$ & $1.0$   \\
				Case 5 & $5.0$ & $0.1$ & $1.0$  & $1.0$ & $1.0$ & $1.0$ & $1.0$ & $1.0$ & $1.0$ & $2.0$ & $2.0$   \\
				Case 6 & $10.0$ & $1.0$ & $1.0$  & $1.0$ & $1.0$ & $1.0$ & $1.0$ & $1.0$ & $1.0$ & $1.0$ & $1.0$   \\
				Case 7 & $3.0$ & $0.1$ & $0.1$  & $1.0$ & $1.0$ & $1.0$ & $1.0$ & $1.0$ & $1.0$ & $1.0$ & $1.0$   \\
				Case 8 & $3.0$ & $1.0$ & $1.0$  & $0.1$ & $0.1$ & $0.1$ & $0.1$ & $0.1$ & $0.1$ & $1.0$ & $1.0$   \\
				Case 9 & $3.0$ & $1.0$ & $1.0$  & $10.0$ & $10.0$ & $10.0$ & $10.0$ & $10.0$ & $10.0$ & $1.0$ & $1.0$  \\
				Case 10 & $3.0$ & $1.0$ & $1.0$  & $1.0$ & $1.0$ & $1.0$ & $1.0$ & $1.0$ & $1.0$ & $0.1$ & $0.1$  \\
				Case 11 & $3.0$ & $1.0$ & $1.0$  & $1.0$ & $1.0$ & $1.0$ & $1.0$ & $1.0$ & $1.0$ & $10.0$ & $10.0$  \\
				Case 12 & $3.0$ & $1.0$ & $1.0$  & $1.0$ & $1.0$ & $1.0$ & $1.0$ & $1.0$ & $1.0$ & $0.5$ & $0.5$  \\
				\bottomrule
			\end{tabular}
		\end{adjustbox}
	\end{table}
	
	\section{Concluding Remarks}
	\label{sec:conclusion}
	
	We presented a Bayesian model-based methodology to simultaneously address interference, treatment nonadherence, and missing outcomes in two-stage randomized experiments. These three complications are significant drivers of unstable and biased causal inferences in modern experiments. Existing methodologies are lacking in their abilities to address all these issues simultaneously. There are three novel contributions of our Bayesian methodology. First, it provides a set of assumptions for performing valid causal inference in two-stage randomized experiments in the presence of these complications. We extended existing assumptions on the structure of interference to handle missing outcomes and also clarified and formalized assumptions about adherence behaviors within and across clusters. Second, it defines new causal estimands, including the overall effects of intervention and interpretable spillover effects, that can be inferred by means of our flexible Bayesian models. Our Bayesian methodologies address the issues of identifiability under two-sided nonadherence. Finally, our methodology can enable more precise causal inferences compared to existing methods for complex, non-standard data-generating mechanisms. We illustrated the utility of our methodology in this respect via simulation studies and the RSBY case study. In the latter study our methodology was able to uncover more definitive evidence of spillover and overall effects of the intervention that were not recognized in the previous analyses by \citet{Imai2021}. Our results are further validated by sensitivity analyses to prior distribution specifications and consideration of the case of MNAR missing outcomes.
	
	Of great interest for future research on our Bayesian methodology is the relaxation of the assumptions of the interference structure. In certain real-life contexts it may be too restrictive to employ the two-stage randomized design and assume stratified interference. A great deal of research has been conducted on inferring causal effects without using special designs such as the two-stage randomized design or clustered encouragement design. The work of \citet{Aronow2017} and \citet{Fredrik2021} provide one possible path for future research based on the network structure and exposure mapping.

	\bibliographystyle{ba}
	\bibliography{causal_inference_interference}
	
	\clearpage
	\appendix
	\begin{supplement}
	\label{sec:supplementary}
		\section{Derivation of $\Bar{D}_{i,j}(z,a)$ and $\Bar{Y}_{i,j}(a)$}
		\label{sec:derivation_d_bar}
		Under Assumptions \ref{asmp:partial} - \ref{asmp:stratified_Y}, we have 
		\begin{equation*}
		\begin{split}
		\Bar{D}_{i,j}(z,a) & = \sum_{\mathbf{z}_{-i,j} \in \mathcal{Z}_{-i,j}} D_{i,j}(z, \mathbf{z}_{-i,j}) p(\mathbf{Z}_{-i,j}=\mathbf{z}_{-i,j} \mid Z_{i,j} = z, A_j=a) \\
		& = \sum_{\mathbf{z}_{-i,j} \in \mathcal{Z}_{-i,j}, Z_{i,j}=z, A_j=a} D_{i,j}(z, a)p(\mathbf{Z}_{-i,j}=\mathbf{z}_{-i,j} | Z_{i,j}=z,A_j=a) \\
		& = \sum_{\mathbf{z}_{-i,j} \in \mathcal{Z}_{-i,j}, Z_{i,j}=z, A_j=a} D_{i,j}(z, a) {N_j - 1 \choose K_j(a)-z}^{-1} = D_{i,j}(z, a),
		\end{split}
		\end{equation*}
		where the second line follows from Assumption \ref{asmp:stratified} and \ref{asmp:stratified_Y} and the third line follows from a simple probability calculation. We also have that
		\begin{equation*}
		\begin{split}
		\Bar{Y}_{i,j}(a) & = 
		\sum_{\mathbf{z}_{j} \in \mathcal{Z}_{j}} Y_{i,j}(\mathbf{z})
		p(\mathbf{Z}_{j} = \mathbf{z}_{j} \mid A_j=a) = \sum_{\mathbf{z}_{j} \in \mathcal{Z}_{j}, A_j=a} Y_{i,j}(\mathbf{z}_{j})
		p(\mathbf{Z}_{j}=\mathbf{z}_{j} |A_j=a) \\
		& = \sum_{\mathbf{z}_{-i,j} \in \mathcal{Z}_{-i,j}, Z_{i,j}=1, A_j=a} Y_{i,j}(Z_{i,j}=1,\mathbf{Z}_{-i,j}=\mathbf{z}_{-i,j})
		p(Z_{i,j}=1,\mathbf{Z}_{-i,j}=\mathbf{z}_{-i,j} |A_j=a) \\
		& + \sum_{\mathbf{z}_{-i,j} \in \mathcal{Z}_{-i,j}, Z_{i,j}=0, A_j=a} Y_{i,j}(Z_{i,j}=0,\mathbf{Z}_{-i,j}=\mathbf{z}_{-i,j})
		p(Z_{i,j}=0,\mathbf{Z}_{-i,j}=\mathbf{z}_{-i,j} |A_j=a) \\
		& = \sum_{\mathbf{z}_{-i,j} \in \mathcal{Z}_{-i,j}, Z_{i,j}=1, A_j=a} Y_{i,j}(Z_{i,j}=1,\mathbf{Z}_{-i,j}=\mathbf{z}_{-i,j})
		p(\mathbf{Z}_{-i,j}=\mathbf{z}_{-i,j} |Z_{i,j}=1,A_j=a)p(Z_{i,j}=1|A_j=a) \\
		& + \sum_{\mathbf{z}_{-i,j} \in \mathcal{Z}_{-i,j}, Z_{i,j}=0, A_j=a} Y_{i,j}(Z_{i,j}=0,\mathbf{Z}_{-i,j}=\mathbf{z}_{-i,j})
		p(\mathbf{Z}_{-i,j}=\mathbf{z}_{-i,j} |Z_{i,j}=0,A_j=a)p(Z_{i,j}=0|A_j=a) \\
		& = \sum_{\mathbf{z}_{-i,j} \in \mathcal{Z}_{-i,j}, Z_{i,j}=1, A_j=a} Y_{i,j}(Z_{i,j}=1,\mathbf{Z}_{-i,j}=\mathbf{z}_{-i,j})
		p(\mathbf{Z}_{-i,j}=\mathbf{z}_{-i,j} |Z_{i,j}=1,A_j=a)\frac{K_j(a)}{N_j} \\
		& + \sum_{\mathbf{z}_{-i,j} \in \mathcal{Z}_{-i,j}, Z_{i,j}=0, A_j=a} Y_{i,j}(Z_{i,j}=0,\mathbf{Z}_{-i,j}=\mathbf{z}_{-i,j})
		p(\mathbf{Z}_{-i,j}=\mathbf{z}_{-i,j} |Z_{i,j}=0,A_j=a)\frac{N_j-K_j(a)}{N_j} \\
		& = \frac{K_j(a)}{N_j} \sum_{\mathbf{z}_{-i,j} \in \mathcal{Z}_{-i,j}, Z_{i,j}=1, A_j=a} Y_{i,j}(Z_{i,j}=1,\mathbf{Z}_{-i,j}=\mathbf{z}_{-i,j})
		{N_j-1\choose K_j(a)-1}^{-1} \\
		& + \frac{N_j-K_j(a)}{N_j} \sum_{\mathbf{z}_{-i,j} \in \mathcal{Z}_{-i,j}, Z_{i,j}=0, A_j=a} Y_{i,j}(Z_{i,j}=0,\mathbf{Z}_{-i,j}=\mathbf{z}_{-i,j})
		{N_j-1\choose K_j(a)}^{-1} \\
		& = \frac{K_j(a)}{N_j} \Bar{Y}_{i,j}(1,a) + \frac{N_j-K_j(a)}{N_j} \Bar{Y}_{i,j}(0,a). \\
		\end{split}
		\end{equation*}
		
		\section{On Exchangeability}
		\label{sec:definetti}
		
		Exchangeability can be justified for the two-stage randomized experiment in the presence of interference between units. First, consider two units that belong to different clusters. Assumption \ref{asmp:partial} and \ref{asmp:partial_Y} imply that they do not interfere with each other. As such, it is plausible that their outcomes are independent. Also, as covariates are independent of treatment assignments, the treatments received and potential outcomes under different treatment assignments would be identical if the units' cluster labels were permuted. Thus exchanging units that belong to different clusters in this manner does not affect the joint distribution for all units, and so they are exchangeable. We next consider two units $i_1$ and $i_2$ in the same cluster $j$. These units are not independent because their outcomes can be affected by interference. If both $i_1$ and $i_2$ are assigned treatment, then as Assumption \ref{asmp:stratified} and \ref{asmp:stratified_Y} imply that interference is determined by the total number of treated units within cluster $j$, the joint distribution of all units in the cluster doesn't change even if their labels are permuted because the number of treated units stays the same. Hence, the units are exchangeable. The same argument applies to the case in which both $i_1$ and $i_2$ are assigned control. Finally, if one is assigned treatment and the other is assigned control, then the total number of treated units in cluster $j$ still remains the same, and so the effect on the rest of the units within cluster $j$ would not change. The joint distribution of $i_1$ and $i_2$ will not change because only one of them is assigned treatment and the other is assigned control. Thus, exchangeability holds in such a case as well. 
		
		Ultimately, Assumptions \ref{asmp:partial} - \ref{asmp:stratified_Y} justify de Finetti's Theorem in the two-stage randomized experiment and the use of parametric models under our methodology. If these assumptions did not hold, unit exchangeability would be questionable because the interference structure may collapse if the labels of the experimental units were permuted, which would affect the joint distribution of all units. 
		
		\section{Derivations of Gibbs Samplers}
		This section presents the algorithm in Section \ref{sec:Gibbs_sampling}. If tractable prior distributions and models are used, which is the case in this manuscript, it will be possible to derive all the conditional posterior distributions involved in the Gibbs sampler in closed-form. Otherwise, Metropolis-within-Gibbs or Hamiltonian Monte Carlo-within-Gibbs steps will need to be considered in the algorithm.
		
		\noindent  In Step 2(a), we sample from $p(G_{i,j}^{(t+1)} \mid X_{i,j}, A_j^{\mathrm{o}},Z_{i,j}^{\mathrm{o}},D_{i,j}^{\mathrm{o}},M_{i,j}^{\mathrm{o}},Y_{i,j}^{\mathrm{obs}}, \boldsymbol{\psi}^{(t)}, \boldsymbol{\phi}^{(t)},\boldsymbol{\theta}^{(t)})$ following Section \ref{sec:imputation_principal_strata}. For example, for the units with $(A_j^{\mathrm{o}}, Z_{ij}^{\mathrm{o}}, D_{ij}^{\mathrm{o}}, M_{ij}^{\mathrm{o}}) = (a_0,0,0,0)$, there are five possible compliance behaviors for them, i.e., $G_{ij} \in \mathcal{G}(a_0,0,0) \equiv \{(\mathcal{c},\mathcal{c}),(\mathcal{n},\mathcal{n}),(\mathcal{c},\mathcal{a}),(\mathcal{n},\mathcal{c}),(\mathcal{n},\mathcal{a})\}$. Recall that $\mathcal{G}(a,z,d)$ is a set of possible compliance types for units with $(A_j^{\mathrm{o}}, Z_{ij}^{\mathrm{o}}, D_{ij}^{\mathrm{o}}) = (a,z,d)$. We derive the Gibbs sampling algorithm for the log-normal data generating mechanism that is defined in Section \ref{sec:simulationstudies} with a slight modification to handle missing outcomes. Specifically, we posit the following models. $M_{i,j}(z,a) \sim \mathrm{Bern}(\rho_{z, a, g})$. For $M_{i,j}(z,a)=0$, $G_{i,j} \sim \mathrm{Multinomial} \left ( \pi_{(\mathcal{n},\mathcal{n})}, \pi_{(\mathcal{c},\mathcal{c})}, \pi_{(\mathcal{a},\mathcal{a})}, \pi_{(\mathcal{n},\mathcal{c})}, \pi_{(\mathcal{n},\mathcal{a})}, \pi_{(\mathcal{c},\mathcal{a})} \right )$ and $Y_{i,j}(z,a) = \left \{ 1 - W_{i,j}(z,a) \right \}\tilde{Y}_{i,j}(z,a)$ where $W_{i,j}(z, a) \sim \mathrm{Bernoulli}(p_{z, a, G_{i,j}})$ and $\tilde{Y}_{i,j}(z,a) \sim$ Log-Normal$(\mu_{z, a, G_{i,j}}, \sigma_{z, a, G_{i,j}}^2)$. Recall that we defined $w_{i,j}^{g}$ as the probability mass function of $G_{i,j}$, $\rho_{i,j}^{g,a,z}$ as the probability mass function of $M_{i,j}$, and $f_{i,j}^{g,a,z}$ as the probability mass/density function of $Y_{i,j}(a,z)$ for $g$ in Section \ref{sec:bayesianinference}. So, $w_{i,j}^{g}= \pi_{g}$, $\rho_{i,j}^{g,a,z}=\rho_{z, a, g}$
		and $f_{i,j}^{g,a,z}=(p_{z, a, g})^{\mathbbm{1}(y=0)} \big( \left (1-p_{z, a, g} \right) h  (y \mid \mu_{z, a, g}, \sigma_{z, a, g}^2 )  \big)^{\mathbbm{1}(y>0)}$
		where  $h  (y \mid \mu_{z, a, g}, \sigma_{z, a, g}^2 )$ 
		is the probability density function for the log-normal distribution with parameters $\mu_{z, a, g}$ and $\sigma_{z, a, g}$. For observations  $(i,j) \in \mathcal{O}^{\mathrm{obs}}=(A_j^{\mathrm{o}}, Z_{i,j}^{\mathrm{o}}, D_{i,j}^{\mathrm{o}},M_{i,j}^{\mathrm{o}}, Y_{i,j}^{\mathrm{obs}}) = (a,z,d,0,y)$, we have
		\begin{equation*}
		\begin{split}
		& p(G_{i,j}^{(t+1)} =g \:|\: \mathcal{O}^{\mathrm{obs}},\boldsymbol{\psi}^{(t)},\boldsymbol{\theta}^{(t)})\\
		&= \frac{(1-\rho^{(t)}_{z, a, g}) \: \pi_g^{(t)} \: (p_{z, a, g}^{(t)})^{\mathbbm{1}(y=0)}\big((1-p_{z, a, g}^{(t)}) h  (y \mid \mu_{z, a, g}^{(t)}, (\sigma_{z, a, g}^{(t)})^{2} )  \big)^{\mathbbm{1}(y>0)} }{\sum_{g \in \mathcal{G}(a,z,d)} (1-\rho^{(t)}_{z, a, g}) \: \pi_g^{(t)} \: (p_{z, a, g}^{(t)})^{\mathbbm{1}(y=0)}\big((1-p_{z, a, g}^{(t)}) h  (y \mid \mu_{z, a, g}^{(t)}, (\sigma_{z, a, g}^{(t)})^{2} )  \big)^{\mathbbm{1}(y>0)} }.
		\end{split}
		\end{equation*}
		For observations  $(i,j) \in \mathcal{O}^{\mathrm{obs}} = (a,z,d,1,*)$, we have
		\begin{equation*}
		\begin{split}
		& p(G_{i,j}^{(t+1)} =g \:|\: \mathcal{O}^{\mathrm{obs}},\boldsymbol{\psi}^{(t)},\boldsymbol{\theta}^{(t)})
		= \frac{\rho^{(t)}_{z, a, g} \: \pi_g^{(t)} }{\sum_{g \in \mathcal{G}(a,z,d)} \rho^{(t)}_{z, a, g} \: \pi_g^{(t)}}
		\end{split}
		\end{equation*}
		for $g \in \mathcal{G}(a,z,d)$. Note that when $M_{i,j}^{\mathrm{o}}=1$, $Y_{i,j}^{\mathrm{obs}}$ is missing, which is denoted by $Y_{i,j}^{\mathrm{obs}}=*$.
		
		In Step 2(b), we impute the missing potential outcomes for unit $i$ in cluster $j$ using the parameter $\boldsymbol{\theta}^{(t)}$, the compliance behaviors $G_{i,j}^{(t+1)}=g$ and the observed variables $(A_j^{\mathrm{o}}, Z_{i,j}^{\mathrm{o}}, D_{i,j}^{\mathrm{o}},M_{i,j}^{\mathrm{o}}) = (a,z,d,m)$. First, we consider $M_{i,j}^{\mathrm{o}}=0$. We need to impute the missing values such that Assumption \ref{asmp:exclusion} holds. The imputation is two-fold. First, we sample $W_{i,j}^{(t+1)} \sim \mathrm{Bernoulli}(p_{z^{'},a^{'},g}^{(t)})$. Then, if $W_{ij}^{(t+1)}=0$, we sample $\tilde{Y}_{ij}^{(t+1)} \sim \mathrm{LogNormal}(\mu_{z^{'},a^{'},g}^{(t)}, (\sigma_{z^{'}, a^{'}, g}^{(t)})^{2})$ for missing potential outcomes where $(a,z) \neq (a^{'},z^{'})$. For example, if we observe  $(A_j^{\mathrm{o}}, Z_{ij}^{\mathrm{o}}, D_{ij}^{\mathrm{o}}) = (a_0,0,0)$ and are given $G_{ij}^{(t+1)}=(\mathcal{c},\mathcal{c})$ in the previous step of the Gibbs sampler, we need to impute three missing potential outcomes $Y_{ij}(1,a_0),Y_{ij}(0,a_1),Y_{ij}(1,a_1)$ using corresponding parameters $( p_{1,a_0,g}^{(t)}, \mu_{1,a_0,(\mathcal{c},\mathcal{c})}^{(t)}, (\sigma_{1,a_0,(\mathcal{c},\mathcal{c})}^{{(t)}})^{2})$, $( p_{0,a_1,g}^{(t)},\mu_{0,a_1,(\mathcal{c},\mathcal{c})}^{(t)}, (\sigma_{0,a_1,(\mathcal{c},\mathcal{c})}^{{(t)}})^{2})$ and $(p_{1,a_1,g}^{(t)}, \mu_{1,a_1,(\mathcal{c},\mathcal{c})}^{(t)}, (\sigma_{1,a_1,(\mathcal{c},\mathcal{c})}^{{(t)}})^{2})$. 
		If $G_{ij}^{(t+1)} \in \{(\mathcal{a},\mathcal{a}),(\mathcal{n},\mathcal{n}),(\mathcal{n},\mathcal{a})\}$, we can invoke Assumption \ref{asmp:exclusion} and only need to impute either $Y_{ij}(0,a_0)$ or $Y_{ij}(0,a_1)$, depending on the observed treatment assignment for the unit. If $A_j^{\mathrm{o}} = a_0$, we impute $Y_{ij}(0,a_1)$, otherwise, we impute $Y_{ij}(0,a_0)$.
		If $G_{i,j}^{(t+1)}=(\mathcal{n},\mathcal{c})$ and  $A_j^{\mathrm{o}} = a_0$, we need to impute two missing potential outcomes $Y_{ij}(0,a_1),Y_{ij}(1,a_1)$. $Y_{i,j}(0,a_0)=Y_{i,j}(1,a_0)$ from Assumption \ref{asmp:exclusion}. If  $(A_j^{\mathrm{o}}, Z_{i,j}^{\mathrm{o}}) = (a_1,0)$, we impute $Y_{ij}(0,a_0)$ and $Y_{ij}(1,a_1)$. If  $(A_j^{\mathrm{o}}, Z_{i,j}^{\mathrm{o}}) = (a_1,1)$, we impute $Y_{ij}(0,a_0)$ and $Y_{ij}(0,a_1)$.  
		If $G_{i,j}^{(t+1)}=(\mathcal{c},\mathcal{a})$ and  $A_j^{\mathrm{o}} = a_1$, we need to impute two missing potential outcomes $Y_{ij}(0,a_0),Y_{ij}(1,a_0)$. $Y_{i,j}(0,a_1)=Y_{i,j}(1,a_1)$ from Assumption \ref{asmp:exclusion}. If  $(A_j^{\mathrm{o}}, Z_{i,j}^{\mathrm{o}}) = (a_0,0)$, we impute $Y_{ij}(1,a_0)$ and $Y_{ij}(0,a_1)$. If  $(A_j^{\mathrm{o}}, Z_{i,j}^{\mathrm{o}}) = (a_0,1)$, we impute $Y_{ij}(0,a_0)$ and $Y_{ij}(0,a_1)$. The same arguments apply to the units with $M_{i,j}^{\mathrm{o}}=1$ except that their potential outcomes are completely missing. Therefore, we need to impute all of four potential outcomes, $Y_{ij}(0,a_0)$, $Y_{ij}(1,a_0)$, $Y_{ij}(0,a_1)$ and $Y_{ij}(1,a_1)$,  for the units.
		
		Finally, in Step 2(d), we use conjugate prior distributions for all parameters. Had we observed the full compliance types of the units, the resulting complete-data likelihood is 
		\begin{equation*}
		\begin{split}
		&\mathcal{L}_{comp}(\boldsymbol{\psi},\boldsymbol{\phi},\boldsymbol{\theta} \:|\: \mathbf{A}^{\mathrm{o}},\mathbf{Z}^{\mathrm{o}},\mathbf{D}^{\mathrm{o}},\mathbf{M}^{\mathrm{o}},\mathbf{Y}^{\mathrm{obs}}, \mathbf{G}) \\
		= &
		\prod_{(i,j):G_{i,j}=(\mathcal{c},\mathcal{c}), Z_{i,j}=0, A_j=a_0,M_{i,j}=0} (1-\rho_{i,j}^{(\mathcal{c},\mathcal{c}),a_0,0})f_{i,j}^{(\mathcal{c},\mathcal{c}),a_0,0}
		\prod_{(i,j):G_{i,j}=(\mathcal{c},\mathcal{c}), Z_{i,j}=1, A_j=a_0,M_{i,j}=0} (1-\rho_{i,j}^{(\mathcal{c},\mathcal{c}),a_0,1})f_{i,j}^{(\mathcal{c},\mathcal{c}),a_0,1}\\
		&\prod_{(i,j):G_{i,j}=(\mathcal{c},\mathcal{c}), Z_{i,j}=0, A_j=a_1,M_{i,j}=0} (1-\rho_{i,j}^{(\mathcal{c},\mathcal{c}),a_1,0})f_{i,j}^{(\mathcal{c},\mathcal{c}),a_1,0}
		\prod_{(i,j):G_{i,j}=(\mathcal{c},\mathcal{c}), Z_{i,j}=1, A_j=a_1,M_{i,j}=0} (1-\rho_{i,j}^{(\mathcal{c},\mathcal{c}),a_1,1})f_{i,j}^{(\mathcal{c},\mathcal{c}),a_1,1}\\
		&\prod_{(i,j):G_{i,j}=(\mathcal{a},\mathcal{a}), A_j=a_0,M_{i,j}=0} (1-\rho_{i,j}^{(\mathcal{a},\mathcal{a}),a_0,0})f_{i,j}^{(\mathcal{a},\mathcal{a}),a_0,0}
		\prod_{(i,j):G_{i,j}=(\mathcal{a},\mathcal{a}), A_j=a_1,M_{i,j}=0} (1-\rho_{i,j}^{(\mathcal{a},\mathcal{a}),a_1,0})f_{i,j}^{(\mathcal{a},\mathcal{a}),a_1,0}\\
		&\prod_{(i,j):G_{i,j}=(\mathcal{n},\mathcal{n}), A_j=a_0,M_{i,j}=0} (1-\rho_{i,j}^{(\mathcal{n},\mathcal{n}),a_0,0})f_{i,j}^{(\mathcal{n},\mathcal{n}),a_0,0}
		\prod_{(i,j):G_{i,j}=(\mathcal{n},\mathcal{n}), A_j=a_1,M_{i,j}=0} (1-\rho_{i,j}^{(\mathcal{n},\mathcal{n}),a_1,0})f_{i,j}^{(\mathcal{n},\mathcal{n}),a_1,0}\\
		&\prod_{(i,j):G_{i,j}=(\mathcal{c},\mathcal{a}), Z_{i,j}=0, A_j=a_0,M_{i,j}=0} (1-\rho_{i,j}^{(\mathcal{c},\mathcal{a}),a_0,0})f_{i,j}^{(\mathcal{c},\mathcal{a}),a_0,0}
		\prod_{(i,j):G_{i,j}=(\mathcal{c},\mathcal{a}), Z_{i,j}=1, A_j=a_0,M_{i,j}=0} (1-\rho_{i,j}^{(\mathcal{c},\mathcal{a}),a_0,1})f_{i,j}^{(\mathcal{c},\mathcal{a}),a_0,1}\\
		&\prod_{(i,j):G_{i,j}=(\mathcal{c},\mathcal{a}), A_j=a_1,M_{i,j}=0} (1-\rho_{i,j}^{(\mathcal{c},\mathcal{a}),a_1,0})f_{i,j}^{(\mathcal{c},\mathcal{a}),a_1,0}
		\prod_{(i,j):G_{i,j}=(\mathcal{n},\mathcal{c}), A_j=a_0,M_{i,j}=0} (1-\rho_{i,j}^{(\mathcal{n},\mathcal{c}),a_0,0})f_{i,j}^{(\mathcal{n},\mathcal{c}),a_0,0}\\
		&\prod_{(i,j):G_{i,j}=(\mathcal{n},\mathcal{c}), Z_{i,j}=0, A_j=a_0,M_{i,j}=0} (1-\rho_{i,j}^{(\mathcal{n},\mathcal{c}),a_1,0})f_{i,j}^{(\mathcal{n},\mathcal{c}),a_1,0}
		\prod_{(i,j):G_{i,j}=(\mathcal{n},\mathcal{c}), Z_{i,j}=1, A_j=a_1,M_{i,j}=0} (1-\rho_{i,j}^{(\mathcal{n},\mathcal{c}),a_1,1})f_{i,j}^{(\mathcal{n},\mathcal{c}),a_1,1}\\
		&\prod_{(i,j):G_{i,j}=(\mathcal{n},\mathcal{a}), A_j=a_0,M_{i,j}=0} (1-\rho_{i,j}^{(\mathcal{n},\mathcal{a}),a_0,0})f_{i,j}^{(\mathcal{n},\mathcal{a}),a_0,0}
		\prod_{(i,j):G_{i,j}=(\mathcal{n},\mathcal{a}), A_j=a_1,M_{i,j}=0} (1-\rho_{i,j}^{(\mathcal{n},\mathcal{a}),a_1,0})f_{i,j}^{(\mathcal{n},\mathcal{a}),a_1,0}\\
		&\prod_{(i,j):G_{i,j}=(\mathcal{c},\mathcal{c}), Z_{i,j}=0, A_j=a_0,M_{i,j}=1} \rho_{i,j}^{(\mathcal{c},\mathcal{c}),a_0,0}
		\prod_{(i,j):G_{i,j}=(\mathcal{c},\mathcal{c}), Z_{i,j}=1, A_j=a_0,M_{i,j}=1} \rho_{i,j}^{(\mathcal{c},\mathcal{c}),a_0,1}\\
		&\prod_{(i,j):G_{i,j}=(\mathcal{c},\mathcal{c}), Z_{i,j}=0, A_j=a_1,M_{i,j}=1} \rho_{i,j}^{(\mathcal{c},\mathcal{c}),a_1,0}
		\prod_{(i,j):G_{i,j}=(\mathcal{c},\mathcal{c}), Z_{i,j}=1, A_j=a_1,M_{i,j}=1} \rho_{i,j}^{(\mathcal{c},\mathcal{c}),a_1,1}\\
		&\prod_{(i,j):G_{i,j}=(\mathcal{a},\mathcal{a}), A_j=a_0,M_{i,j}=1} \rho_{i,j}^{(\mathcal{a},\mathcal{a}),a_0,0}
		\prod_{(i,j):G_{i,j}=(\mathcal{a},\mathcal{a}), A_j=a_1,M_{i,j}=1} \rho_{i,j}^{(\mathcal{a},\mathcal{a}),a_1,0}\\
		&\prod_{(i,j):G_{i,j}=(\mathcal{n},\mathcal{n}), A_j=a_0,M_{i,j}=1} \rho_{i,j}^{(\mathcal{n},\mathcal{n}),a_0,0}
		\prod_{(i,j):G_{i,j}=(\mathcal{n},\mathcal{n}), A_j=a_1,M_{i,j}=1} \rho_{i,j}^{(\mathcal{n},\mathcal{n}),a_1,0}\\
		&\prod_{(i,j):G_{i,j}=(\mathcal{c},\mathcal{a}), Z_{i,j}=0, A_j=a_0,M_{i,j}=1} \rho_{i,j}^{(\mathcal{c},\mathcal{a}),a_0,0}
		\prod_{(i,j):G_{i,j}=(\mathcal{c},\mathcal{a}), Z_{i,j}=1, A_j=a_0,M_{i,j}=1} \rho_{i,j}^{(\mathcal{c},\mathcal{a}),a_0,1}\\
		&\prod_{(i,j):G_{i,j}=(\mathcal{c},\mathcal{a}), A_j=a_1,M_{i,j}=1} \rho_{i,j}^{(\mathcal{c},\mathcal{a}),a_1,0}
		\prod_{(i,j):G_{i,j}=(\mathcal{n},\mathcal{c}), A_j=a_0,M_{i,j}=1} \rho_{i,j}^{(\mathcal{n},\mathcal{c}),a_0,0}\\
		&\prod_{(i,j):G_{i,j}=(\mathcal{n},\mathcal{c}), Z_{i,j}=0, A_j=a_0,M_{i,j}=1} \rho_{i,j}^{(\mathcal{n},\mathcal{c}),a_1,0}
		\prod_{(i,j):G_{i,j}=(\mathcal{n},\mathcal{c}), Z_{i,j}=1, A_j=a_1,M_{i,j}=1} \rho_{i,j}^{(\mathcal{n},\mathcal{c}),a_1,1}\\
		&\prod_{(i,j):G_{i,j}=(\mathcal{n},\mathcal{a}), A_j=a_0,M_{i,j}=1} \rho_{i,j}^{(\mathcal{n},\mathcal{a}),a_0,0}
		\prod_{(i,j):G_{i,j}=(\mathcal{n},\mathcal{a}), A_j=a_1,M_{i,j}=1} \rho_{i,j}^{(\mathcal{n},\mathcal{a}),a_1,0}\\
		&\prod_{(i,j):G_{i,j}=(\mathcal{c},\mathcal{c})}w_{i,j}^{(\mathcal{c},\mathcal{c})}
		\prod_{(i,j):G_{i,j}=(\mathcal{a},\mathcal{a})}w_{i,j}^{(\mathcal{a},\mathcal{a})}
		\prod_{(i,j):G_{i,j}=(\mathcal{n},\mathcal{n})}w_{i,j}^{(\mathcal{n},\mathcal{n})}
		\\
		&
		\prod_{(i,j):G_{i,j}=(\mathcal{c},\mathcal{a})}w_{i,j}^{(\mathcal{c},\mathcal{a})}
		\prod_{(i,j):G_{i,j}=(\mathcal{n},\mathcal{c})}w_{i,j}^{(\mathcal{n},\mathcal{c})}
		\prod_{(i,j):G_{i,j}=(\mathcal{n},\mathcal{a})}w_{i,j}^{(\mathcal{n},\mathcal{a})}\\
		\end{split}
		\end{equation*}
		Now, assuming appropriate conjugate prior distributions for each parameter, we use the corresponding factor of the complete-data likelihood function to derive the posterior update for each parameter.
		
		\noindent\underline{Update of $\boldsymbol{\pi}$: }
		Consider a prior distribution $(\pi_{(\mathcal{c},\mathcal{c})},\pi_{(\mathcal{a},\mathcal{a})},\pi_{(\mathcal{n},\mathcal{n})},\pi_{(\mathcal{c},\mathcal{a})},\pi_{(\mathcal{n},\mathcal{c})},\pi_{(\mathcal{n},\mathcal{a})}) \sim \mathrm{Dirichlet}(\alpha)$ where $\alpha=(1,1,1,1,1,1)$. Then, using the corresponding factor of $\mathcal{L}_{comp}$, we draw from
		$\pi^{(t+1)} \sim  \mathrm{Dirichlet}(N_{(\mathcal{c},\mathcal{c})}^{(t)}+1,N_{(\mathcal{a},\mathcal{a})}^{(t)}+1,N_{(\mathcal{n},\mathcal{n})}^{(t)}+1,N_{(\mathcal{c},\mathcal{a})}^{(t)}+1,N_{(\mathcal{n},\mathcal{c})}^{(t)}+1,N_{(\mathcal{n},\mathcal{a})}^{(t)}+1)
		$
		where $N_g^{(t)}$ is the number of compliance type $g$ at the $t$-th MCMC step.
		
		\noindent\underline{Update of $\boldsymbol{p}$: }
		Consider a prior distribution $p_{g}^{z,a} \sim \mathrm{Beta}(1,1)$.
		Then, using the corresponding factor of $\mathcal{L}_{comp}$, we have the posterior distribution such that, for $z=0,1$ and $a=a_0,a_1$,
		\begin{equation*}
		\begin{split}
		&p_{z,a,(\mathcal{c},\mathcal{c})}^{(t+1)} \sim  \mathrm{Beta}\bigg(K_{0}^{(\mathcal{c},\mathcal{c}),z,a,0}+1,K_{1}^{(\mathcal{c},\mathcal{c}),z,a,0}+1\bigg) \\
		&p_{0,a,(\mathcal{a},\mathcal{a})}^{(t+1)} \sim  \mathrm{Beta}\bigg(K_{0}^{(\mathcal{a},\mathcal{a}),0,a,0}+K_{0}^{(\mathcal{a},\mathcal{a}),1,a,0}+1,K_{1}^{(\mathcal{a},\mathcal{a}),0,a,0}+K_{1}^{(\mathcal{a},\mathcal{a}),1,a,0}+1\bigg) \\
		&p_{0,a,(\mathcal{n},\mathcal{n})}^{(t+1)} \sim  \mathrm{Beta}\bigg(K_{0}^{(\mathcal{n},\mathcal{n}),0,a,0}+K_{0}^{(\mathcal{n},\mathcal{n}),1,a,0}+1,K_{1}^{(\mathcal{n},\mathcal{n}),0,a,0}+K_{1}^{(\mathcal{n},\mathcal{n}),1,a,0}+1\bigg) \\
		&p_{z,a_0,(\mathcal{c},\mathcal{a})}^{(t+1)} \sim  \mathrm{Beta}\bigg(K_{0}^{(\mathcal{c},\mathcal{a}),z,a_0,0}+1,K_{1}^{(\mathcal{c},\mathcal{a}),z,a_0,0}+1\bigg) \\
		&p_{0,a_1,(\mathcal{c},\mathcal{a})}^{(t+1)} \sim  \mathrm{Beta}\bigg(K_{0}^{(\mathcal{c},\mathcal{a}),0,a_1,0}+K_{0}^{(\mathcal{c},\mathcal{a}),1,a_1,0}+1,K_{1}^{(\mathcal{c},\mathcal{a}),0,a_1,0}+K_{1}^{(\mathcal{c},\mathcal{a}),1,a_1,0}+1\bigg) \\
		&p_{0,a_0,(\mathcal{n},\mathcal{c})}^{(t+1)} \sim  \mathrm{Beta}\bigg(K_{0}^{(\mathcal{n},\mathcal{c}),0,a_0,0}+K_{0}^{(\mathcal{n},\mathcal{c}),1,a_0,0}+1,K_{1}^{(\mathcal{n},\mathcal{c}),0,a_0,0}+K_{1}^{(\mathcal{n},\mathcal{c}),1,a_0,0}+1\bigg) \\
		&p_{z,a_1,(\mathcal{n},\mathcal{c})}^{(t+1)} \sim  \mathrm{Beta}\bigg(K_{0}^{(\mathcal{n},\mathcal{c}),z,a_1,0}+1,K_{1}^{(\mathcal{n},\mathcal{c}),z,a_1,0}+1\bigg) \\
		&p_{0,a,(\mathcal{n},\mathcal{a})}^{(t+1)} \sim  \mathrm{Beta}\bigg(K_{0}^{(\mathcal{n},\mathcal{a}),0,a,0}+K_{0}^{(\mathcal{n},\mathcal{a}),1,a,0}+1,K_{1}^{(\mathcal{n},\mathcal{a}),0,a,0}+K_{1}^{(\mathcal{n},\mathcal{a}),1,a,0}+1\bigg) \\
		\end{split}
		\end{equation*}
		
		where 
		\begin{equation*}
		\begin{split}
		&K_{0}^{g,z,a,m} =\sum_{(i,j):G_{i,j}^{(t+1)}=g,Z_{i,j}=z,A_j=a,M_{i,j}=m} \mathbbm{1}(Y_{i,j}^{\mathrm{obs}} = 0) \\
		&K_{1}^{g,z,a,m} =\sum_{(i,j):G_{i,j}^{(t+1)}=g,Z_{i,j}=z,A_j=a,M_{i,j}=m} \mathbbm{1}(Y_{i,j}^{\mathrm{obs}} \neq 0) 
		\end{split}
		\end{equation*}
		Note that, by Assumption \ref{asmp:exclusion}, the number of parameters to be estimated varies with compliance behaviors. This can be seen in the complete-data likelihood as well. That is, if the parameter is not defined due to Assumption \ref{asmp:exclusion}, the corresponding observations are used for the update of the counterpart parameter. For example, $p_{1,a_0,(\mathcal{a},\mathcal{a})}$ is not defined due to Assumption \ref{asmp:exclusion}. In such a case, all the observations with $A_j^{\mathrm{o}}=a_0$ are used for updating $p_{(0,a_0,\mathcal{a},\mathcal{a})}$ regardless of the observed value of $Z_{i,j}^{\mathrm{o}}$.
		
		\noindent\underline{Update of $\boldsymbol{\rho}$: }
		Consider a prior distribution $\rho_{z,a,g} \sim \mathrm{Beta}(1,1)$
		Then, using the corresponding factor of $\mathcal{L}_{comp}$, we have the posterior distribution such that, for $z=0,1$ and $a=a_0,a_1$,
		
		\begin{equation*}
		\begin{split}
		&\rho_{z,a,(\mathcal{c},\mathcal{c})}^{(t+1)} \sim    \mathrm{Beta}\bigg(J_{0}^{(\mathcal{c},\mathcal{c}),z,a}+1,J_{1}^{(\mathcal{c},\mathcal{c}),z,a}+1\bigg) \\
		&\rho_{0,a,(\mathcal{a},\mathcal{a})}^{(t+1)} \sim    \mathrm{Beta}\bigg(J_{0}^{(\mathcal{a},\mathcal{a}),0,a}+J_{0}^{(\mathcal{a},\mathcal{a}),1,a}+1,J_{1}^{(\mathcal{a},\mathcal{a}),0,a}+J_{1}^{(\mathcal{a},\mathcal{a}),1,a}+1\bigg) \\
		&\rho_{0,a,(\mathcal{n},\mathcal{n})}^{(t+1)} \sim    \mathrm{Beta}\bigg(J_{0}^{(\mathcal{n},\mathcal{n}),0,a}+J_{0}^{(\mathcal{n},\mathcal{n}),1,a}+1,J_{1}^{(\mathcal{n},\mathcal{n}),0,a}+J_{1}^{(\mathcal{n},\mathcal{n}),1,a}+1\bigg) \\
		&\rho_{z,a_0,(\mathcal{c},\mathcal{a})}^{(t+1)} \sim  \mathrm{Beta}\bigg(J_{0}^{(\mathcal{c},\mathcal{a}),z,a_0}+1,J_{1}^{(\mathcal{c},\mathcal{a}),z,a_0}+1\bigg) \\
		&\rho_{0,a_1,(\mathcal{c},\mathcal{a})}^{(t+1)} \sim  \mathrm{Beta}\bigg(J_{0}^{(\mathcal{c},\mathcal{a}),0,a_1}+J_{0}^{(\mathcal{c},\mathcal{a}),1,a_1}+1,J_{1}^{(\mathcal{c},\mathcal{a}),0,a_1}+J_{1}^{(\mathcal{c},\mathcal{a}),1,a_1}+1\bigg) \\
		&\rho_{0,a_0,(\mathcal{n},\mathcal{c})}^{(t+1)} \sim  \mathrm{Beta}\bigg(J_{0}^{(\mathcal{n},\mathcal{c}),0,a_0}+J_{0}^{(\mathcal{n},\mathcal{c}),1,a_0}+1,J_{1}^{(\mathcal{n},\mathcal{c}),0,a_0}+J_{1}^{(\mathcal{n},\mathcal{c}),1,a_0}+1\bigg) \\
		&\rho_{z,a_1,(\mathcal{n},\mathcal{c})}^{(t+1)} \sim  \mathrm{Beta}\bigg(J_{0}^{(\mathcal{n},\mathcal{c}),z,a_1}+1,J_{1}^{(\mathcal{n},\mathcal{c}),z,a_1}+1\bigg) \\
		&\rho_{0,a,(\mathcal{n},\mathcal{a})}^{(t+1)} \sim    \mathrm{Beta}\bigg(J_{0}^{(\mathcal{n},\mathcal{a}),0,a}+J_{0}^{(\mathcal{n},\mathcal{a}),1,a}+1,J_{1}^{(\mathcal{n},\mathcal{a}),0,a}+J_{1}^{(\mathcal{n},\mathcal{a}),1,a}+1\bigg) \\
		\end{split}
		\end{equation*}
		
		where 
		\begin{equation*}
		\begin{split}
		&J_{0}^{g,z,a} =\sum_{(i,j):G_{i,j}^{(t+1)}=g,Z_{i,j}=z,A_j=a} \mathbbm{1}(M_{i,j}^{\mathrm{o}} = 0) \\
		&J_{1}^{g,z,a} =\sum_{(i,j):G_{i,j}^{(t+1)}=g,Z_{i,j}=z,A_j=a} \mathbbm{1}(M_{i,j}^{\mathrm{o}} = 1) 
		\end{split}
		\end{equation*}
		
		\noindent\underline{Update of $\boldsymbol{\mu}$ and $\boldsymbol{\sigma}$: }
		We only show the update of $\mu_{0,a_0,(\mathcal{c},\mathcal{c})}$ and $\sigma_{0,a_0,(\mathcal{c},\mathcal{c})}$ but we can follow the same procedures for the other compliance types and observed values. For the notational convenience, $\sigma_{0,a_0,(\mathcal{c},\mathcal{c})}$ denotes the variance parameter. We should note the fact that the number of target units varies with compliance behavior, which we saw in the updates of $p$ and $\rho$. For conjugate priors $\mu_{z,a,g} \sim N(0, \sqrt{5}^2)$ and $\sigma_{z,a,g} \sim IG(0.1, 1)$ for $\forall g,z,a$, we have
		\begin{equation*}
		\begin{split}
		\sigma_{0,a_0,(\mathcal{c},\mathcal{c})}^{(t+1)} &\sim IG\bigg(0.1+S_0, \: 1+S_1\bigg)\\
		\mu_{0,a_0,(\mathcal{c},\mathcal{c})}^{{(t+1)}} &\sim N\bigg(\frac{S_2}{S_0+\sigma_{0,a_0,(\mathcal{c},\mathcal{c})}^{{(t+1)}}/5} , \: \frac{\sigma_{0,a_0,(\mathcal{c},\mathcal{c})}^{{(t+1)}}}{S_0+\sigma_{0,a_0,(\mathcal{c},\mathcal{c})}^{{(t+1)}}/5}\bigg)
		\end{split}
		\end{equation*}
		where
		\begin{equation*}
		\begin{split}
		& S_0 = \sum_{(i,j):G_{i,j}^{(t+1)}=(\mathcal{c},\mathcal{c}),Z_{i,j}=0,A_j=a_0} \mathbbm{1}(Y_{i,j}^{\mathrm{obs}}>0)\\
		&S_1 = \frac{\sum_{(i,j):G_{ij}^{(t+1)}=(\mathcal{c},\mathcal{c}),Z_{i,j}=0,A_j=a_0,Y_{i,j}^{\mathrm{obs}}>0}\bigg(\log Y_{i,j}^{\mathrm{obs}}-\mu_{0,a_0,(\mathcal{c},\mathcal{c})}^{{(t)}}\bigg)^2}{2}\\
		&S_2 = \sum_{(i,j):G_{ij}^{(t+1)}=(\mathcal{c},\mathcal{c}),Z_{i,j}=0,A_j=a_0,Y_{i,j}^{\mathrm{obs}}>0}\log Y_{i,j}^{\mathrm{obs}}\\
		\end{split}
		\end{equation*}

		\section{Simulation Studies for Different Cluster Sizes.}
		\label{sec:simulation_diff_J}
		Table \ref{tab:simulation_J} contains the simulation results for different values of $J$. We simulate for each fixed size $J = 10, 20, 50$, and also for unequal sizes of clusters. The number of experimental units is fixed at $N = 10000$. The unequal-sized clusters are generated as: $100$ clusters with $J = 10$, $50$ clusters with $J = 20$, $40$ clusters with $J = 50$, $20$ clusters with $J = 100$, and $20$ clusters with $J = 20$. The other simulation conditions are the same as in \ref{sec:simulation_studies_overview}. We see that both our methodology and the IJM method are fairly robust to different values of $J$ in the sense that the results do not substantially differ from the ones in Table \ref{tab:simulation_fn_lognormal}.

	\begin{table*}
    \centering
    \caption{Evaluation metrics for our Bayesian methodology versus the method of \citet{Imai2021} (abbreviated as ``IJM'') under the Log-Normal data-generating process and the finite-population perspective for various values of $J$.}
    \begin{adjustbox}{width=13.3cm}
        \begin{tabular}{lrrrrrrrrr}
            \toprule
            & & \multicolumn{2}{c}{Coverage}   & \multicolumn{2}{c}{Bias}   & \multicolumn{2}{c}{MSE}   & \multicolumn{2}{c}{Interval Width} \\
            \cmidrule(lr){3-4} \cmidrule(lr){5-6} \cmidrule(lr){7-8} \cmidrule(lr){9-10} 
            &$J$  & IJM & Bayes & IJM & Bayes & IJM & Bayes & IJM & Bayes \\ \hline
            &10  & 95\% & 98\% & 1.47E+02 & 3.44E+02 & 1.50E+07 & 1.20E+06 & 1.36E+04 & 5.56E+03 \\ 
            $\mathrm{CADE}(0,a_0)$&20 & 93\% & 98\% & 2.01E+02 & 3.77E+02 & 1.71E+07 & 1.44E+06 & 1.37E+04 & 6.00E+03 \\ 
            &50 & 94\% & 99\% & -1.56E+02 & 4.07E+02 & 1.88E+07 & 1.91E+06 & 1.38E+04 & 5.68E+03 \\
            &Unequal & 95\% & 99\% & -1.58E+02 & 3.55E+02 & 1.83E+07 & 9.91E+05 & 1.37E+04 & 5.88E+03 \\
            \hline
            &10  & 91\% & 98\% & -3.04E+02 & 9.19E+02 & 1.21E+08 & 5.06E+06 & 3.39E+04 & 1.07E+04 \\ 
            $\mathrm{CADE}(1,a_1)$&20 & 90\% & 97\% & -1.97E+01 & 1.14E+03 & 1.35E+08 & 7.60E+06 & 3.31E+04& 1.15E+04 \\ 
            &50 & 95\% & 98\% & -1.47E+03 & 7.59E+02 & 9.97E+07 & 3.96E+06 & 3.44E+04 &1.07E+04 \\ 
            &Unequal & 94\% & 97\% & -5.73E+02 & 9.76E+02 & 1.17E+08 & 4.93E+06 & 3.43E+04 & 1.08E+04 \\
            \hline
            &10  & 96\% & 98\% & -9.01E+01 & 1.67E+02 & 3.75E+06 & 3.24E+05 & 6.87E+03 &2.86E+03 \\ 
            $\mathrm{ITT}_{\mathrm{Y},\cdot,\cdot}(a_0)$&20 & 93\% & 98\% & 9.62E+01 & 1.87E+02 & 4.38E+06 & 3.94E+05 & 6.99E+03 & 3.10E+03 \\ 
            &50 & 94\% & 98\% & 8.60E+01 & 2.04E+02 & 4.76E+06 & 5.16E+05 & 6.98E+03 &2.91E+03 \\ 
            &Unequal & 94\% & 99\% & -7.55E+01 & 1.81E+02 & 4.77E+06 & 2.73E+05 & 7.05E+03 & 3.00E+03 \\
            \hline
            & 10  & 91\% & 97\% & -1.37E+02 & 4.57E+02 & 2.45E+07 & 1.15E+06 & 1.53E+04 & 5.24E+03 \\ 
            $\mathrm{ITT}_{\mathrm{Y},\cdot,\cdot}(a_1)$& 20 & 92\% & 97\% & 1.72E+01 & 5.65E+02 & 2.68E+07 & 1.73E+06 &  1.49E+04 &5.39E+03 \\ 
            &50 & 95\% & 97\% & -6.59E+02 & 3.76E+02 & 2.00E+07 & 9.06E+05 & 1.54E+04 &4.99E+03 \\ 
            &Unequal & 95\% & 97\% & -2.14E+02 & 4.77E+02 & 2.33E+07 & 1.05E+06 & 1.56E+04 & 5.09E+03 \\
            \hline
            &10  & 92\% & 94\% & -3.93E-03 & 1.97E-03 & 1.76E-04 & 1.24E-04 &4.78E-02& 4.04E-02 \\ 
            $\mathrm{ITT}_{\mathrm{D},\cdot,\cdot}(a_0)$&20 & 98\% & 97\% & -3.50E-03 & -1.56E-03 & 1.58E-04 & 1.06E-04 & 4.91E-02 &4.09E-02 \\ 
            &50 & 95\% & 95\% & -1.08E-03 & -1.11E-03 & 1.64E-04 & 1.11E-04 & 4.89E-02 &4.04E-02 \\ 
            &Unequal & 95\% & 94\% & -6.67E-03 & 2.85E-04 & 1.88E-03 & 1.18E-04 & 1.70E-01 & 4.10E-02 \\
            \hline
            &10  & 95\% & 95\% & -5.17E-03 & 4.46E-03 & 2.38E-04 & 1.95E-04 & 6.06E-02 & 4.61E-02 \\ 
            $\mathrm{ITT}_{\mathrm{D},\cdot,\cdot}(a_1)$&20 & 94\% & 93\% & -4.51E-03 & 4.82E-03 & 3.20E-04 & 2.49E-04 & 6.24E-02 &4.68E-02\\ 
            &50 & 94\% & 92\% & -5.71E-03 & 2.86E-03 & 3.04E-04 & 2.26E-04 & 6.38E-02 &4.63E-02 \\ 
            &Unequal & 95\% & 95\% & -1.67E-03 & 3.81E-03 & 1.72E-03 & 2.13E-04 & 1.59E-01 & 5.53E-02 \\
            \bottomrule
        \end{tabular}
    \end{adjustbox}
    \label{tab:simulation_J}
\end{table*}

\section{Simulation Studies under the Gamma data-generating Process}
We present an additional simulation study using the Gamma distribution for the outcome model. The outcomes are specified as: $\tilde{Y}_{i,j}(z,a) \sim$ Gamma$(\alpha_{z,a, G_{i,j}},\theta_{z,a, G_{i,j}})$, where $\alpha$ and $\theta$ are the shape and scale parameters with prior distributions $\alpha_{z,a, G_{i,j}}=\tilde{\alpha}_{z,a, G_{i,j}}\mathbbm{1}(\tilde{\alpha}_{z,a, G_{i,j}}>0)$ with $\tilde{\alpha}_{z,a, G_{i,j}} \sim \mathrm{Normal}(1,\sqrt{1000}^2)$ and $\theta_{z,a, G_{i,j}} \sim \mathrm{InverseGamma}(1,100)$ respectively.  This additional scenario serves to check if our approach consistently provides comparable results with the frequentist approach under different data generating processes. Existing frequentist approaches are expected to perform well in this case as the Gamma distribution with this parameterization is not heavily skewed and does not yield many outliers. 

Table \ref{tab:simulation_fn_gamma} summarizes the results of our simulation study in the case of the Gamma distribution. The Bayesian approach is well-calibrated in the sense that it yields intervals that exhibit nearly $95\%$ coverage across all conditions. It also has the same level of bias as the frequentist model, although it does sometime yield slightly larger bias for the same reasons as in the Log-Normal case. The Bayesian approach performs better than the frequentist approach in terms of MSE, even though the Gamma distribution does not produce as many outliers as the Log-Normal distribution.

\begin{table*}
		\centering
		\caption{Evaluation metrics for our Bayesian methodology versus the method of \citet{Imai2021} (abbreviated as ``IJM'') under the Gamma data-generating process and the finite-population perspective.}
		\begin{adjustbox}{width=13.3cm}
			\begin{tabular}{lrrrrrrrrr}
				\toprule
				& & \multicolumn{2}{c}{Coverage}   & \multicolumn{2}{c}{Bias}   & \multicolumn{2}{c}{MSE}   & \multicolumn{2}{c}{Interval Width}  \\
				\cmidrule(lr){3-4} \cmidrule(lr){5-6} \cmidrule(lr){7-8} \cmidrule(lr){9-10}
				&N  & IJM & Bayes & IJM & Bayes & IJM & Bayes & IJM & Bayes \\ \hline
				&5000  & 90\% & 95\% & 3.26E+01 & -1.91E+00 & 3.25E+03 & 1.76E+03 & 1.82E+02 & 1.67E+02\\ 
				$\mathrm{CADE}(0,a_0)$&10000 & 92\% & 96\% & 1.54E+01 & -2.37E+00 & 1.34E+03 & 9.64E+02 & 1.28E+02 & 1.16E+02\\ 
				&50000 & 90\% & 94\% & -6.21E+00 & -1.24E+00 & 2.83E+02 & 1.89E+02 & 5.72E+01 & 5.11E+01\\
				\hline
				&5000  & 94\% & 93\% & -1.32E+01 & 1.25E+01 & 3.69E+03 & 3.18E+03 & 2.38E+02 & 2.24E+02\\ 
				$\mathrm{CADE}(1,a_1)$&10000 & 96\% & 96\% & -3.10E+00 & 7.05E+00 & 1.95E+03 & 1.75E+03 & 1.69E+02 & 1.55E+02 \\ 
				&50000 & 94\% & 93\% & 3.20E+00 & 1.51E+00 & 4.15E+02 & 3.47E+02 & 7.58E+01 & 6.88E+01\\ 
				\hline
				&5000  & 91\% & 95\% & -1.39E+01 & -2.40E+00 & 6.14E+02 & 3.31E+02 & 8.17E+01 & 7.35E+01\\ 
				$\mathrm{ITT}_{\mathrm{Y},\cdot,\cdot}(a_0)$&10000 & 92\% & 94\% & 6.96E+00 & -1.42E+00 & 2.78E+02 & 2.02E+02 & 5.80E+01 & 5.20E+00\\ 
				& 50000 & 90\% & 94\% & -3.46E+00 & 7.52E-01 & 6.25E+01 & 5.98E+01 & 2.58E+01 & 2.31E+01\\ 
				\hline
				&5000  & 93\% & 94\% & 7.80E+00 & 4.43E+00 & 7.18E+02 & 5.85E+02 & 1.02E+02 & 7.35E+01\\ 
				$\mathrm{ITT}_{\mathrm{Y},\cdot,\cdot}(a_1)$& 10000 & 95\% & 96\% & -3.85E+00 & 2.29E+00 & 3.39E+02 & 2.88E+02 &  7.17E+01 & 6.57E+01\\ 
				& 50000 & 95\% & 94\% & 7.12E-01 & 7.51E-01 & 6.87E+01 & 5.98E+01 & 3.23E+01 & 2.94E+01\\ 
				\hline
				&5000  & 97\% & 92\% & -1.86E-04 & -8.26E-03 & 2.82E-04 & 2.94E-04 &7.16E-02& 6.17E-02\\ 
				$\mathrm{ITT}_{\mathrm{D},\cdot,\cdot}(a_0)$&10000 & 93\% & 93\% & 1.17E-03 & -2.51E-03 & 1.78E-04 & 1.62E-04 & 5.13E-02 &4.47E-02\\ 
				&50000 & 91\% & 94\% & -3.92E-03 & -9.61E-04 & 4.98E-05 & 2.90E-05 & 2.26E-02 & 2.04E-02\\ 
				\hline
				&5000  & 95\% & 91\% & 7.29E-03 & -1.93E-03 & 6.66E-04 & 4.93E-04 & 9.25E-02 & 7.62E-02\\ 
				$\mathrm{ITT}_{\mathrm{D},\cdot,\cdot}(a_1)$&10000 & 93\% & 94\% & -7.28E-03 & -1.19E-03 & 3.29E-04 & 2.21E-04 & 6.52E-02 &5.69E-02\\ 
				&50000 & 95\% & 96\% & -1.63E-03 & 4.39E-04 & 5.51E-05 & 4.50E-05 & 2.94E-02 &2.69E-02\\ 
				\bottomrule
			\end{tabular}
		\end{adjustbox}
		\label{tab:simulation_fn_gamma}
	\end{table*}
    
		\section{Parameters for Simulation Studies and Super-population Evaluation}
		\label{sec:sim_robustness}
		\begin{table*}[t]
			\centering
			
			\caption{Parameters for simulation studies for the Log-Normal daga-generating process.}
			
			\label{tab:parameters_simulation}
			
			\begin{adjustbox}{width=13.3cm}
				\begin{tabular}{ l|rrrr }
					
					& $Z_{ij}=0,\;A_j=a_0$ & $Z_{ij}=1,\;A_j=a_0$ & $Z_{ij}=0,\;A_j=a_1$ & $Z_{ij}=1,\;A_j=a_1$ \\
					\hline
					$\mu_{cc}$& $5$ & $7.5$ & $5$ & $7.5$ \\  
					$\mu_{aa}$& $10$ & - & $10$ & - \\  
					$\mu_{nn}$& $3$ & - & $3$ & - \\  
					$\mu_{ca}$& $5$ & $8$ & $10$ & - \\  
					$\mu_{nc}$& $2$ & - & $4$ & $8$ \\  
					$\mu_{na}$& $2$ & - & $10$ & - \\  
					$\sigma_{cc}^2$& $1.5$ & $2.5$ & $1.5$ & $2.5$ \\  
					$\sigma_{aa}^2$& $2$ & - & $2.5$ & - \\  
					$\sigma_{nn}^2$& $2$ & - & $2.5$ & - \\  
					$\sigma_{ca}^2$& $1.5$ & $1.5$ & $2.5$ & - \\  
					$\sigma_{nc}^2$& $2$ & - & $2.5$ & $2.5$ \\  
					$\sigma_{na}^2$& $1.5$ & - & $1.5$ & - \\  
					$p_{cc}$& $0.1$ & $0.2$ & $0.1$ & $0.2$ \\  
					$p_{aa}$& $0.05$ & - & $0.05$ & - \\  
					$p_{nn}$& $0.03$ & - & $0.03$ & - \\  
					$p_{ca}$& $0.1$ & $0.2$ & $0.1$ & - \\  
					$p_{nc}$& $0.02$ & - & $0.08$ & $0.18$ \\  
					$p_{na}$& $0.04$ & - & $0.06$ & - \\ 
					$\pi_{cc}$&\multicolumn{4}{c}{0.4} \\ 
					$\pi_{aa}$&\multicolumn{4}{c}{0.2} \\ 
					$\pi_{nn}$&\multicolumn{4}{c}{0.2} \\ 
					$\pi_{ca}$&\multicolumn{4}{c}{0.1} \\ 
					$\pi_{nc}$&\multicolumn{4}{c}{0.05} \\ 
					$\pi_{na}$&\multicolumn{4}{c}{0.05} \\ 
				\end{tabular}
			\end{adjustbox}
		\end{table*}
		
		\begin{table*}[t]
			\centering
			
			\caption{Parameters for simulation studies for the Gamma data-generating process.}
			
			\label{tab:parameters_simulation_gamma}
			
			\begin{adjustbox}{width=13.3cm}
				\begin{tabular}{ l|rrrr }
					
					& $Z_{ij}=0,\;A_j=a_0$ & $Z_{ij}=1,\;A_j=a_0$ & $Z_{ij}=0,\;A_j=a_1$ & $Z_{ij}=1,\;A_j=a_1$ \\
					\hline
					$\alpha_{cc}$& $10$ & $12$ & $10$ & $12$ \\  
					$\alpha_{aa}$& $13$ & - & $13$ & - \\  
					$\alpha_{nn}$& $8$ & - & $8$ & - \\  
					$\alpha_{ca}$& $10$ & $11$ & $12$ & - \\  
					$\alpha_{nc}$& $9$ & - & $10$ & $13$ \\  
					$\alpha_{na}$& $8.5$ & - & $13$ & - \\  
					$\theta_{cc}$& $100$ & $125$ & $100$ & $125$ \\  
					$\theta_{aa}$& $100$ & - & $100$ & - \\  
					$\theta_{nn}$& $90$ & - & $90$ & - \\  
					$\theta_{ca}$& $100$ & $100$ & $85$ & - \\  
					$\theta_{nc}$& $90$ & - & $100$ & $125$ \\  
					$\theta_{na}$& $100$ & - & $125$ & - \\  
					$p_{cc}$& $0.1$ & $0.2$ & $0.1$ & $0.2$ \\  
					$p_{aa}$& $0.05$ & - & $0.05$ & - \\  
					$p_{nn}$& $0.05$ & - & $0.05$ & - \\  
					$p_{ca}$& $0.1$ & $0.2$ & $0.1$ & - \\  
					$p_{nc}$& $0.03$ & - & $0.1$ & $0.15$ \\  
					$p_{na}$& $0.04$ & - & $0.08$ & - \\ 
					$\pi_{cc}$&\multicolumn{4}{c}{0.3} \\ 
					$\pi_{aa}$&\multicolumn{4}{c}{0.2} \\ 
					$\pi_{nn}$&\multicolumn{4}{c}{0.17} \\ 
					$\pi_{ca}$&\multicolumn{4}{c}{0.13} \\ 
					$\pi_{nc}$&\multicolumn{4}{c}{0.1} \\ 
					$\pi_{na}$&\multicolumn{4}{c}{0.1} \\ 
				\end{tabular}
			\end{adjustbox}
		\end{table*}
		
		Table \ref{tab:parameters_simulation} presents the parameters for simulation studies. For the frequentist evaluations, we also consider the super-population versions of the causal estimands that we defined in Sections \ref{sec:estimand1} and \ref{sec:estimands2}. This is done to eliminate a source of variation in our simulations corresponding to the different values of the finite-population estimands across the simulated datasets. The specific super-population estimands that we focus in our evaluations are $\mathrm{CADE}_{\mathrm{sp}}(a_0) = \E \left [ Y_{i,j}(1,a_0) - Y_{i,j}(0,a_0) \mid G_{i,j} \in \{(\mathcal{c}, \mathcal{c}), (\mathcal{c}, \mathcal{a}) \} \right ] $, 
		$\mathrm{CADE}_{\mathrm{sp}}(a_1) = \E \left [ Y_{i,j}(1,a_1) - Y_{i,j}(0,a_1) \mid G_{i,j} \in \{(\mathcal{c}, \mathcal{c}), (\mathcal{n}, \mathcal{c})\} \right ]$, 
		$\mathrm{ITT}_{Y, \mathrm{sp}}(a_0) = \E \left [ Y_{i,j}(1,a_0) - Y_{i,j}(0,a_0) \right ]$, 
		$\mathrm{ITT}_{Y, \mathrm{sp}}(a_1) = \E \left [ Y_{i,j}(1,a_1) - Y_{i,j}(0,a_1) \right \}$, 
		$ \mathrm{ITT}_{D, \mathrm{sp}}(a_0) = \E \left \{ D_{i,j}(1,a_0) - D_{i,j}(0,a_0) \right ]$, 
		
		\noindent $\mathrm{ITT}_{D, \mathrm{sp}}(a_1) = \E \left [ D_{i,j}(1,a_1) - D_{i,j}(0,a_1) \right ]$.
		\begin{equation*}
		\begin{split}
		&CADE_{sp}(0,a_0) = \E[Y_{i,j}(1,a_0) - Y_{i,j}(0,a_0) \:|\: G_{i,j} \in \{(\mathcal{c},\mathcal{c}),(\mathcal{c},\mathcal{a})\}] \\
		&= (1-\E[W_{i,j}(1,a_0)\:|\: G_{i,j} \in \{(\mathcal{c},\mathcal{c}),(\mathcal{c},\mathcal{a})\}])\:\E[\tilde{Y}{i,j}(1,a_0)\:|\: G_{i,j} \in \{(\mathcal{c},\mathcal{c}),(\mathcal{c},\mathcal{a})\}] \\
		&- (1-\E[W_{i,j}(0,a_0)\:|\: G_{i,j} \in \{(\mathcal{c},\mathcal{c}),(\mathcal{c},\mathcal{a})\}])\:\E[\tilde{Y}{i,j}(0,a_0)\:|\: G_{i,j} \in \{(\mathcal{c},\mathcal{c}),(\mathcal{c},\mathcal{a})\}] \\
		&= \bigg( \frac{\pi_{(\mathcal{c},\mathcal{c})}}{\pi_{(\mathcal{c},\mathcal{c})} + \pi_{(\mathcal{c},\mathcal{a})}}(1-p_{(\mathcal{c},\mathcal{c})}^{1,0}) + \frac{\pi_{(\mathcal{c},\mathcal{a})}}{\pi_{(\mathcal{c},\mathcal{c})} + \pi_{(\mathcal{c},\mathcal{a})}}(1-p_{(\mathcal{c},\mathcal{a})}^{1,0})\bigg)\\
		& \times \bigg( \frac{\pi_{(\mathcal{c},\mathcal{c})}}{\pi_{(\mathcal{c},\mathcal{c})} + \pi_{(\mathcal{c},\mathcal{a})}}\exp{(\mu_{(\mathcal{c},\mathcal{c})}^{1,0}+}\frac{\sigma_{(\mathcal{c},\mathcal{c})}^{1,0^2}}{2}) + \frac{\pi_{(\mathcal{c},\mathcal{a})}}{\pi_{(\mathcal{c},\mathcal{c})} + \pi_{(\mathcal{c},\mathcal{a})}}\exp{(\mu_{(\mathcal{c},\mathcal{a})}^{1,0}+}\frac{\sigma_{(\mathcal{c},\mathcal{a})}^{1,0^2}}{2})\bigg)\\
		& - \bigg( \frac{\pi_{(\mathcal{c},\mathcal{c})}}{\pi_{(\mathcal{c},\mathcal{c})} + \pi_{(\mathcal{c},\mathcal{a})}}(1-p_{(\mathcal{c},\mathcal{c})}^{0,0}) + \frac{\pi_{(\mathcal{c},\mathcal{a})}}{\pi_{(\mathcal{c},\mathcal{c})} + \pi_{(\mathcal{c},\mathcal{a})}}(1-p_{(\mathcal{c},\mathcal{a})}^{0,0})\bigg) \\
		& \times \bigg( \frac{\pi_{(\mathcal{c},\mathcal{c})}}{\pi_{(\mathcal{c},\mathcal{c})} + \pi_{(\mathcal{c},\mathcal{a})}}\exp{(\mu_{(\mathcal{c},\mathcal{c})}^{0,0}+}\frac{\sigma_{(\mathcal{c},\mathcal{c})}^{0,0^2}}{2}) + \frac{\pi_{(\mathcal{c},\mathcal{a})}}{\pi_{(\mathcal{c},\mathcal{c})} + \pi_{(\mathcal{c},\mathcal{a})}}\exp{(\mu_{(\mathcal{c},\mathcal{a})}^{0,0}+}\frac{\sigma_{(\mathcal{c},\mathcal{a})}^{0,0^2}}{2})\bigg)\\
		= & 4765.78
		\end{split}
		\end{equation*}
		The rest can be calculated in the same way. $\mathrm{CADE}_{sp}(1,a_1) = 5156.41$, $
		\mathrm{ITT}_{Y, \mathrm{sp}}(a_0) = 2382.89$,$
		\mathrm{ITT}_{Y, \mathrm{sp}}(a_1)= 2324.13$, $
		\mathrm{ITT}_{D, \mathrm{sp}}(a_0) = \pi_{cc} + \pi_{ca} = 0.5$, $
		\mathrm{ITT}_{D, \mathrm{sp}}(a_1) = \pi_{cc} + \pi_{nc} = 0.45$. Table \ref{tab:simulation_lognormal_sp} and \ref{tab:simulation_gamma_sp} present the simulation results under the super-population perspective. We see that the results change only slightly from the finite-population evaluation.
		
		\begin{table*}
			\centering
			\caption{Evaluation for the Log-Normal data-generating processes under the super-population perspective.}
			\begin{adjustbox}{width=13.3cm}
				\begin{tabular}{lrrrrrrrrr}
					\toprule
					& & \multicolumn{2}{c}{Coverage}   & \multicolumn{2}{c}{Bias}   & \multicolumn{2}{c}{MSE}   & \multicolumn{2}{c}{Interval Width}  \\
					\cmidrule(lr){3-4} \cmidrule(lr){5-6} \cmidrule(lr){7-8} \cmidrule(lr){9-10} 
					&N  & IJM & Bayes & IJM & Bayes & IJM & Bayes & IJM & Bayes \\ \hline 
					&5000  & 96\% & 98\% & 2.40E+02 & 5.18E+02 & 3.60E+07 & 4.59E+06 & 1.91E+04 & 9.11E+03 \\ 
					$\mathrm{CADE}_{\mathrm{sp}}(0,a_0)$&10000 & 97\% & 98\% & -2.72E+02 & 2.72E+02 & 1.39E+07 & 1.91E+06 & 1.36E+04 & 5.41E+03 \\ 
					&50000 & 94\% & 95\% & -6.51E+01 & 1.68E+02 & 3.58E+06 & 2.72E+05 & 6.61E+04 & 2.09E+03 \\
					\hline
					&5000  & 92\% & 98\% & -8.67E+02 & 1.22E+03 & 2.28E+08 & 5.87E+06 & 4.52E+04 & 1.67E+04\\ 
					$\mathrm{CADE}_{\mathrm{sp}}(1,a_1)$&10000 & 93\% & 98\% & 7.40E+01 & 8.74E+02 & 1.29E+08 & 3.54E+06 & 3.38E+04& 1.13E+04 \\ 
					&50000 & 95\% & 98\% & 3.28E+02 & 3.12E+02 & 2.18E+07 & 7.77E+05 & 1.68E+05 &4.81E+03 \\ 
					\hline
					&5000  & 96\% & 98\% & 1.42E+02 & 2.54E+02 & 9.13E+06 & 1.20E+06 & 9.63E+03 &4.68E+03 \\ 
					$\mathrm{ITT}_{Y, \mathrm{sp}}(a_0)$&10000 & 96\% & 98\% & -1.31E+02 & 1.32E+02 & 3.50E+06 & 4.81E+05 & 6.86E+03 & 2.80E+03 \\ 
					& 50000 & 94\% & 95\% & -3.05E+01 & 8.54E+01 & 9.06E+05 & 7.50E+04 & 3.32E+04 &1.08E+03 \\ 
					\hline
					& 5000  & 92\% & 98\% & -3.41E+02 & 6.01E+02 & 4.62E+07 & 1.38E+06 & 2.04E+04 & 7.76E+03 \\ 
					$\mathrm{ITT}_{Y, \mathrm{sp}}(a_1)$& 10000 & 93\% & 97\% & 6.08E+01 & 4.28E+02 & 2.64E+07 & 8.15E+05 &  1.52E+04 &5.30E+03 \\ 
					& 50000 & 95\% & 98\% & 1.47E+02 & 1.45E+02 & 4.42E+06 & 1.69E+05 & 7.54E+04&2.23E+03 \\ 
					\hline
					&5000  & 94\% & 92\% & -4.17E-04 & 2.68E-03 & 3.20E-04 & 2.67E-04 &6.96E-02& 5.70E-02 \\ 
					$\mathrm{ITT}_{D, \mathrm{sp}}(a_0)$&10000 & 94\% & 89\% & -3.31E-04 & -1.70E-03 & 1.74E-04 & 1.60E-04 & 4.92E-02 &4.06E-02 \\ 
					&50000 & 94\% & 92\% & -2.79E-05 & -8.00E-06 & 3.27E-05 & 2.70E-05 & 2.20E-02 &1.83E-02 \\ 
					\hline
					&5000  & 96\% & 94\% & -1.58E-03 & 4.34E-03 & 4.80E-04 & 4.30E-04 & 9.06E-02 & 7.64E-02 \\ 
					$\mathrm{ITT}_{D, \mathrm{sp}}(a_1)$&10000 & 93\% & 92\% & -7.21E-04 & 3.50E-03 & 2.80E-04 & 2.50E-04 & 6.40E-02 &5.51E-02 \\ 
					&50000 & 96\% & 94\% & -1.34E-04 & 5.48E-04 & 4.74E-05 & 4.30E-05 & 2.86E-02 &2.53E-02 \\ 
					\bottomrule
				\end{tabular}
				
			\end{adjustbox}
			\label{tab:simulation_lognormal_sp}
		\end{table*}
		
		\begin{table*}
			\centering
			\caption{Evaluation for the Gamma data-generating processes under the super-population perspective.}
			
			\begin{adjustbox}{width=13.3cm}
				\begin{tabular}{lrrrrrrrrr}
					\toprule
					& & \multicolumn{2}{c}{Coverage}   & \multicolumn{2}{c}{Bias}   & \multicolumn{2}{c}{MSE}   & \multicolumn{2}{c}{Interval Width}  \\
					\cmidrule(lr){3-4} \cmidrule(lr){5-6} \cmidrule(lr){7-8} \cmidrule(lr){9-10} 
					&N  & IJM & Bayes & IJM & Bayes & IJM & Bayes & IJM & Bayes \\ \hline 
					&5000   & 94\% & 93\% & 2.90E+00 & -9.78E-01 & 2.20E+03 & 2.25E+03 & 1.82E+02 & 1.67E+02\\ 
					$\mathrm{CADE}_{\mathrm{sp}}(0,a_0)$&10000 & 96\% & 94\% & -1.15E+00 & -3.24E+00 & 1.10E+03 & 1.12E+03 & 1.28E+02 & 1.16E+02\\ 
					&50000 & 92\% & 91\% & -8.56E-01 & -1.23E+00 & 2.45E+02 & 2.40E+02 & 5.72E+01 & 5.11E+01\\
					\hline
					&5000  & 96\% & 92\% & -1.15E+00 & 1.15E+01 & 3.51E+03 & 3.49E+03 & 2.38E+02 & 2.24E+02\\ 
					$\mathrm{CADE}_{\mathrm{sp}}(1,a_1)$&10000 & 95\% & 94\% & 2.90E+00 & 8.71E+00 & 1.95E+03 & 1.91E+03 & 1.69E+02 & 1.55E+02 \\ 
					&50000 & 95\% & 91\% & 1.82E+00 & 3.29E+00 & 4.08E+02 & 4.15E+02 & 7.58E+01 & 6.88E+01\\ 
					\hline
					&5000 & 95\% & 93\% & 6.15E-01 & -2.13E+00 & 4.22E+02 & 4.21E+02 & 8.17E+01 & 7.35E+01\\ 
					$\mathrm{ITT}_{Y, \mathrm{sp}}(a_0)$& 10000 & 94\% & 92\% & -2.50E-01 & -1.78E+00 & 2.29E+02 & 2.30E+02 & 5.80E+01 & 5.20E+00\\ 
					& 50000 & 92\% & 90\% & -3.94E-01 & -6.83E-01 & 5.07E+01 & 4.99E+01 & 2.58E+01 & 2.31E+01\\ 
					\hline
					& 5000 & 95\% & 93\% & -1.17E+00 & 3.31E+00 & 6.58E+02 & 6.34E+02 & 1.02E+02 & 9.30E+01\\ 
					$\mathrm{ITT}_{Y, \mathrm{sp}}(a_1)$& 10000 & 96\% & 94\% & 2.90E-01 & 2.42E+00 & 3.23E+02 & 3.10E+02 &  7.17E+01 & 6.57E+01\\ 
					& 50000 & 96\% & 93\% & 4.64E-01 & 1.08E+00 & 6.84E+01 & 6.96E+01 & 3.23E+01 & 2.94E+01\\ 
					\hline
					&5000 & 97\% & 91\% & -3.34E-03 & 8.90E-03 & 2.93E-04 & 3.35E-04 &7.16E-02& 6.17E-02\\ 
					$\mathrm{ITT}_{D, \mathrm{sp}}(a_0)$&10000 & 93\% & 91\% & 6.71E-04 & -2.41E-03 & 1.77E-04 & 1.77E-04 & 5.13E-02 &4.47E-02\\ 
					&50000 & 93\% & 92\% & -2.36E-04 & -7.88E-04 & 3.45E-05 & 3.60E-05 & 2.26E-02 & 2.04E-02\\ 
					\hline
					&5000 & 94\% & 87\% & -1.12E-03 & 2.61E-03 & 6.14E-04 & 5.86E-04 & 9.25E-02 & 7.62E-02\\ 
					$\mathrm{ITT}_{D, \mathrm{sp}}(a_1)$&10000 & 95\% & 92\% & -9.77E-04 & 1.38E-03 & 2.77E-04 & 2.55E-04 & 6.52E-02 &5.69E-02\\ 
					&50000 & 95\% & 95\% & -6.25E-04 & 6.77E-04 & 5.28E-05 & 5.10E-05 & 2.94E-02 &2.69E-02\\ 
					\bottomrule
				\end{tabular}
			\end{adjustbox}
			\label{tab:simulation_gamma_sp}
		\end{table*}
		
		\section{Parameters for Misspecification and MNAR evaluation.}
		The parameters that we used in the simulation in Section \ref{sec:simulation_misspecification} and \ref{sec:MNAR_evaluation} are provided in Table \ref{tab:params_misspecification} and \ref{tab:params_missingrate}.
			\begin{table*}[t]
			\centering
			
			\caption{Additional parameters for simulation studies under misspecification in Section \ref{sec:simulation_misspecification}}
			
			\label{tab:params_misspecification}
			
			\begin{adjustbox}{width=13.3cm}
				\begin{tabular}{ l|rrrr }
					
					& $Z_{ij}=0,\;A_j=a_0$ & $Z_{ij}=1,\;A_j=a_0$ & $Z_{ij}=0,\;A_j=a_1$ & $Z_{ij}=1,\;A_j=a_1$ \\
					\hline
					$C_{cc}$& $10$ & $15$ & $10$ & $15$ \\  
					$C_{aa}$& $15$ & - & $15$ & - \\  
					$C_{nn}$& $9$ & - & $9$ & - \\  
					$C_{ca}$& $10$ & $15$ & $15$ & - \\  
					$C_{nc}$& $9$ & - & $10$ & $15$ \\  
					$C_{na}$& $9$ & - & $15$ & - \\  
				\end{tabular}
			\end{adjustbox}
		\end{table*}
			
			\begin{table*}[t]
			\centering
			
			\caption{Additional parameters for simulation studies under the MNAR mechanism in Section \ref{sec:MNAR_evaluation}}
			
			\label{tab:params_missingrate}
			
			\begin{adjustbox}{width=13.3cm}
				\begin{tabular}{ l|rrrr }
					
					& $Z_{ij}=0,\;A_j=a_0$ & $Z_{ij}=1,\;A_j=a_0$ & $Z_{ij}=0,\;A_j=a_1$ & $Z_{ij}=1,\;A_j=a_1$ \\
					\hline
					$\rho_{cc}$& $0.1$ & $0.1$ & $0.1$ & $0.1$ \\  
					$\rho_{aa}$& $0.05$ & - & $0.05$ & - \\  
					$\rho_{nn}$& $0.2$ & - & $0.2$ & - \\  
					$\rho_{ca}$& $0.1$ & $0.1$ & $0.05$ & - \\  
					$\rho_{nc}$& $0.2$ & - & $0.1$ & $0.1$ \\  
					$\rho_{na}$& $0.2$ & - & $0.05$ & - \\  
				\end{tabular}
			\end{adjustbox}
		\end{table*}
	
	\section{Additional Simulation Studies under MNAR with various missing rates.}
	Table \ref{tab:simulation_MNAR_different_missingrates} presents additional simulation results under MNAR. This simulation aims to compare the performance of our Bayesian methodology with the IJM method for different missing rates of outcomes. We fix $N=10000$ and $J=100$. The missingness column ``None'' means that we do not have any missing outcomes. ``Moderate'' means that the outcomes are missing based on the probabilities given in Table \ref{tab:params_missingrate}. ``Severe'' corresponds to the case where the outcomes are missing with higher probabilities. The missing probabilities are provided in Table \ref{tab:params_missingrate_severe}.  We see that the higher the missing rate is, the more poorly the IJM method performs. Our method successfully accommodates the underlying missingness mechanism so the results do not change much.

	\begin{table*}[t]
			\centering
			
			\caption{Missing probabilities for the severe case.}
			
			\label{tab:params_missingrate_severe}
			
			\begin{adjustbox}{width=13.3cm}
				\begin{tabular}{ l|rrrr }
					
					& $Z_{ij}=0,\;A_j=a_0$ & $Z_{ij}=1,\;A_j=a_0$ & $Z_{ij}=0,\;A_j=a_1$ & $Z_{ij}=1,\;A_j=a_1$ \\
					\hline
					$\rho_{cc}$& $0.5$ & $0.4$ & $0.5$ & $0.4$ \\  
					$\rho_{aa}$& $0.1$ & - & $0.1$ & - \\  
					$\rho_{nn}$& $0.65$ & - & $0.65$ & - \\  
					$\rho_{ca}$& $0.5$ & $0.4$ & $0.1$ & - \\  
					$\rho_{nc}$& $0.65$ & - & $0.5$ & $0.4$ \\  
					$\rho_{na}$& $0.65$ & - & $0.1$ & - \\  
				\end{tabular}
			\end{adjustbox}
		\end{table*}

	\begin{table*}
			\centering
			\caption{Performance comparison under MNAR with different missing rates.  The IJM is evaluated after applying listwise deletion of rows with missing outcome.}
			\label{tab:simulation_MNAR_different_missingrates}
			\begin{adjustbox}{width=13.3cm}
				\begin{tabular}{lrrrrrrrrr}
					\toprule
					& & \multicolumn{2}{c}{Coverage}   & \multicolumn{2}{c}{Bias}   & \multicolumn{2}{c}{MSE}   & \multicolumn{2}{c}{Interval Width}  \\
					\cmidrule(lr){3-4} \cmidrule(lr){5-6} \cmidrule(lr){7-8} \cmidrule(lr){9-10} 
					&Missingness  & IJM & Bayes & IJM & Bayes & IJM & Bayes & IJM & Bayes \\ \hline 
					&None   & 96\% & 98\% & -2.52E+02 & 2.99E+02 & 1.39E+07 & 1.94E+06 & 1.36E+04 & 5.41E+03 \\ 
					$\mathrm{CADE}(0,a_0)$&Moderate & 98\% & 98\% & -2.52E+02 & 2.02E+02 & 1.58E+07 & 1.12E+06 & 1.48E+04 & 1.41E+04\\ 
					&Severe & 95\% & 93\% & -3.33E+03 & -1.13E+03 & 5.29E+07 & 2.02E+06 & 2.40E+04 & 5.14E+03\\
					\hline
					&None  & 93\% & 97\% & 2.12E+02 & 8.92E+02 & 1.29E+08 & 3.61E+06 & 3.38E+04& 1.13E+04 \\ 
					$\mathrm{CADE}(1,a_1)$&Moderate & 93\% & 97\% & 6.10E+02 & 1.25E+03  & 1.47E+08 & 6.07E+06 & 3.71E+04 & 3.92E+04\\ 
					&Severe & 97\% & 94\% & -4.73E+03 & -5.40E+02 & 4.18E+08 & 2.79E+06 & 6.08E+04 & 6.36E+03\\ 
					\hline
					&None &  96\% & 98\% & -1.21E+02 & 1.46E+02 & 3.49E+06 & 4.86E+05 & 6.86E+03 & 2.80E+03 \\ 
					$\mathrm{ITT}_{Y, \cdot,\cdot}(a_0)$& Moderate & 97\% & 98\% & 1.94E+02 & 9.60E+01 & 3.83E+06 & 3.02E+05 & 7.30E+03 &  7.06E+03\\ 
					&Severe & 89\% & 94\% & -1.83E+03 & -5.75E+02 & 9.69E+06 & 5.27E+05 & 9.26E+03 & 2.63E+03 \\
					\hline
					&None & 94\% & 97\% & 1.18E+02 & 4.41E+02 & 2.64E+07 & 8.34E+05 &  1.52E+04 &5.30E+03 \\ 
					$\mathrm{ITT}_{Y, \cdot,\cdot}(a_1)$& Moderate & 93\% & 97\% & 2.66E+02 & 6.20E+02 & 2.83E+07 & 1.40E+06 & 1.63E+04 & 1.80E+04\\ 
					&Severe & 96\% & 93\% & -2.06E+03 & -2.35E+02 & 5.71E+07 & 6.62E+05 & 2.22E+04 & 3.04E+03\\ 
					\hline
					&None & 93\% & 94\% & -2.26E-04 & -1.47E-03 & 1.74E-04 & 1.21E-04 & 4.92E-02 &4.06E-02 \\ 
					$\mathrm{ITT}_{D, \cdot,\cdot}(a_0)$&Moderate & 72\% & 92\% & -1.71E-02 &  -2.36E-03 & 4.72E-04 &  1.62E-04 & 5.04E-02 &  4.23E-02\\ 
					&Severe & 0\% & 90\% & -1.17E-01 & -1.89E-03 & 1.40E-02 & 1.63E-04 & 5.33E-02 & 4.34E-02\\ 
					\hline
					&None &  93\% & 93\% & -2.34E-03 & 3.76E-03 & 2.85E-04 & 2.24E-04 & 6.40E-02 &5.51E-02\\ 
					$\mathrm{ITT}_{D, \cdot,\cdot}(a_1)$&Moderate & 89\% & 91\% & -1.09E-02 & 5.62E-03 & 4.42E-04 & 2.50E-04 & 6.90E-02  &  5.43E-02\\ 
					&Severe & 6\% & 89\% & -7.96E-02 & -2.89E-03 & 6.89E-03 & 2.70E-04 & 9.08E-02 & 5.69E-02\\ 
					\bottomrule
				\end{tabular}
			\end{adjustbox}
		\end{table*}

	\end{supplement}

\end{document}